\documentclass[a4paper,fleqn,usenatbib]{mnras}

\usepackage[T1]{fontenc}
\usepackage{ae,aecompl}

\usepackage{graphicx}	
\usepackage{caption,xtab,booktabs} 
\usepackage{subfigure} 

\usepackage[dvipsnames]{xcolor}

\title[The 6 -- 9~$\mu$m PAH bands in starburst-dominated galaxies]{Profile comparison of the 6 -- 9~$\mu$m polycyclic aromatic hydrocarbon bands in starburst-dominated galaxies}

\author[Carla M. Canelo et al.]{
Carla M. Canelo,$^{1}$\thanks{E-mail: camcanelo@gmail.com}
Dinalva A. Sales$^{2}$,
Am\^ancio C. S. Fria\c{c}a$^{1}$,
Miriani Pastoriza$^{3}$,
\newauthor Kar\'in Men\'endez-Delmestre$^{4}$
\\
$^{1}$Departamento de Astronomia, Instituto de Astronomia, Geof\'isica e Ci\^encias Atmosf\'ericas, Universidade de S\~ao Paulo, S\~ao Paulo, Brazil\\
$^{2}$Instituto de Matem\'atica, Estat\'istica e F\'isica, Universidade Federal do Rio Grande, Rio Grande do Sul, Brazil \\
$^{3}$Departamento de Astronomia, Instituto de F\'isica, Universidade Federal do Rio Grande do Sul, Rio Grande do Sul, Brazil \\
$^{4}$Observat\'orio do Valongo, Universidade Federal do Rio de Janeiro, Ladeira Pedro Ant\^onio 43, Rio de Janeiro, Brazil, 20080-090
}

\date{Accepted XXX. Received YYY; in original form ZZZ}

\pubyear{2021}

\begin{document}
\label{firstpage}
\pagerange{\pageref{firstpage}--\pageref{lastpage}}
\maketitle

\begin{abstract}
Polycyclic aromatic hydrocarbons (PAHs) are of great astrochemical and astrobiological interest due to their potential to form prebiotic molecules. We analyse the 7.7 and 8.6~$\mu$m PAH bands in 126 predominantly starburst-dominated galaxies extracted from the Spitzer/IRS ATLAS project. Based on the peak positions of these bands, we classify them into the different A, B, and C Peeters' classes, which allows us to address the potential characteristics of the PAH emitting population. We compare this analysis with previous work focused on the 6.2~$\mu$m PAH band for the same sample. For the first time in the literature, this statistical analysis is performed on a  sample of galaxies. In our sample, the 7.7~$\mu$m complex is equally distributed in A and B object's class while the 8.6~$\mu$m band presents more class B sources. Moreover, 39 per cent of the galaxies were distributed into A class  objects for both 6.2 and 7.7~$\mu$m bands and only 18 per cent received the same A classification for the three bands. The ``A A A'' galaxies presented higher temperatures and less dust in their interstellar medium. Considering the redshift range covered by our sample, the distribution of the three bands into the different Peeters' classes reveals a potential cosmological evolution in the molecular nature of the PAHs that dominate the interstellar medium in these galaxies, where B class  objects seem to be more frequent at higher redshifts and, therefore, further studies have to be addressed.

\end{abstract}




\section{Introduction}
\label{sec:intro}

The main reservoir of molecular organic material in space is in the form of polycyclic aromatic hydrocarbons (PAHs) \citep{Eh06}. Their emission in the interstellar medium (ISM) belongs to a molecular class normally referred as the Aromatic Infrared Bands \citep[AIB,][]{job92}, in which other classes of organics and inorganics contribute on a tiny scale to the emitting material \citep{allamandola99}. Due to their high luminosity, the AIBs dominate the mid-infrared (MIR) emission of many objects including those at high redshift \citep{Papovich06,  Teplitz07}. PAHs can be responsible for up to 50 per cent of the MIR luminosity, with  major bands peaking at 3.3, 6.2, 7.7, 8.6, 11.3 and 12.7~$\mu$m \citep{Li04, Smith07}, and being observed in the ISM of galactical and extra-galactic environments \citep[e.g.][]{sales10,sales13,ruschel-dutra14}.

Together with other aromatic macromolecules, they are the most abundant class of molecular species that must have been transported to the planets by comets, meteorites and interplanetary dust deposition \citep{Eh02}. Because of their stable molecular structure, they have been delivered almost intact to planets such as Earth and Mars despite of having been produced in other parts of the Solar System or Galaxy. They are of great astrobiological interest due to their potential to form prebiotic molecules and to have played a fundamental role in the origins of life in the stages preceding the RNA World \citep[PAH World model,][]{Eh06}.  

For instance, the substitution of a carbon for a nitrogen atom creates a polycyclic aromatic nitrogen heterocycle (PANH), which can be a precursor for prebiotic nitrogen heterocycle molecules. \citet{Hud05} suggested that a significant fraction of the nitrogen in the ISM is depleted into PANHs and, moreover, these molecules could be causing the shift in the position of the 6.2~$\mu$m PAH band to sligthly shorter wavelengths. 

\citet{Peeters02} studied the profile variations among the PAH bands in several astrophysical objects. Considering the 6 -- 9~$\mu$m spectral region, their sample could be separated into three different classes -- A, B and C --  depending on the band peak positions. This region is composed by three main features -- a band at 6.2~$\mu$m, a complex of overlapping bands at 7.7~$\mu$m with  two components at 7.6 and 7.8~$\mu$m, and a band at 8.6~$\mu$m \citep{Ricca18}. For the 6.2~$\mu$m band, the profile A peaks at shorter wavelengths compared to B and C profiles. On the other hand, the classes A and B differ in the relative strength of the 7.6 and 7.8~$\mu$m features, which seem to be shifted to 8.2~$\mu$m for class C objects \citep{Tielens08}.

The PAH features are often present in star-forming systems, diminished and modified in high-intensity starbursts and, eventually, disappear in active galactic nuclei (AGN) systems \citep{yan07}. In spite of this, the 11.3~$\mu$m PAH band can be observed in nuclear regions of galaxies as close as dozen parsecs from the AGN and for Seyfert-like AGN luminosities, suggesting a dusty material such as a nuclear tori or discs that allow the survival of PAH molecules in nuclear environments \citep{sales13,Alonso-Herrero14,Alonso-Herrero16, Monfredini19}.
Starburst spectra are dominated by strong emission of these features, not only in the continuum shape \citep{Genzel00} but in the 5 -- 8~$\mu$m spectral range with the 6.2~$\mu$m band and the blue wing of the 7.7~$\mu$m PAH complex \citep{Brandl06} as well. In fact, starburst galaxies and most ULIRGs \citep[Ultra-Luminous Infrared Galaxies, ][]{yan05} present the MIR spectra dominated by dust grain emission and absorption features.

In general, the 6.2, 7.7 and 8.6~$\mu$m band emissions arise from the contribution of ionised PAHs \citep[e.g.][]{Tielens08}. Nevertheless, the variability of the PAH profiles (and their peak positions) in different astrophysical environments have been attributed, for example, to the local physical conditions and to the PAH molecules' size, charge, geometry and heterogeneity \citep[e.g][]{Draine01, Draine07, Smith07, sales13}. Despite displaying considerable diversity, the 6.2 and 7.7~$\mu$m features are produced by  CC vibration modes and are prominently evident even in relatively low-resolution data \citep{Tielens08}. Although these CC modes allow these bands to be connected to each other in some cases, mainly for class A \citep{died04}, the 8.6~$\mu$m band CH vibration modes vary less and the profile variation may not be necessarily connected to that of the others \citep{died04, Tielens08, Candian15}. Despite this, the 6.2, 7.7 and 8.6~$\mu$m bands are tightly correlated \citep{Peeters17} and the classes depend on the type of the source. In special, class A sources are generally linked to interstellar material illuminated by ultraviolet (UV) radiation, X-rays, cosmic rays and shock regions, which include HII regions, reflection nebulae, and the general ISM of the Milky Way and other galaxies \citep[e.g.][]{Tielens08,Shannon19}. Therefore, class A sources are associated with processed dust material, indicating the presence, for instance, of ionised PAHs and heteroatom substitution, such as PANHs. 

In this sense, the analysis of the 6.2, 7.7 and 8.6~$\mu$m bands could reveal an overview of the physical and chemical conditions of the sources, and also the potential presence of PAHNs molecules due to the kind of astrophysical environments. More specifically, the observed class A 6.2~$\mu$m band, for example, has only been well reproduced by an inner carbon replaced by nitrogen within the aromatic rings \citep{Hud05}. \citet{Canelo18} analysed the 6.2~$\mu$m band profile of 155 starburst-dominated galaxies with redshift $\leq$ 2.5 and distributed 67 per cent of the objects into the class A, suggesting a dominance of the PAHN emission in this band. The vibration mode association, specially for the 7.7~$\mu$m complex, could furnish another strategy for deriving the variations of the 6.2~$\mu$m band in an indirect way by analysing if both features present the same classification for the same object. In addition, the advantage of this kind of study provides important insights to the behaviour of PAH molecules through the ISM galaxy evolution in the Universe.   

With this in mind, we here analyse and classify the 7.7 and 8.6~$\mu$m features of 126 galaxies observed with the Spitzer telescope according to the Peeters' classes. We also compare the results with the previous study of the 6.2~$\mu$m band performed by \citet{Canelo18}. We  present here  for  the first time a statistical analysis on PAH profiles based on a sample  of  extragalactic  sources and considering the Peeters' classification. Similar work has only been performed to the NGC~1808 galaxy using high spatial resolution spectroscopy data  \citep{sales13}.


This paper is structured as follows: the selection of our sample is explained in Section \ref{sec:data} and the data analysis performed in the spectroscopic data is described in Section \ref{sec:analysis}. Section \ref{sec:results} discusses the results and Section \ref{sec:conclusion} presents the summary and conclusion. 


\section{Data selection}
\label{sec:data}

Starburst galaxies are ideal targets for PAH studies since they carry different bursts of young stellar populations and, consequently, present strong PAH emission in the MIR spectral wavelengths, particularly in the 6 -- 9~$\mu$m region. Furthermore, in order to continue the analysis of \citet{Canelo18}, the same data sample was considered (hereafter, MIR\_SB sample). It is a sub-sample originally extracted from the ATLAS MIR starburst-dominated galaxies sample of the Spitzer/IRS ATLAS project\footnote{http://www.denebola.org/atlas/} \citet{caballero}. 

The ATLAS project possesses spectra of several types of extragalactic objects, such as Seyfert, radiogalaxies and submillimeter galaxies. The limit set between AGN- and starburst-dominated sources was previously classified by \citet{caballero} and was based on the fraction of a PDR (photo-dissociation region) component at r$_{PDR}$ = 0.15, corresponding to equivalent widths (EW) of EW$_{6.2}$ = 0.2~$\mu$m or EW$_{11.3}$ = 0.2~$\mu$m as an alternative boundary. Although  EW$_{6.2}$ can also be low when starlight dominates the continuum, the diagram with EW$_{6.2}$  and the strength of the 10~$\mu$m silicate feature is a standard tool to distinguish between AGN- and starburst-dominated sources \citep[e.g.][]{Spoon2007,Willett2010}, from which the classification with the r$_{PDR}$ is derived \citep{caballero}.

\begin{figure}
\centering
\includegraphics[scale=0.47]{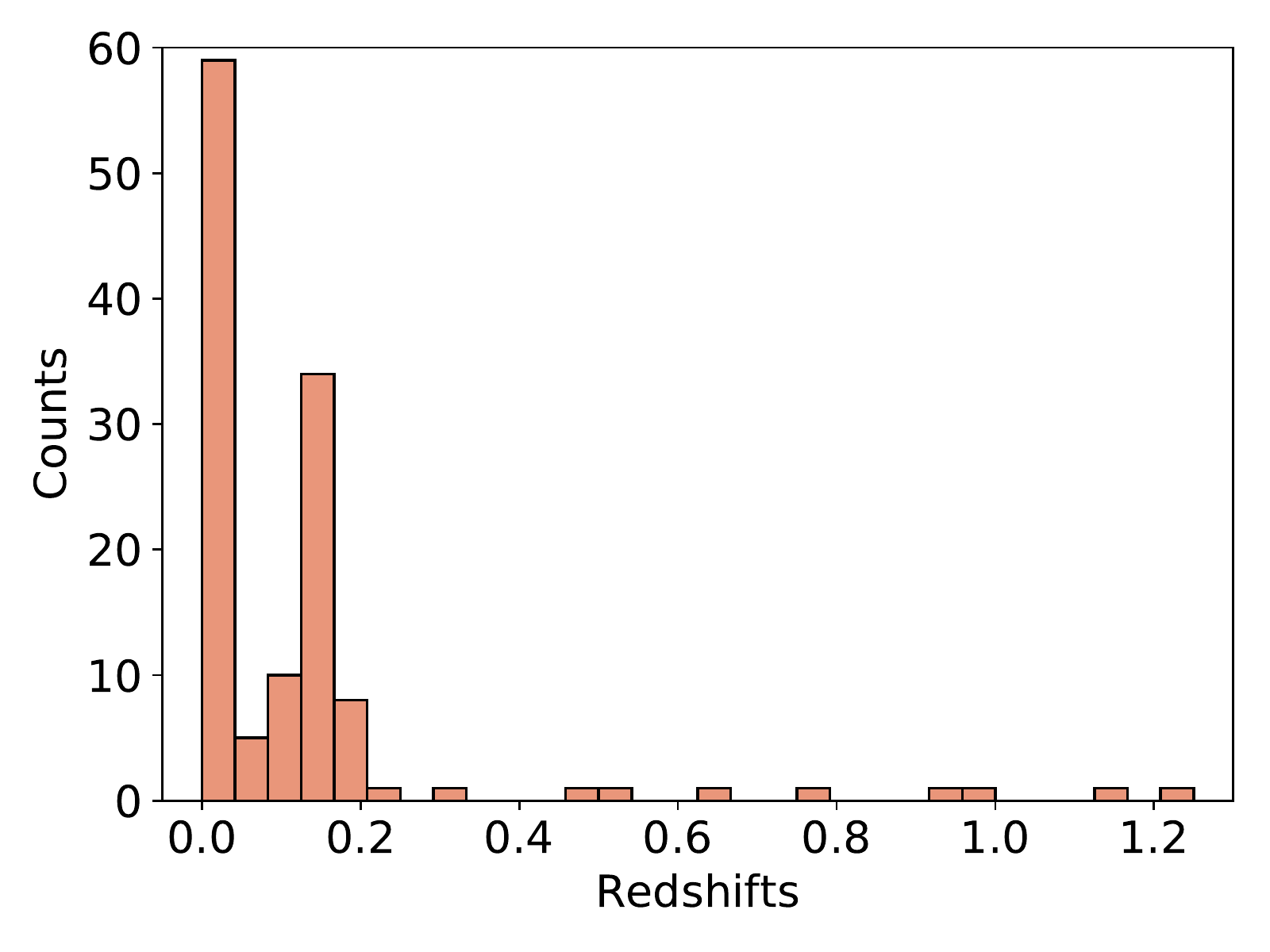}
\caption{Histogram of the 126 redshifts of our galaxy's sample.}
\label{fig:hist1}
\end{figure}

The sources were observed in low resolution by the Infrared Spectrograph \citep[IRS, ][]{Houck04} of the Spitzer Space Telescope \citep{Werner04} and their reduced spectra were extracted from Post-script figures uploaded to the arxiv.org preprint service by their authors. Regardless of the limited accuracy of the recovered wavelength and flux values from the Post-script figures, the resulting introduced uncertainty provides a negligible impact in statistical analysis \citep{caballero}, which is the scope of our work. Most of the galaxies in the ATLAS were observed as point-like sources, including our sample. Although the data are not at the rest wavelength, the MIR\_SB sample have already been corrected with best redshift values supplied by the ATLAS project, which were selected from the literature and checked with NED (NASA Extragalactic Database).

From the 155 sources used by \citet{Canelo18}, 29 objects present  observational uncertainties in the 7--9~$\mu$m region much higher than those of the 6.2~$\mu$m band, probably  due to its complexity, such as the blending bands in the 7.7~$\mu$m complex.  These uncertainties prevent the proper fitting of the 7.7 and 8.6~$\mu$m bands and a reliable analysis and comparison of the results for these objects. These sources were removed from the our original sample and 126 galaxies were studied in this work and their information is available in Table~\ref{tab:sources}. The distribution of the redshifts can be seen in Fig. \ref{fig:hist1}. For more details about the sample and the redshift values used here, see \citet{Canelo18} and references therein.

It is important to mentioned that the ATLAS project contains the tools needed to expand the study of PAHs in starburst-dominated galaxies as a continuation of the statistical analysis performed by \citet{Canelo18}. This work aims to furnish an overview of the PAH properties in these sources. For future studies, with a more detailed analysis and reliable pipeline uncertainties,  we suggest the use of the Cornell Atlas of Spitzer/Infrared Spectrograph Sources \citep[CASSIS,][]{cassis}, for instance.


\section{Data Analysis}
\label{sec:analysis}

\subsection{Continuum subtraction}
\label{sec:continuum}

A first step that must be performed to adequately fit the PAH features in our spectra is the subtraction of the underlying continuum emission.  One approach is to decompose the continuum with a spline and subtract it from the spectrum. \citet{Peeters17} define two distinct spline methods according to the anchor points used in the process. The local spline is determined by anchor points at roughly 5.4, 5.8, 6.6, 7.2, 8.2, 9.0, 9.3, 9.9, 10.2, 10.9, 11.7, 12.1, 13.1, 13.9, 14.7 and 15.0~$\mu$m.  On the other hand, the global spline does not consider 8.2~$\mu$m as an anchor point, leaving a broad emission feature to the 7--9~$\mu$m region.

\begin{figure*}
\centering
\includegraphics[width=.47\textwidth]{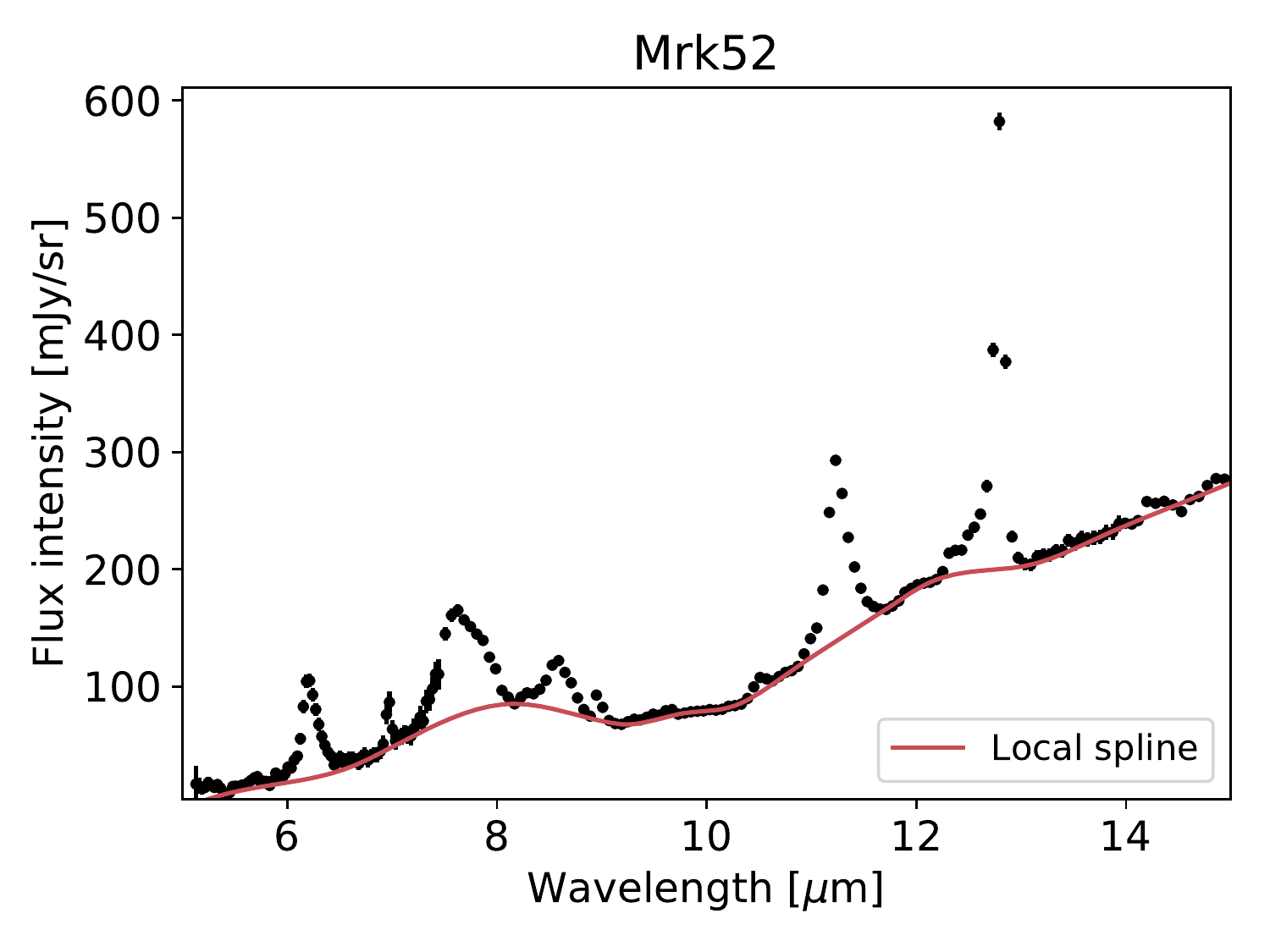}
\includegraphics[width=.47\textwidth]{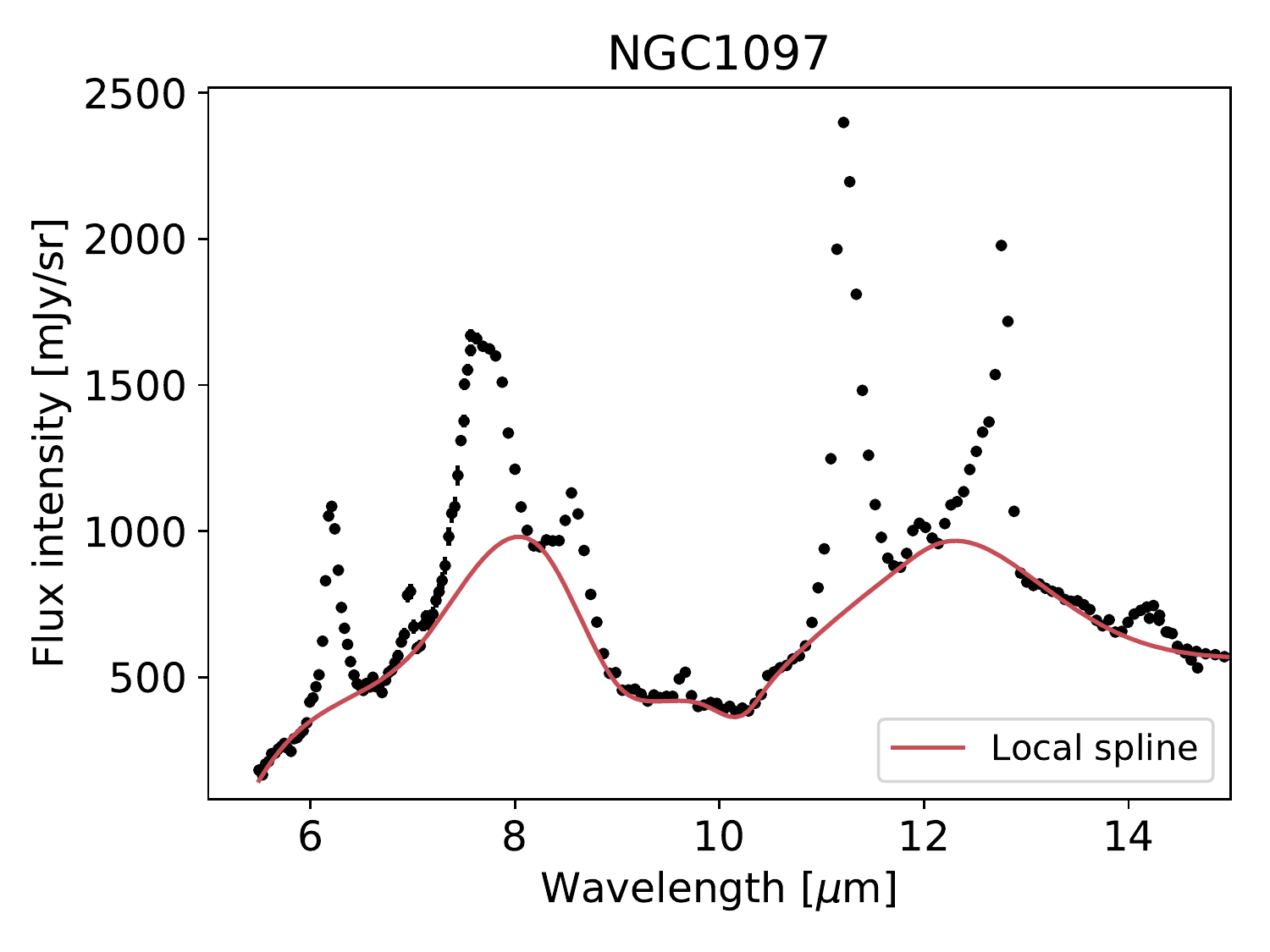}
\includegraphics[width=.47\textwidth]{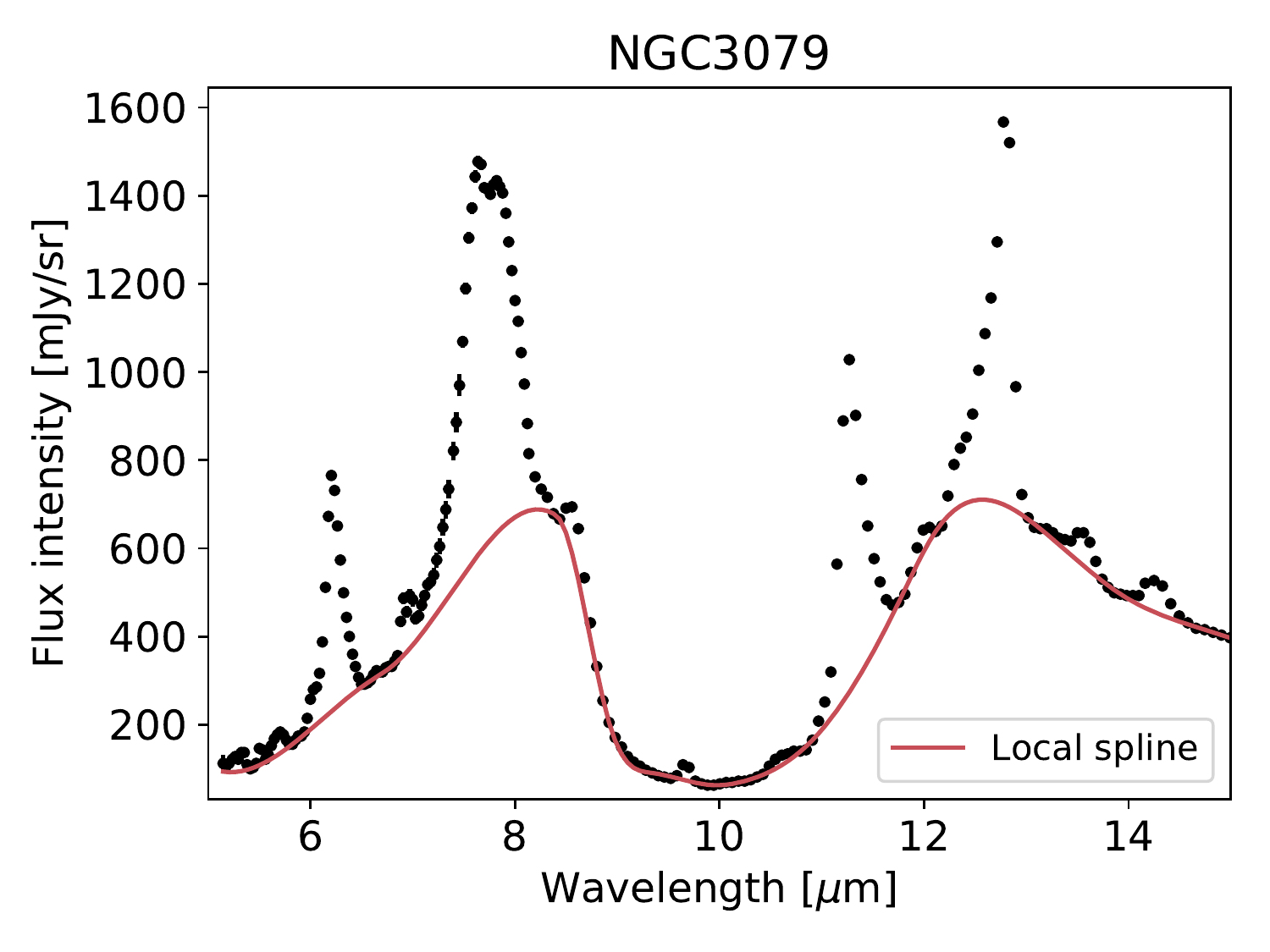}
\includegraphics[width=.47\textwidth]{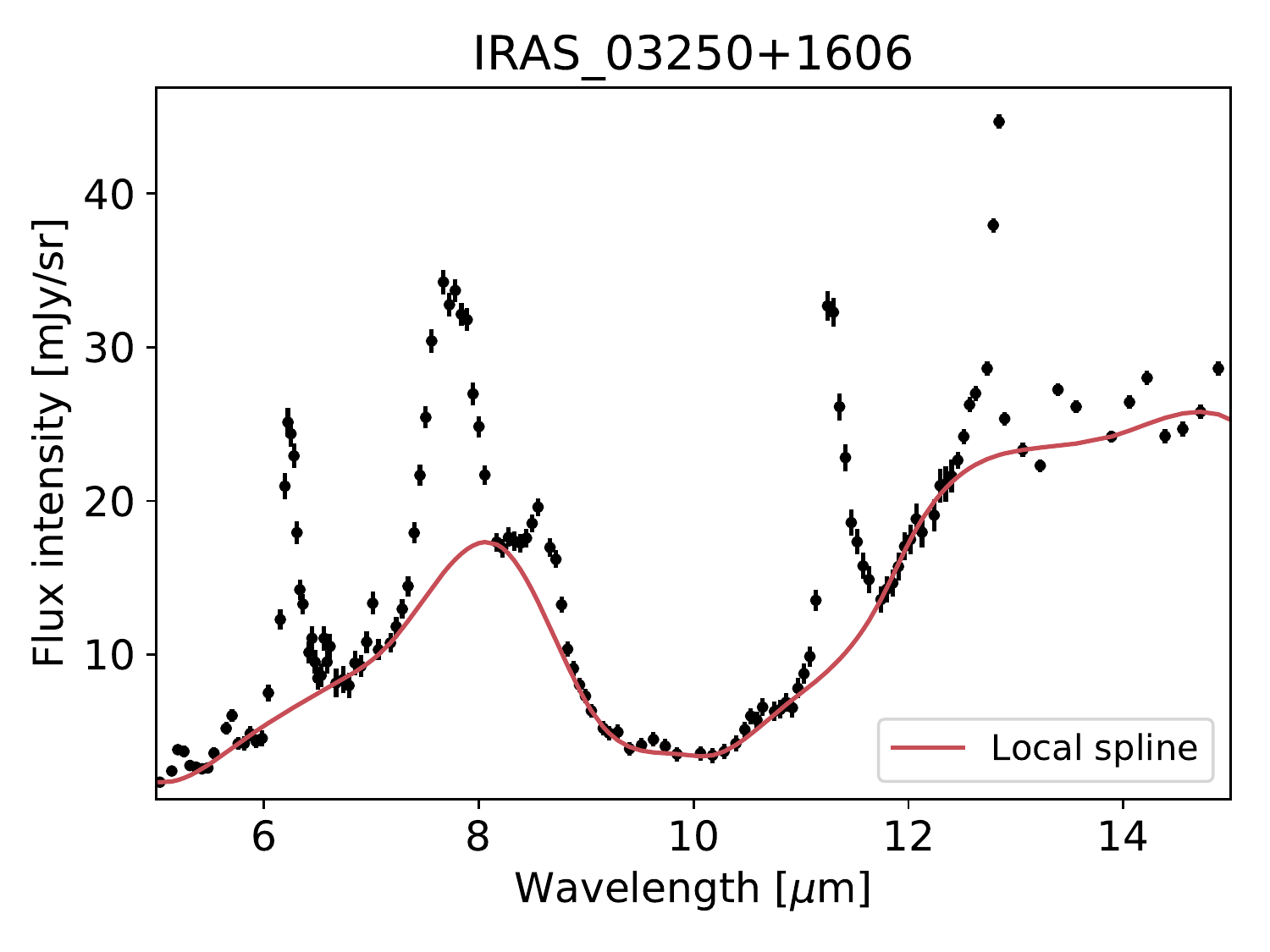}
\caption{Local spline decomposition of the continuum emission represented by the red line for four objects. The data points are represented by the dots with the vertical error-bars as uncertainties.}
\label{fig:splines}
\end{figure*}

Previous works have demonstrated that the overall conclusions on PAH intensity correlations for a large sample of objects are independent of the chosen decomposition approach \citep[e.g.][]{Smith07, Galliano08}. The fit of the 6.2~$\mu$m is independent of the spline decomposition method. However, the chosen spline clearly influences the 7.7~$\mu$m complex intensities \citep{Peeters17}. This influence relies on the 8.2~$\mu$m anchor point that have been previously called of broad emission feature and is located underneath the 7.7 and 8.6~$\mu$m bands. It is prominent in post-AGB (Asymptotic Giant Branch) star spectra, although its remnant could be present in planetary nebula \citep[][and references therein]{Joblin08}. This dust feature is not normally present in star-forming regions but can be very strong in C class objects, such as a few LINER (Low-Ionisation Nuclear Emission-line Region) galaxies \citep{Vega10}.

\citet{Peeters17} considered the 8.2~$\mu$m dust emission as a bump feature in the local spline decomposition and also treated it as a PAH feature in the global spline decomposition, produced by very large PAH molecules with a number of carbon atoms varying from 100 to 150, PAH clusters or very small grains. However, other PAH decomposition methods \citep[such as PAHFIT, ][]{Smith07} do not consider this bump as a PAH feature, instead it is diluted in the wings of the Drude profiles of the PAH bands.  This feature was also reported as a C$^{+}_{60}$ emission detected at a position close to the B star HD 200775 in the NGC7023 reflection nebula \citep{Berne13}.

Both global and local spline method were approached in order to infer which is more feasible for our sample. The anchor points were calculated by finding the minimum value in a small range ($\pm$~0.05~$\mu$m) for each pivot point  \citep[based on ][]{Peeters17} to ensure we are not compromising the PAH fluxes. Each spectrum was analysed individually in case some anchor points needed to be removed,  included or manually changed. This analysis depends mainly on the presence of emission features (ionic, molecular or PAHs) in each spectrum. Moreover, the continuum decomposition in spectral regions up to 6.2~$\mu$m, the 7--9~$\mu$m complex and wavelengths greater than $\approx$~11~$\mu$m present an independent  behaviour. The main sensitive anchor point that can influence the further analysis is at 8.2~$\mu$m. 

In this work, we decided to use previous widely method based on local spline decomposition to focus specifically on the 7.7 and 8.6~$\mu$m PAH bands. This continuum approach allow us to study both bands separately by fitting the PAH features of each region independently of the other \citep{Brandl06}.  Four examples of this method are displayed in Fig. \ref{fig:splines} in which it is possible to see differences in the 8~$\mu$m bump due to the PAH profiles of each galaxy. This procedure is not possible with the global spline because of the blending of the profiles and at least another PAH feature must be considered during the fitting. Therefore, the simultaneous fitting of the four (7.6, 7.8, 8.2 and 8.6~$\mu$m) features after the subtraction of the global spline was also performed. The resulting uncertainties reach up to 70\% in PAH fluxes due to the profile blending, which do not allow for reliable analysis of the PAH bands and their classification simultaneously. This fact reinforces the choice of the local spline method for our study.

Regarding the general anchor points, we allowed a small variation of 0.05 -- 0.1~$\mu$m in their positions in order to choose the minimum values and avoid the unwanted removal of PAH flux. For a sample with 126 sources, random discrepancies in the anchor points are not expected to influence the subsequent analysis. Nevertheless, we varied the spline decomposition of the galaxy NGC~1097 and analysed possible differences in the results. The central wavelength and FWHM (full width at half maximum) remains practically the same with less than 1 per of variation. The amplitude can vary up to 20 per cent for the 6.2 and 8.6 $\mu$m bands. The 7.6 and 7.8 $\mu$m features are less affected and their flux ratio varied up to 5 per cent, which is lower than the respective uncertainty of their ratio. In conclusion, small variations in the anchor points do not interfere with the Peeters' classification. However,  these  variations  may  induce a systematic error in the PAH fluxes and band ratios. Considering  that the impact of these variations would be similar to the entire sample, we note that the specific choice of anchor points does not compromise our analysis and final results. Moreover, despite differences in the PAH fluxes according to the continuum decomposition, the band ratios are independent of the chosen approach \citep[e.g.][]{Galliano08,Peeters17}.

Concerning the extinction, we have not performed a correction for our galaxies once they may present different extinction contributions or even low values. However, some sources do present higher silicate absorption as exemplified by NGC 3079 (Fig. \ref{fig:splines}).  In such cases, the 8.6~$\mu$m band fitting could be compromised if an extinction correction is not applied. \citet{Hensley2020} provided a new constraint on models of interstellar dust in the MIR with an extinction curve that extrapolates smoothly to determinations of the mean Galactic extinction curve \citep{Schlafly2016} at shorter wavelengths and to dust opacities inferred from emission at longer wavelengths. The extinction at longer wavelengths is dominated by the silicate features, specially at  9.7~$\mu$m, which could directly  impact the 8.6~$\mu$m PAH band.

On the other hand, \citet{Hirashita2020}, calculated the evolution of infrared spectral energy distribution (SED) based on a one-zone evolution model of grain size distribution in a galaxy,  and  considering silicate, carbonaceous dust and PAHs. Their results indicate that spatially inhomogeneous dust evolution could be important, and that the emissions from different dust components have different weights for the ISM phases. To better understand this issue, the authors suggest the inclusion of dust evolution models in hydrodynamic simulations, together with the investigation  and modelling of enhanced small-grain production and suppression of PAHs in ionised regions.

These facts illustrate the complexity in simulating the extinction curves and dust emission. A more accurate measurement would require the ability to shape and/or shift the silicate template in order to reproduce the diversity observed within the ATLAS sample \citep{caballero}, which is not the scope of our work. Nevertheless, we selected an spectrum  with obviously low silicate absorption to use as a template to enforce attenuation by applying the extinction curve of \citet{Hensley2020} at various levels. Then, we proceeded with the same data analysis in order to estimate the main effects of the extinction in our sample. The detailed approach is described in Appendix~\ref{sec:ext} and the results are discussed in Section~\ref{sec:results}. 

In addition, \citet{caballero} also performed a silicate analysis and values of the silicate strength (S$_{sil}$) and optical depth at 9.7~$\mu$m ($\tau_{9.7}$) are available in the ATLAS. In this sense,  another way to approach this issue is to compare these values with our results, especially for the 8.6~$\mu$m band, to  evaluate if the absence of an extinction correction influences the final analysis. The main difficulty in estimating the strength of the 9.7~$\mu$m feature is the identification of the underlying continuum, which results in larger uncertainties of the S$_{sil}$ values rather than the $\tau_{9.7}$  values \citep{caballero}. On the other hand, the authors performed a spectral decomposition of the sources by fitting the 5--15~$\mu$m rest-frame range of the spectrum to a parametrised function \citep[see Equation 1,][for more details]{caballero} and considering $\tau(\lambda)$ from the Galactic Centre extinction law \citep{Chiar06}. The $\tau_{9.7}$ parameter, which corresponds to a fixed foreground screen, is calculated using a Levenberg-Marquardt $\chi^2$-minimisation algorithm. Their approach  results in a sensible estimate of the 9.7~$\mu$m optical depth from the wings of the silicate profile and with a tight correlation to the silicate strength. For these reasons, our analysis uses  $\tau_{9.7}$  values instead the silicate strength to compare our results.

\subsection{Gaussian fit of the 7--9~$\mu$m region}
\label{sub-analysis}

To study the PAH profile of starburst-dominated galaxies, we applied the same procedure used for the 6.2~$\mu$m band fitting \citep{Canelo18} in the 7.6, 7.8 and 8.6~$\mu$m features. We constructed a \textsc{python}-based script to estimate their central wavelength, amplitude and  FWHM through the optimisation algorithms from the submodule \textit{scipy.optmize.curve\_fit}. The central wavelength, amplitude and FWHM uncertainties were also derived by this tool with least-squares minimisation from the flux uncertainties provided by the ATLAS. The initial guesses for the parameters were selected from \citet{Smith07}. 

In order to standardise the procedure, we also used Gaussian profiles to fit the features \citep[e.g. ][]{Peeters17, Stoch17,Canelo18}. The H$_2$ emission line at 8.026~$\mu$m, when present, was also subtracted before the PAH fit was performed. There is also the fainter component at 8.33~$\mu$m \citep{Peeters02, Smith07} that is typically negligible when present \citep{Peeters17,Stoch17}. However, some sources of our sample possess this feature and the separation of the fit regions allows us to constrain the 8.6~$\mu$m band to avoid the 8.33~$\mu$m component.

Spectra with local spline decomposition can be separated into two fitting regions -- the 7.7~$\mu$m complex and the 8.6~$\mu$m band. We used three Gaussian profiles to fit the 7.6, 7.8 and 8.6~$\mu$m features, with the latter fitted independently of the others. The fit uncertainties of the complex are expected to be higher than those of the 8.6~$\mu$m band due to the blending of the 7.6 and 7.8~$\mu$m features.  \citet{Stoch17}, for instance, firstly fitted this spectral region with free parameters in order to previously test their fixed-parameter Gaussian decomposition in the HII region W49A. In this situation, their source had its 7.7~$\mu$m complex better fitted with just one component and they decided to constraint the peak positions and widths of the features to recover the 7.6 and 7.7~$\mu$m features. Our goal is to better comprehend a possible variability in the profiles and, therefore, we need to perform a free parameter fit. However, in order to avoid the blending of the features and broader profile's widths than expected by the previous works, we fixed the FWHM values according to \citet{Peeters02} in 0.28~$\mu$m to 7.6~$\mu$m and 0.32~$\mu$m to 7.8~$\mu$m profiles, respectively. 

For the comparison of the PAH features, we also calculated the flux intensities by integration of the fitted Gaussian profiles in the intervals of 6.1 -- 6.35~$\mu$m for the 6.2~$\mu$m band; 7.2 -- 8.2~$\mu$m for the 7.6 and 7.8~$\mu$m components; and 8.2 -- 9~$\mu$m for the 8.6~$\mu$m band. The integrated flux uncertainties were estimated considering the maximum error variation in the fitted Gaussian parameters, in which we altered each fitted parameter up to its  maximum error and then integrated the Gaussian. The difference between both profiles was considered as the error of the flux estimation. 

\begin{table}
	\centering
	\caption{Profile peak positions for each Peeters' classes \citep{Peeters02}. F$_{7.6}$/F$_{7.8}$ represents the flux ratio between 7.6 and 7.8~$\mu$m features and is used to classify the 7.7~$\mu$m complex. }
	\label{tab:classes}
	\begin{tabular}{cccc}
		\hline
		Class & 6.2~$\mu$m & 7.7~$\mu$m & 8.6~$\mu$m\\
		\hline
		A & < 6.23 & $\sim$ 7.6 & < 8.6\\
		 &  & (F$_{7.6}$/F$_{7.8}$ $\geq$ 1) & \\
		\hline
		B & 6.23 < $\lambda$ < 6.29 & $\sim$ 7.8 & > 8.6\\
		 & & (F$_{7.6}$/F$_{7.8}$ < 1) & \\
		\hline
		C & > 6.29 & $\sim$ 8.22 & ------ \\
		\hline
	\end{tabular}
\end{table}

The Gaussian fit results allow us to group the galaxies into the Peeters' classes according to the Table~\ref{tab:classes}. The classification system for the 6.2 and 8.6~$\mu$m bands consists basically in the divergence of the central wavelength bands. The 7.7~$\mu$m complex is comprised of both 7.6 and 7.8~$\mu$m components and class identification relies on their relative strength. This was measured by the flux ratio of the features, hereafter as F$_{7.6}$/F$_{7.8}$. The classification of the complex is based on this bands ratio which will be described in the next sections and used in our analysis instead of the complex peak position's method. The ratio uncertainties were obtained with error propagation derived from  Gaussian  equation  for  normally-distributed  errors. The greater flux indicates the prominent feature of the complex. The decomposition of this complex can be very sensitive to the used method.

Another approach to be tested is to determine the barycenter of the complex by identifying at which wavelength the area under the band is half of the area of the entire complex \citep{Sloan07,Shannon19}. To estimate the barycenter of the complex, the integration was performed for the sum of both Gaussian profiles and in the same range as described before for these features. The Peeters' classification for the sample obtained with this method is only one per cent different from the results of the previous approach. Therefore, we used the flux ratio method in our analysis. 

\begin{figure*}
\centering
\includegraphics[width=.47\textwidth]{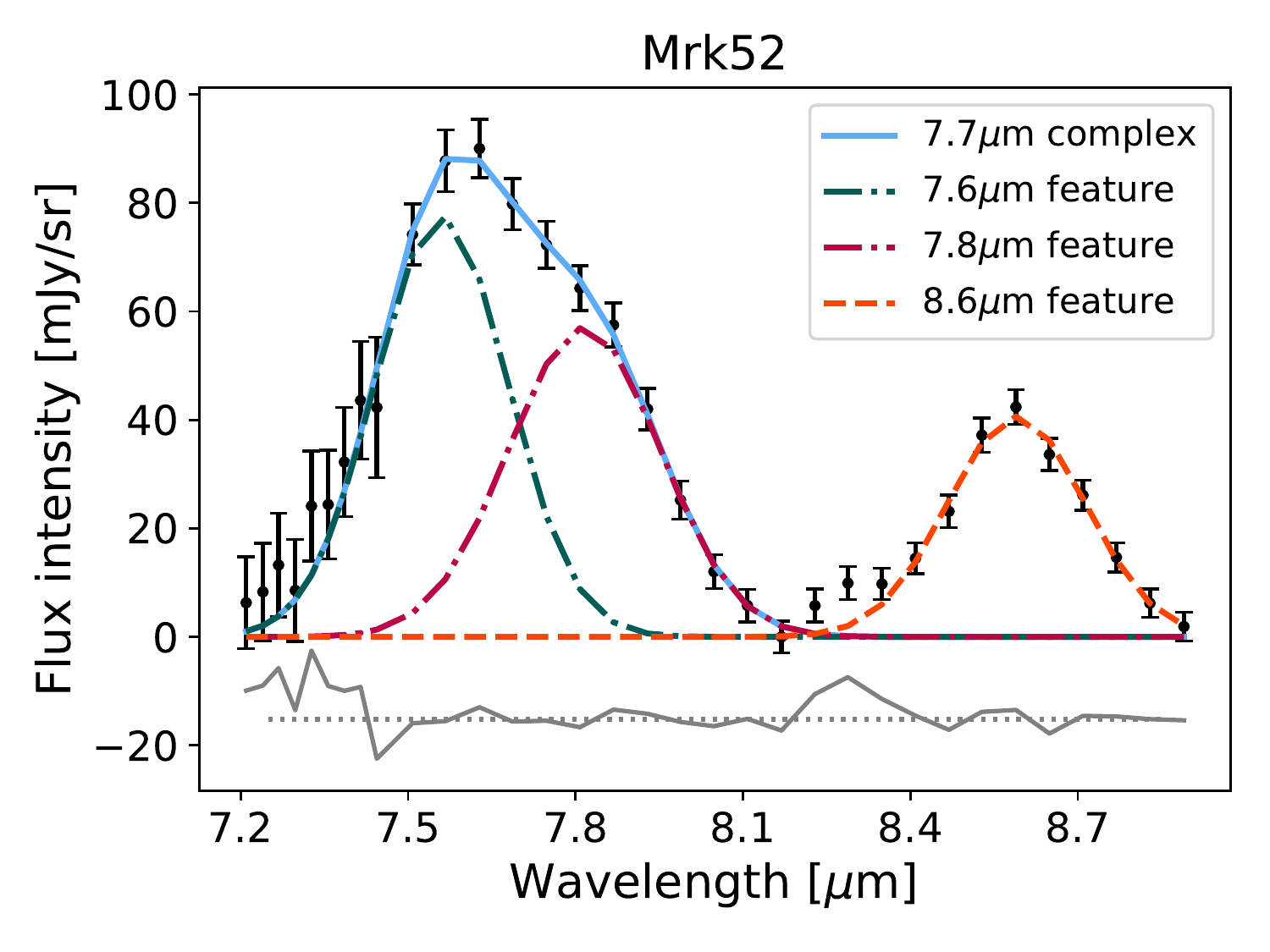}
\includegraphics[width=.47\textwidth]{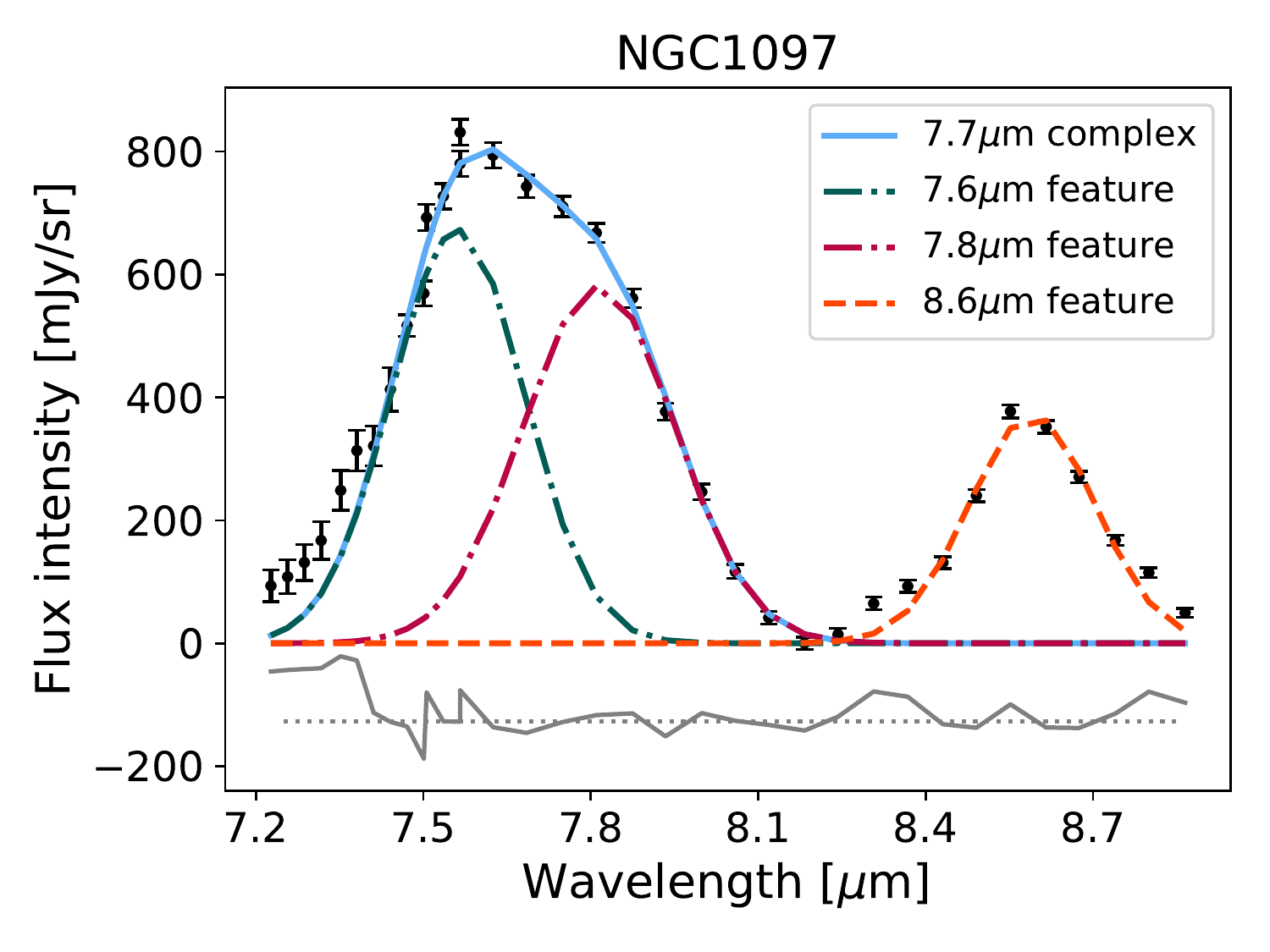}
\includegraphics[width=.47\textwidth]{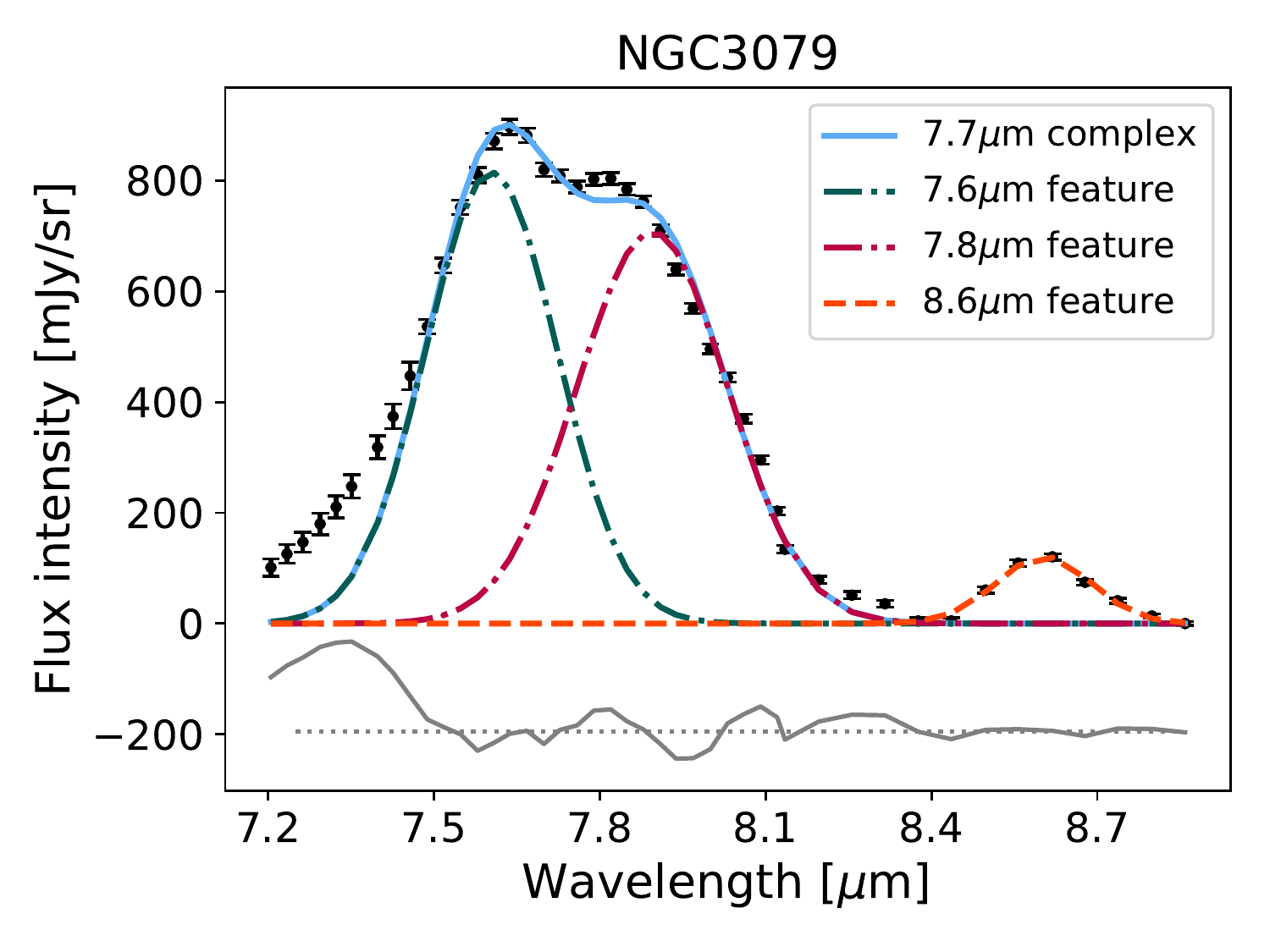}
\includegraphics[width=.47\textwidth]{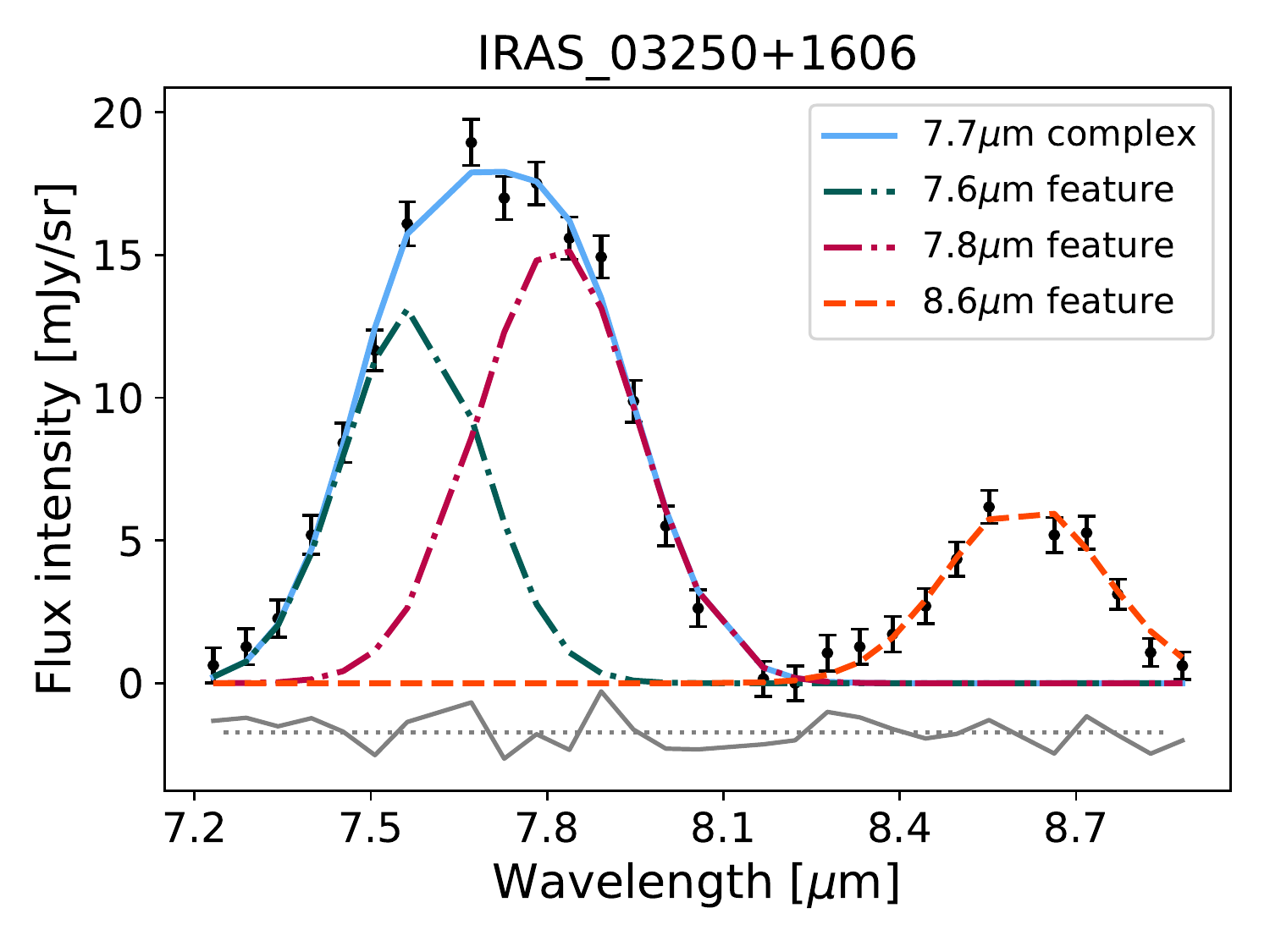}
\caption{Gaussian fit results of four objects. The solid blue line represents the total fit for the 7.7~$\mu$m complex with two Gaussian components respectively identified in the label. The individual Gaussian of the 8.6~$\mu$m band is also shown. The data points are represented by the dots with the vertical error-bars as uncertainties and the solid grey lines represent the fit residuals offset to better visualisation (dotted grey line).}
\label{fig:gauss1}
\end{figure*}

\section{Results and Discussion}
\label{sec:results}

\subsection{The band profiles of starburst-dominated galaxies}

\begin{figure}
\centering
\includegraphics[scale=0.5]{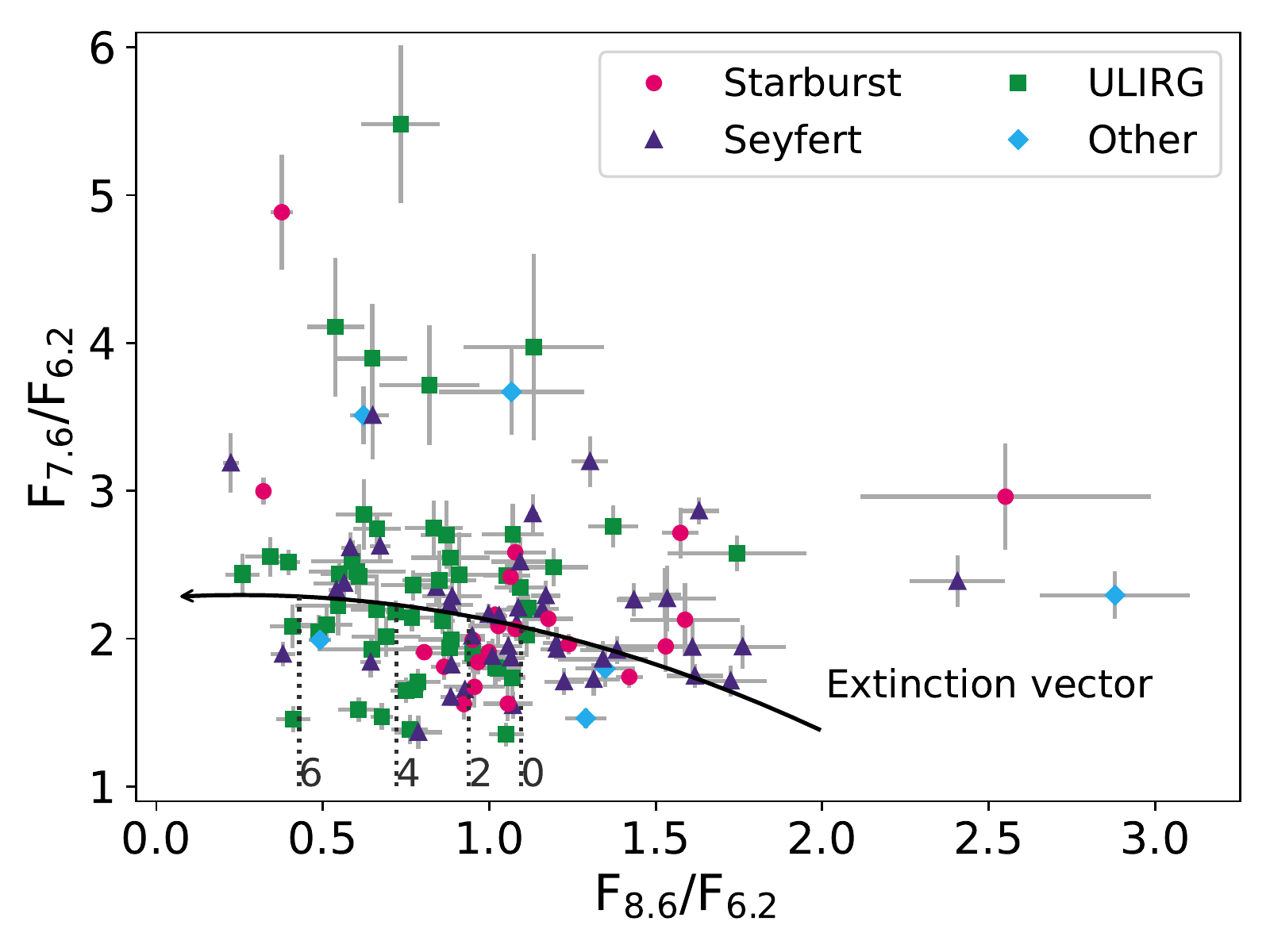}
\caption{Comparison of the flux intensities obtained for each galaxy normalised for the 6.2$\mu$m band. Their types were divided into four main groups with data points represented by the symbols labelled in the legend. The uncertainties are displayed as grey error bars. The theoretical extinction vector, obtained in Appendix~\ref{sec:ext}, is shown with its respective silicate optical depth of $\tau_{9.7}$~= 0, 2, 4 and 6 labelled (dotted line).} 
\label{fig:F62}
\end{figure}

The 6.2~$\mu$m fit results are presented and discussed in \citet{Canelo18}. We compare them here with the statistics obtained for the other two bands. Table \ref{tab:fit7788} contains all the fit results of the 7.7 and 8.6~$\mu$m bands. Although the FWHM values for the 7.6 and 7.8~$\mu$m features were maintained constant along the sources, they failed to adequately reproduce these features in ten galaxies. Smaller values for one or both features needed to be used to accomplish the fitting. They are shown in Table  \ref{tab:fit7788} together with the FWHM values for the 8.6~$\mu$m band that were treated as free parameters for this band.

The same galaxies shown in the previous section with the local spline decomposition (Fig. \ref{fig:splines}) are now presented with the fitted PAH profiles in Fig. \ref{fig:gauss1}. A feature between 7.2 and 7.5~$\mu$m that is not well reproduced by the fitting can be seen in the residual plots. It can be related to the fainter 7.42~$\mu$m PAH emission that belongs to the 7.7~$\mu$m complex \citep{Smith07}. The objects Mrk 52 and NGC 1097 appear to have the 8.33~$\mu$m band, as can be also seen in the residual values. Nevertheless, both features are faint and do not interfere with the analyses of the prominent PAH bands.

In general, the 7.7~$\mu$m features present higher intensities than the 8.6~$\mu$m feature. However, the relative intensities between them vary according to the object as can be seen in Fig. \ref{fig:gauss1}. The starburst galaxy Mrk 52 and NGC 1097 (starburst and Seyfert 1 galaxy) received the same ``A'' classification for the three bands (6.2, 7.7 and 8.6~$\mu$m) and present prominent PAH features, despite of the predominance of the 7.6~$\mu$m component. On the other hand, the starburst galaxy NGC 3079 received the classification of ``A A B'' for the bands, respectively. Finally, the ULIRG + LINER source IRAS\_03250+1606 has its three bands classified as ``B'' object. 

The 8.6~$\mu$m band flux of NGC~3079 could be affected by the extinction that reduces the flux of the band. In comparison to the other galaxies shown if Fig.~\ref{fig:gauss1}, this source presents the lowest relative flux intensity for the 8.6~$\mu$m band and also the deepest silicate absorption at 9.7~$\mu$m. Indeed, if we compare the relative intensities of the 8.6~$\mu$m band of the four presented objects, we note that IRAS\_03250+1606 and NGC~3079 possess lower values than the other two sources. This could also show a strong correlation of band fluxes along the spectra. This well-known correlation in the relative band intensities is expected because of the similar vibration modes (CC and CH) and can also reveal the properties of the emitting PAH population. In fact, variations in the intrinsic relative strength of the CC versus CH modes have been attributed to the effect of ionisation \citep{Tielens08}. For instance, the 6.2, 7.7 and 8.6~$\mu$m bands are generally linked to ionised PAHs in which PAH cations dominate the emission of the 6.2 and 7.6~$\mu$m features \citep{allamandola99,Peeters02, Galliano08}. On the other hand, the 7.8~$\mu$m feature and the 8.6~$\mu$m band may have greater contribution of neutral  and anion PAHs \citep{Ricca12, Peeters17}. 

The size distribution and structure of the PAH population can also contribute to the observed profile variations. \citet[][ and references therein]{Peeters17} attributed the 7.6~$\mu$m emission to compact (and positively charged) PAH with 50--100 carbon atoms while the 7.8~$\mu$m emission to very large PAH with 100--150 carbon atoms or PAH clusters with bay regions or modified duo CH groups. Finally, the 8.6~$\mu$m emission can be attributed to very large, compact and symmetric PAHs with 96--150 carbon atoms. Unfortunately, the relative intensities between the 6.2~$\mu$m band and the 7.7~$\mu$m complex do not track PAH size effectively due to this mixed distribution of sizes throughout this range \citep{Marag18}, although small PAHs emit more strongly at the 6.2 and 7.7~$\mu$m since both features arise from ionised grains \citep[e.g.][]{Draine07}.

\begin{figure}
\centering
\includegraphics[scale=0.5]{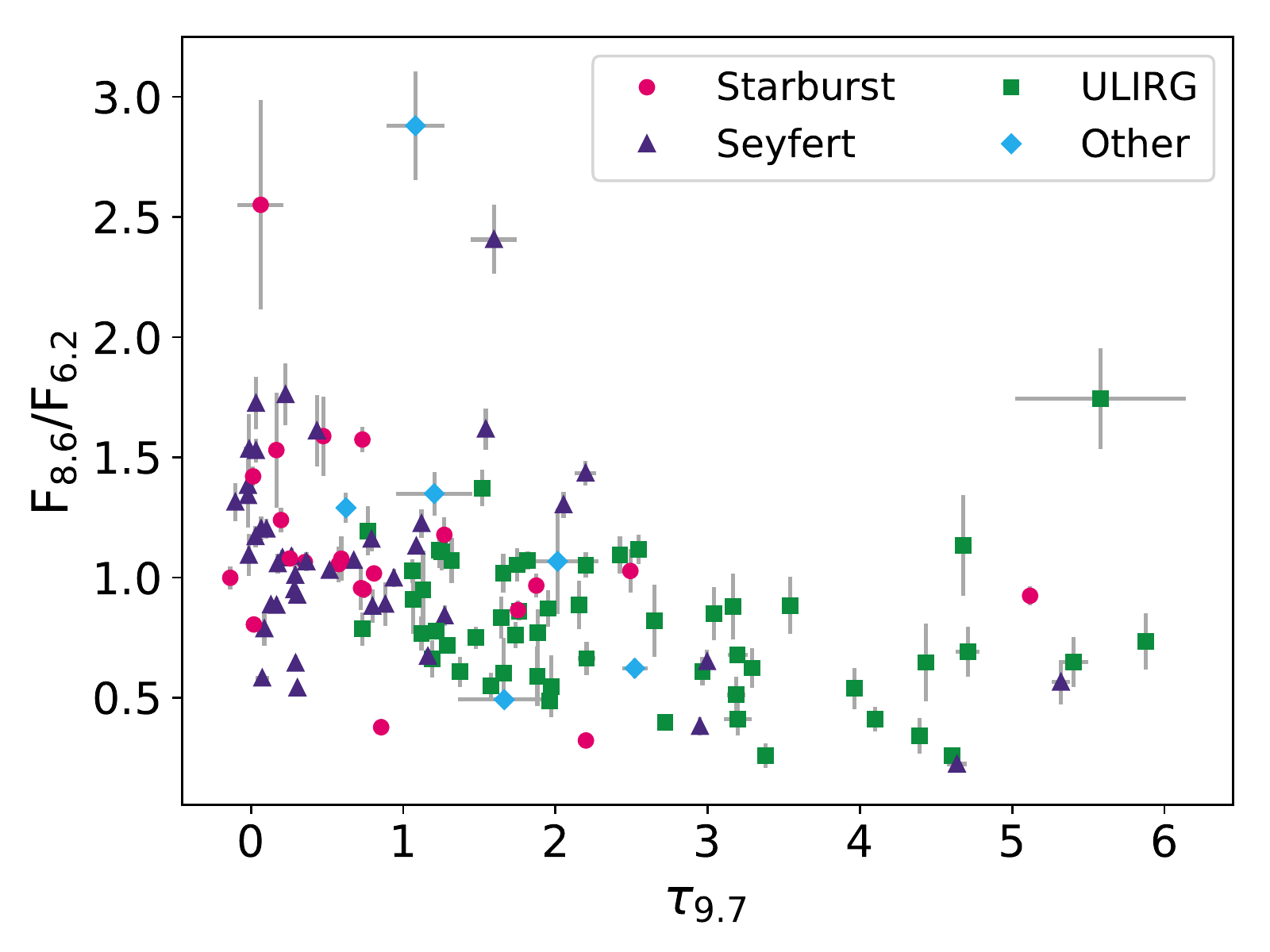}
\includegraphics[scale=0.5]{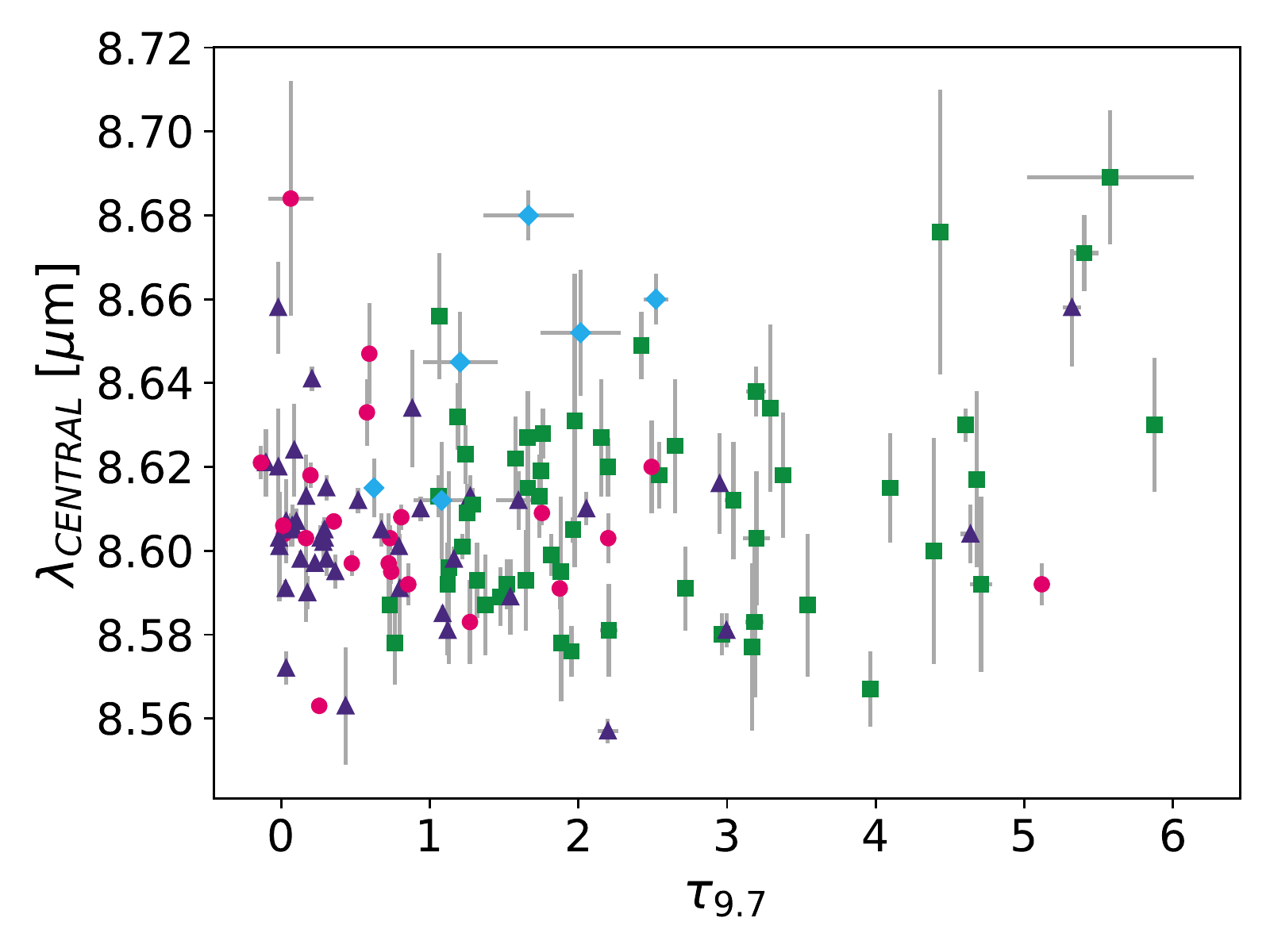}
\caption{Comparison of the flux intensities of F$_{8.6}$/F$_{6.2}$ (top) and central wavelength of the 8.6~$\mu$m band (down) with the optical depths at 9.7~$\mu$m. The type of the sources are labelled in the legend. The uncertainties are displayed as grey error bars.}
\label{fig:tau1}
\end{figure}

In order to verify the well-known correlations between the PAH profiles, Fig.~\ref{fig:F62} compares the flux ratios of F$_{7.6}$/F$_{6.2}$ and F$_{8.6}$/F$_{6.2}$. The band fluxes are available in Table \ref{tab:fluxes}. The different types of objects do not seem to influence the ratios. They are probably connected to physical conditions and the dominant PAH population of the sources. The correlation between these two bands is well established in a variety of environments and within extended sources. \citet{Peeters17} and \citet{Ricca18} presented the ratios for reflection nebulae observations and computed values simulated for an excitation of 8 eV. Some of our results are similar, but we also obtained greater ratios specially for F$_{7.6}$/F$_{6.2}$. According to \citet{Ricca18}, the relative intensity of the 7.7 and 6.2~$\mu$m bands is greater than 1 for PAHs with even number of carbons and less than 1 for odd-carbon PAHs, which we can conclude from our data (Fig.~\ref{fig:F62}) that most of our galaxies may be dominated by even-carbon PAHs.

According to \citet{Galliano08}, the extinction is not expected to explain the majority of the observed band ratio variations. In our results, the effect of extinction is clearly exhibited by the extinction vector in Fig.~\ref{fig:F62} and  has little impact  in the studied wavelength range, corroborating with \citet{Galliano08} claims. We can also see in the figure that ULIRGs are located in a region with lower F$_{8.6}$/F$_{6.2}$ ratio than the other types of galaxies. As a matter of fact, this could be related to the silicate feature at 9.7~$\mu$m that might be suppressing the 8.6~$\mu$m flux and making this band more sensitive to the extinction effects, as  discussed in Appendix~\ref{sec:ext}. To accomplish that, Fig. \ref{fig:tau1} shows the relative intensities of F$_{8.6}$/F$_{6.2}$ compared with $\tau_{9.7}$, which presents a strong correlation with the silicate strength, as discussed in Section~\ref{sec:continuum}.  The ULIRGs present higher $\tau_{9.7}$ values compared to the other galaxies' types, which also correspond to lower F$_{8.6}$/F$_{6.2}$ flux ratio.  It is possible to see in the figure a slight decrease in the flux ratio with higher $\tau_{9.7}$ values,  suggesting that the 8.6~$\mu$m flux may be underestimated due to the silicate strength. We, therefore, investigated the peak position, which is the important parameter to our analysis and is also shown in Figure~\ref{fig:tau1}, and FWHM of the band. They do not seem to be highly influenced by the silicate strength and the extinction, but the band peaks more frequently at wavelengths higher than 8.60~$\mu$m for $\tau_{9.7} \geq$~4, as expected from our analysis of the extinction (see Appendix~\ref{sec:ext} for more details). Nevertheless, just a few sources (a maximum of 10 per cent of our sample) present such high $\tau_{9.7}$ values. We also analysed the spectral contribution of the PDR and AGN (Active Galactic Nuclei) components calculated by \citet{caballero} and our sample presented no correlations.

\subsection{Distribution into the Peeters' classes}
\label{sec:comparison}

In spite of profile variations caused by differences in the astrophysical environments, the classes are normally linked to the type of ionisation source. Class A sources can be associated with interstellar material illuminated by a star, including HII regions, reflection nebulae, and the general ISM of the Milky Way and other galaxies. On the other hand, class B objects can be associated with circumstellar material and include planetary nebula, a variety of post-AGB objects and Herbig AeBe stars. Finally, class C sources are limited to a few extreme carbon-rich post-AGB objects \citep{Peeters02,Tielens08}. 

In the case of our sample, the starburst-dominated emission seems to suppress the possible influence of different type classification of the galaxies. To complement the analysis, the type of the sources were divided into four main groups which comprise the following types: Starburst -- starburst with contribution of HII region, Seyfert and LINER; Seyfert -- Seyferts 1, intermediate, 2 and 3; ULIRG -- ULIRGs and ULIRGs with contribution of HII region and LINER; and Other -- infrared galaxy, Fanaroff-Riley galaxy, LINER, quasi-stellar object and submillimeter galaxy.  

\begin{figure}
    \centering
    \includegraphics[scale=0.5]{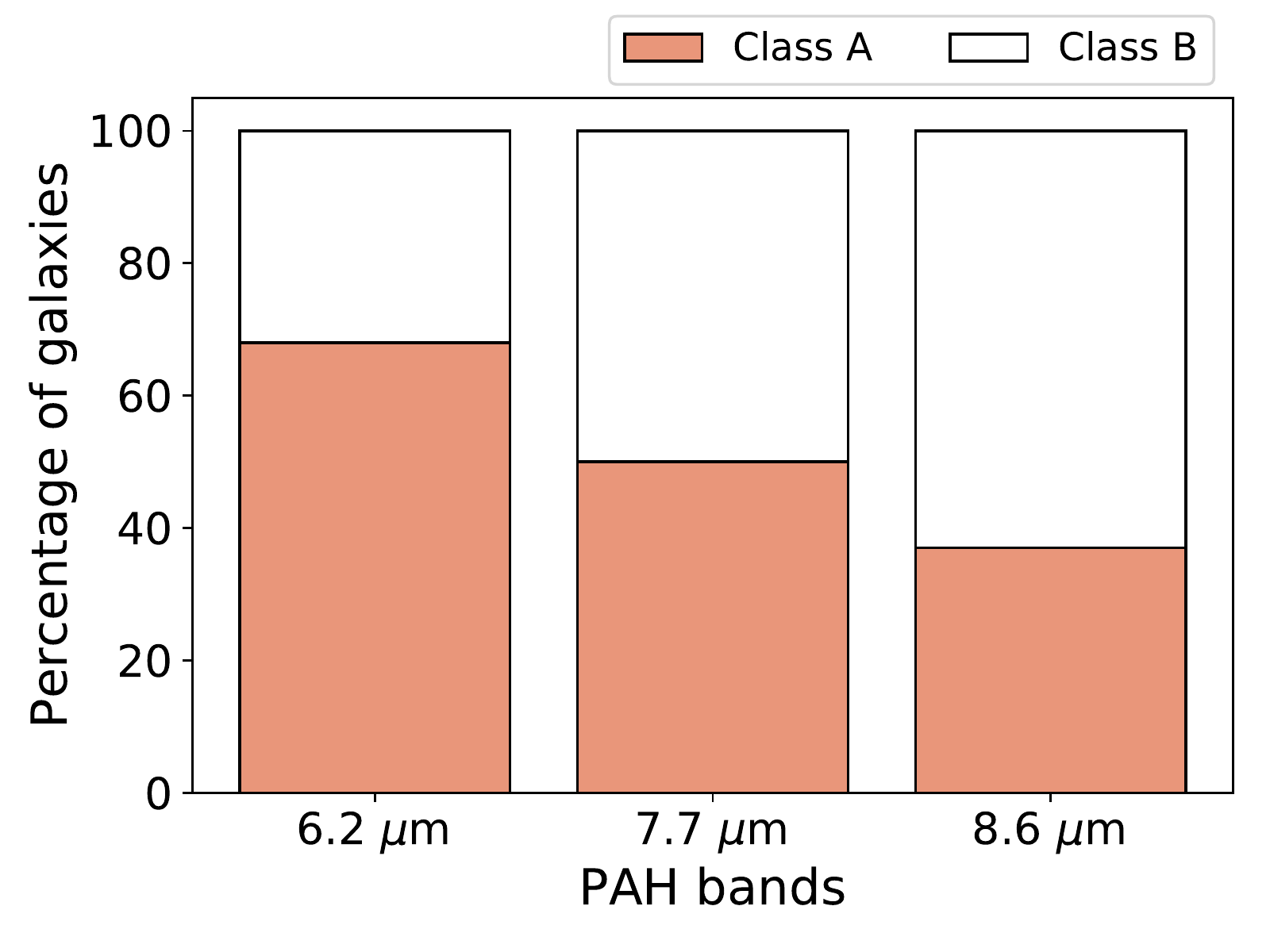}
    \caption{Histogram of the Peeters' classification for the 6.2, 7.7 and 8.6~$\mu$m bands in percentage. In the case of the 6.2~$\mu$m, the 1 per cent of class C sources was add to the class B percentage.}
    \label{fig:hist-class}
\end{figure}

\begin{table}
	\centering
	\caption{Profile distribution of the 126 studied galaxies into the Peeters' class.}
	\label{tab:percent-classes}
	\begin{tabular}{cccc}
		\hline
		\textbf{Band} & \textbf{Class A} & \textbf{Class B} &   \textbf{Class C}\\
        ($\mu$m) & (per cent) &  (per cent) &  (per cent)\\ 		\hline
	    6.2 & 68 & 31 & 1 \\
        7.7 & 50 & 50 & --- \\
        8.6 & 37 & 63 & --- \\
		\hline
	\end{tabular}
\end{table}

\begin{figure}
\centering
\includegraphics[scale=0.55]{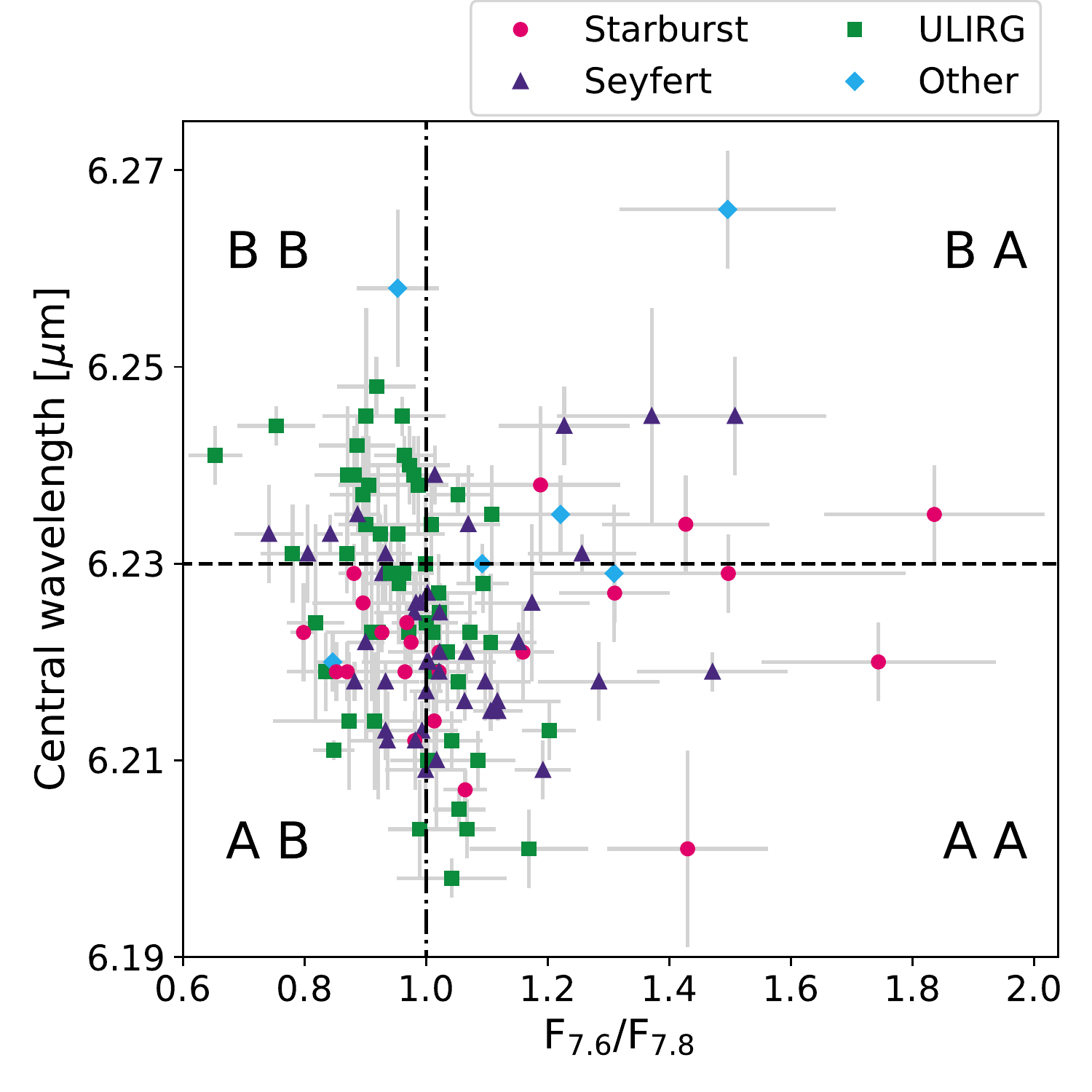}
\includegraphics[scale=0.55]{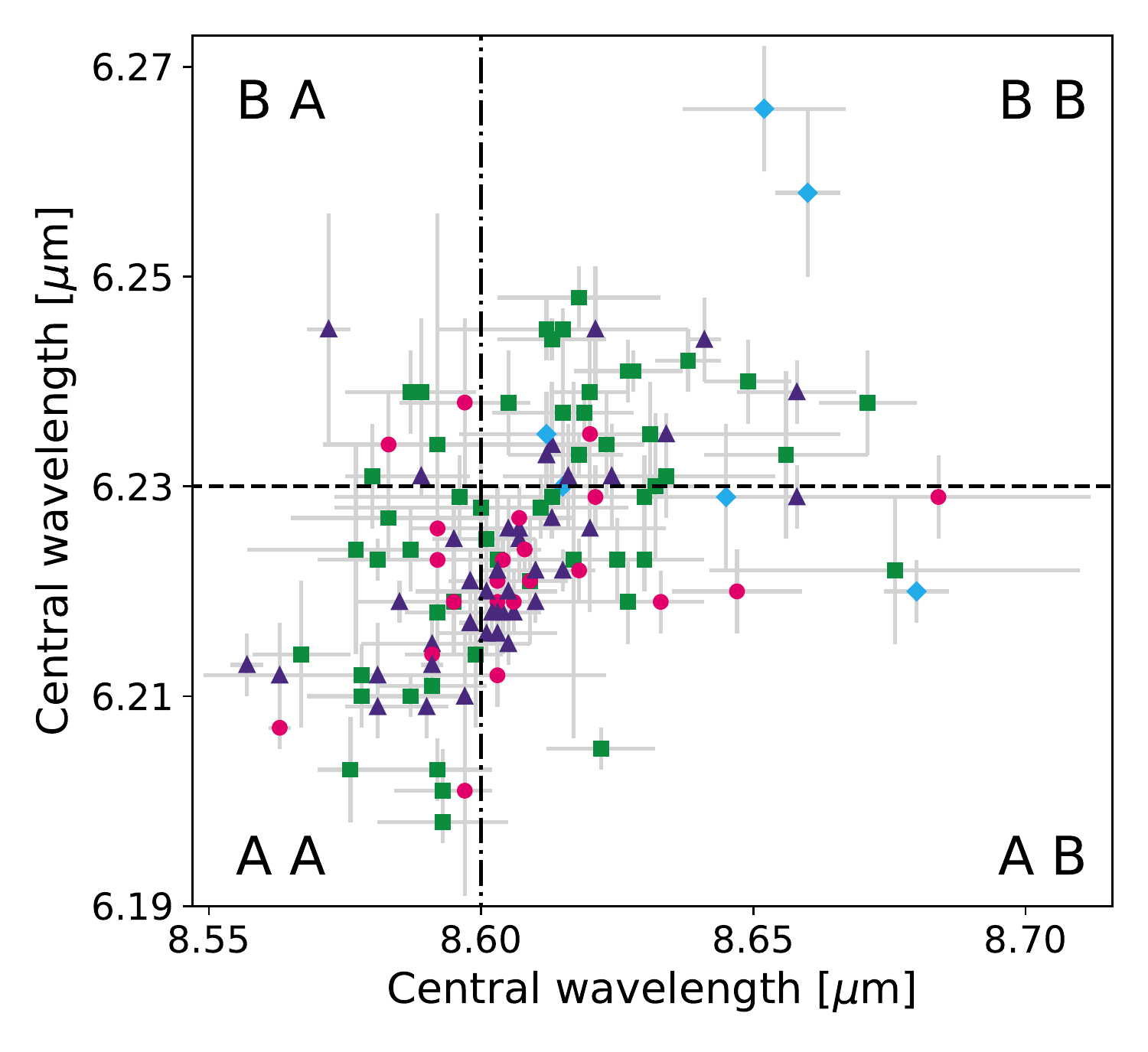}

\caption{Relation between the flux ratio F$_{7.6}$/F$_{7.8}$ (\textit{top)} and the peak position of the 8.6~$\mu$m band (\textit{bottom}) through the peak position of the 6.2~$\mu$m band. Horizontal dashed lines represents the limits of the Peeters' classes which are also indicated by the letters in each quadrant of the figures. The first letter always corresponds to the 6.2~$\mu$m band. The error bars are displayed in grey.}
\label{fig:62x77}
\end{figure}


\begin{figure*}
    \centering
    \includegraphics[scale=0.52]{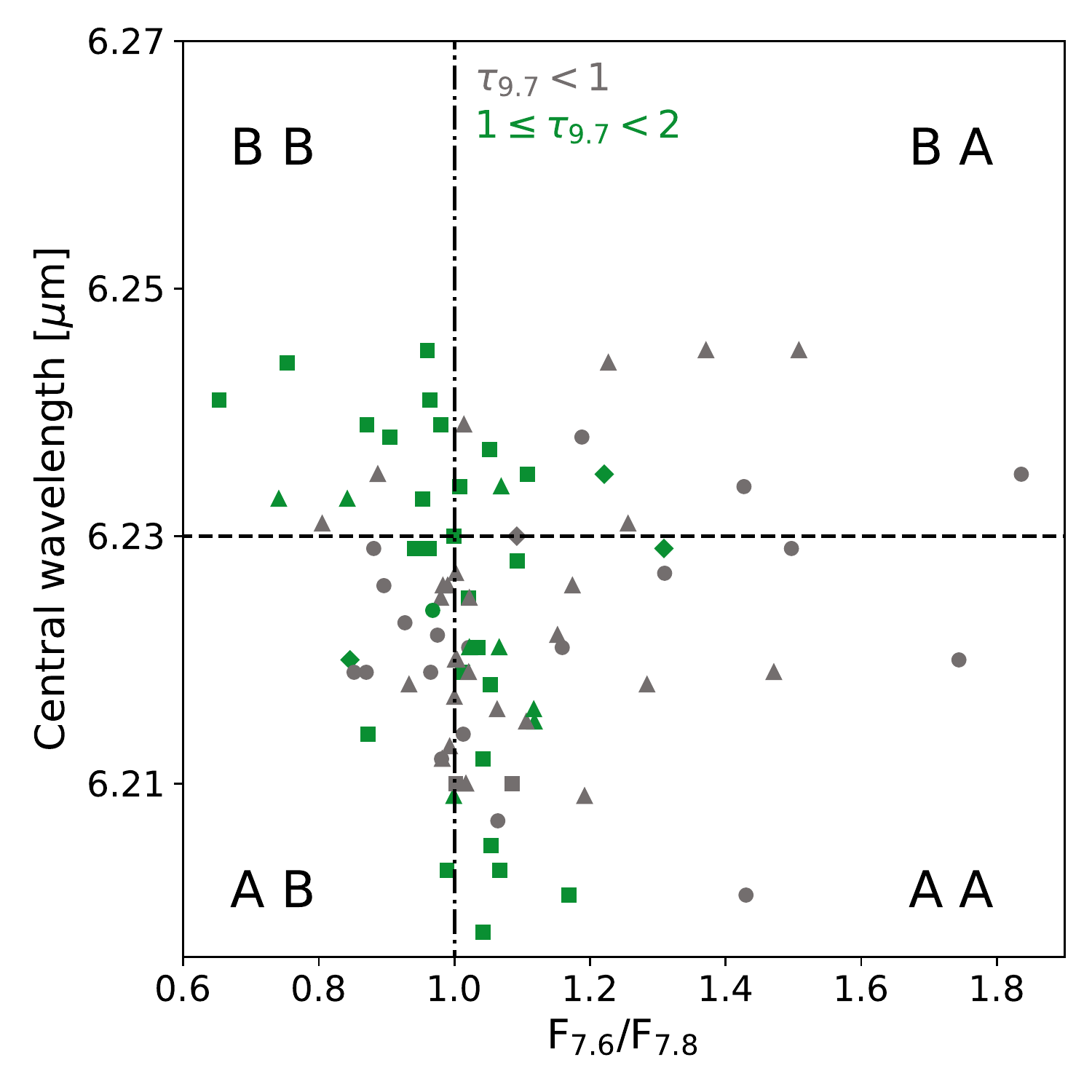}
    \includegraphics[scale=0.52]{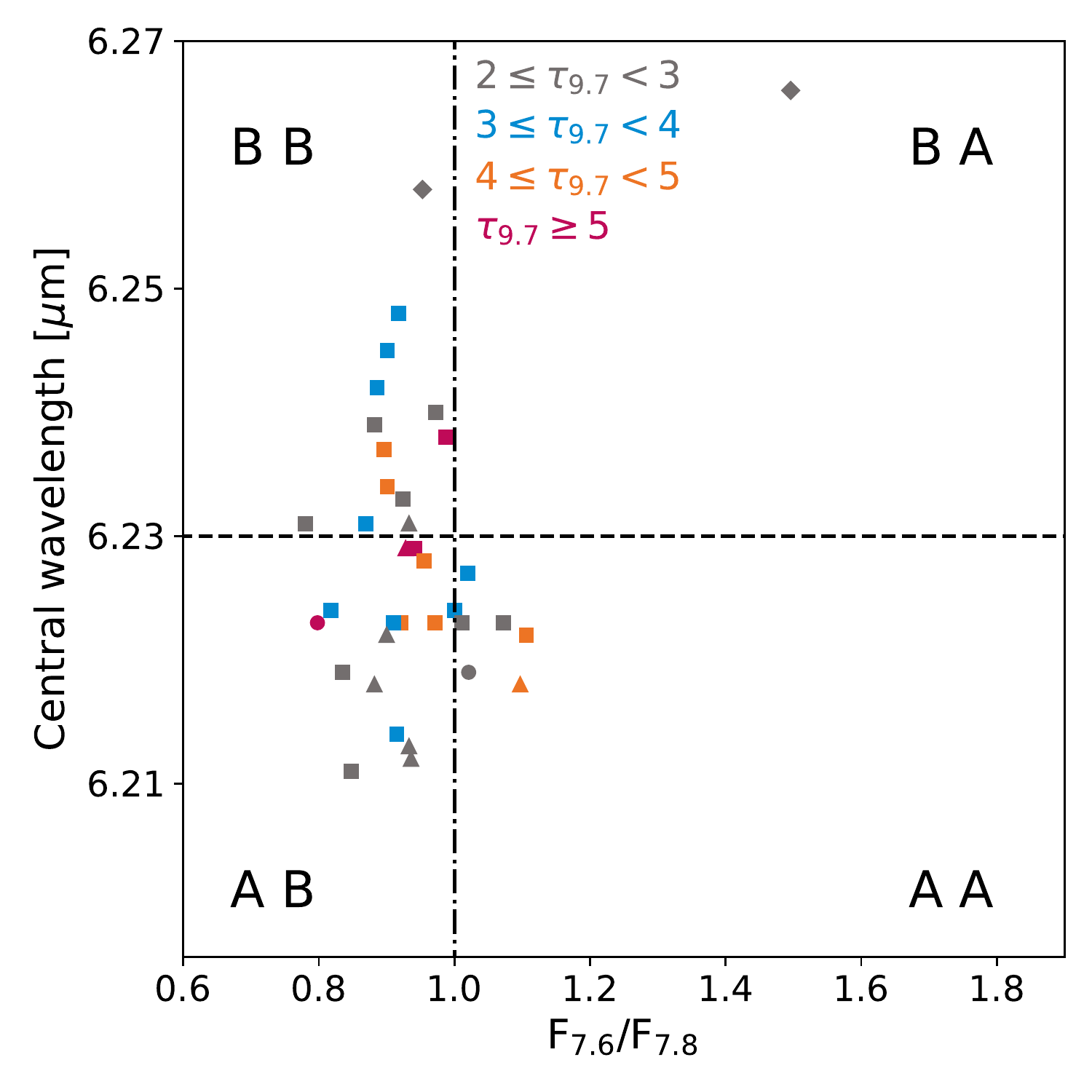}
    \includegraphics[scale=0.52]{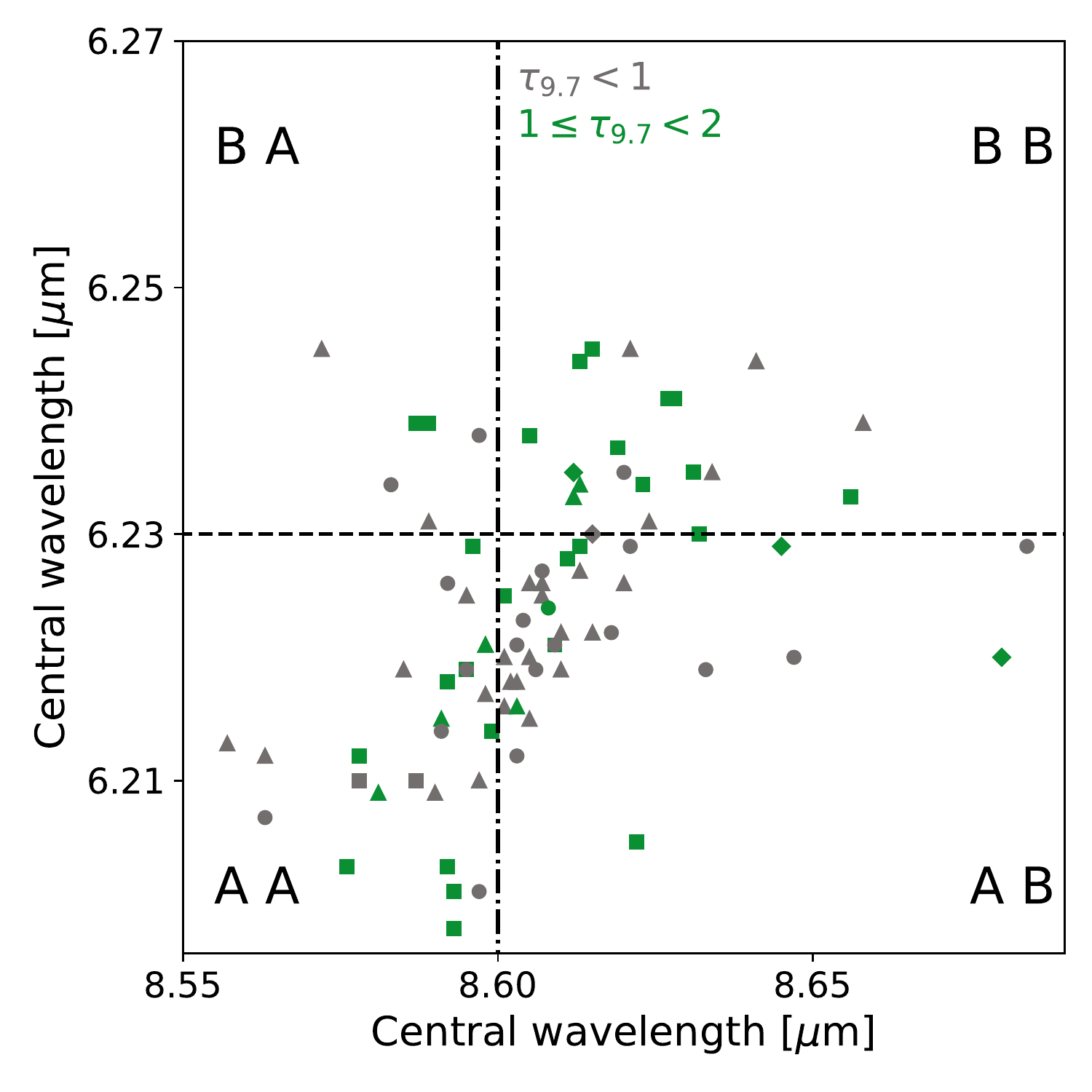}
    \includegraphics[scale=0.52]{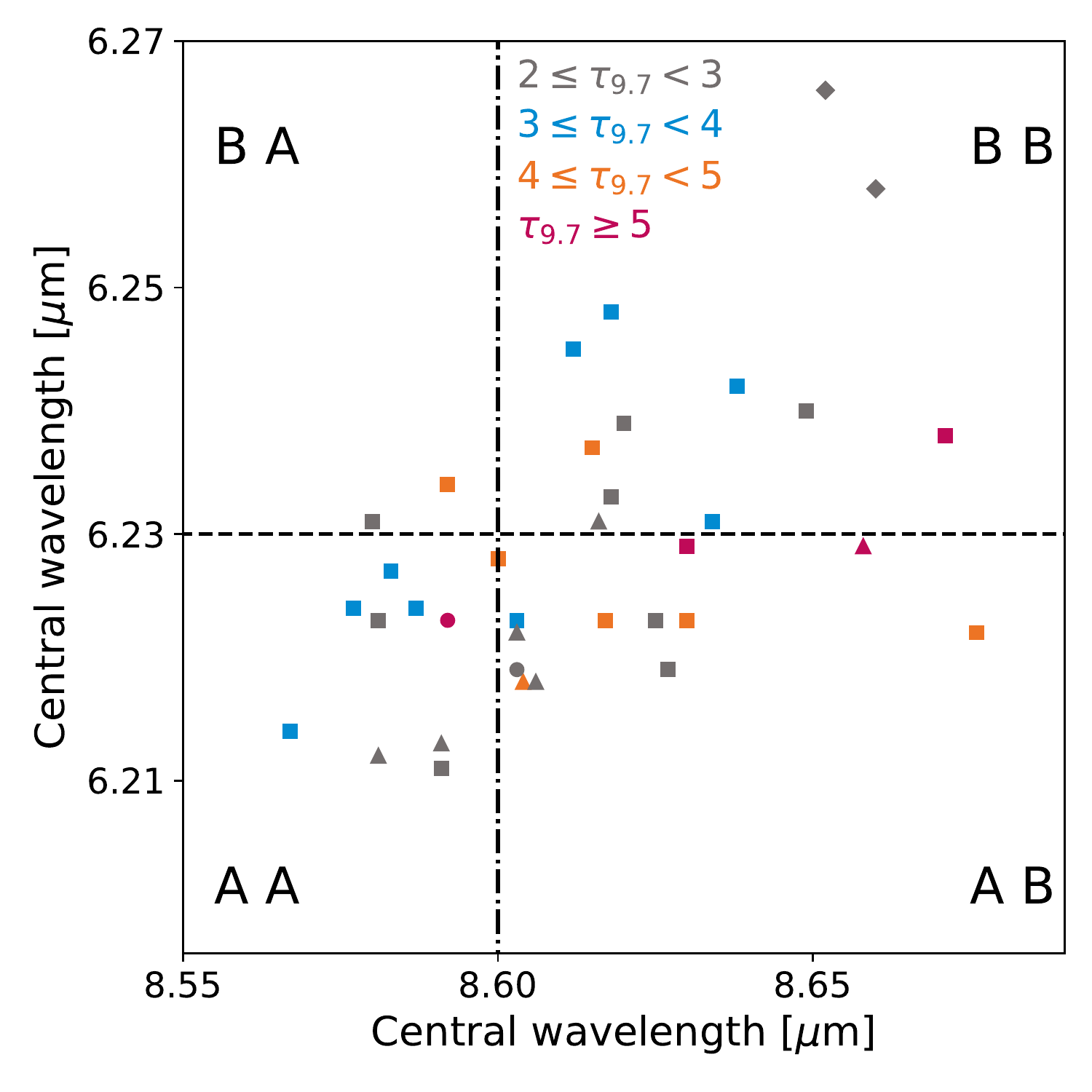}
    \caption{Relation between the flux ratio F$_{7.6}$/F$_{7.8}$ (\textit{top)} and the peak position of the 8.6~$\mu$m band (\textit{bottom}) through the peak position of the 6.2~$\mu$m band. Horizontal dashed lines represents the limits of the Peeters' classes which are also indicated by the letters in each quadrant of the figures. The first letter always corresponds to the 6.2~$\mu$m band. The error bars are not displayed. The types of the galaxies are represented by circle - Starburst, triangle - Seyfert, square - ULIRG and diamond - Others. The colours represent different ranges of $\tau_{9.7}$ values.}
    \label{fig:colormapTau}
\end{figure*}

Table~\ref{tab:percent-classes} and Fig. \ref{fig:hist-class} summarise the classification of the 126 objects into Peeters' classes derived from the fits. The classifications of the whole sample are given in Table~\ref{tab:results-classes}. Class A objects represent 68 per cent of the 6.2~$\mu$m profile. As matter of fact, they are the most common objects in the Universe and embrace several astrophysical sources \citep{pino08}. The 7.7~$\mu$m complex obtained 50 per cent of class A and B objects. Actually, the F$_{7.6}$/F$_{7.8}$ ratios varied around 1.0, which is the limit between the A and B classes. Apparently, fixed FHWM values furnished a balance in the class distribution of the objects. On the other hand, the 8.6~$\mu$m band is 63 per cent represented by class B objects. Class C only appears for the 6.2~$\mu$m profile representing 1 per cent.

From the total of our sample, 39 per cent of the galaxies were distributed into class A objects for both 6.2 and 7.7~$\mu$m bands and 18 per cent maintained the same A classification for the 8.6~$\mu$m band. This suggests an  strong correlation between the 6.2 and 7.7~$\mu$m features, as also the minor connection of the 6.2 and 8.6~$\mu$m bands. Considering the B class, 20 per cent of the sources were distributed into this class for the first two bands and 17 per cent for all three bands.  This fact could indicate that the correlation between the classes is stronger in class A objects as already discussed in \citet{died04}. Nevertheless, this seems to be more restricted to the 6.2 and 7.7~$\mu$m bands while 7.7 and 8.6~$\mu$m bands may be more correlated for B objects. 

\begin{figure*}
\centering
\includegraphics[scale=0.9]{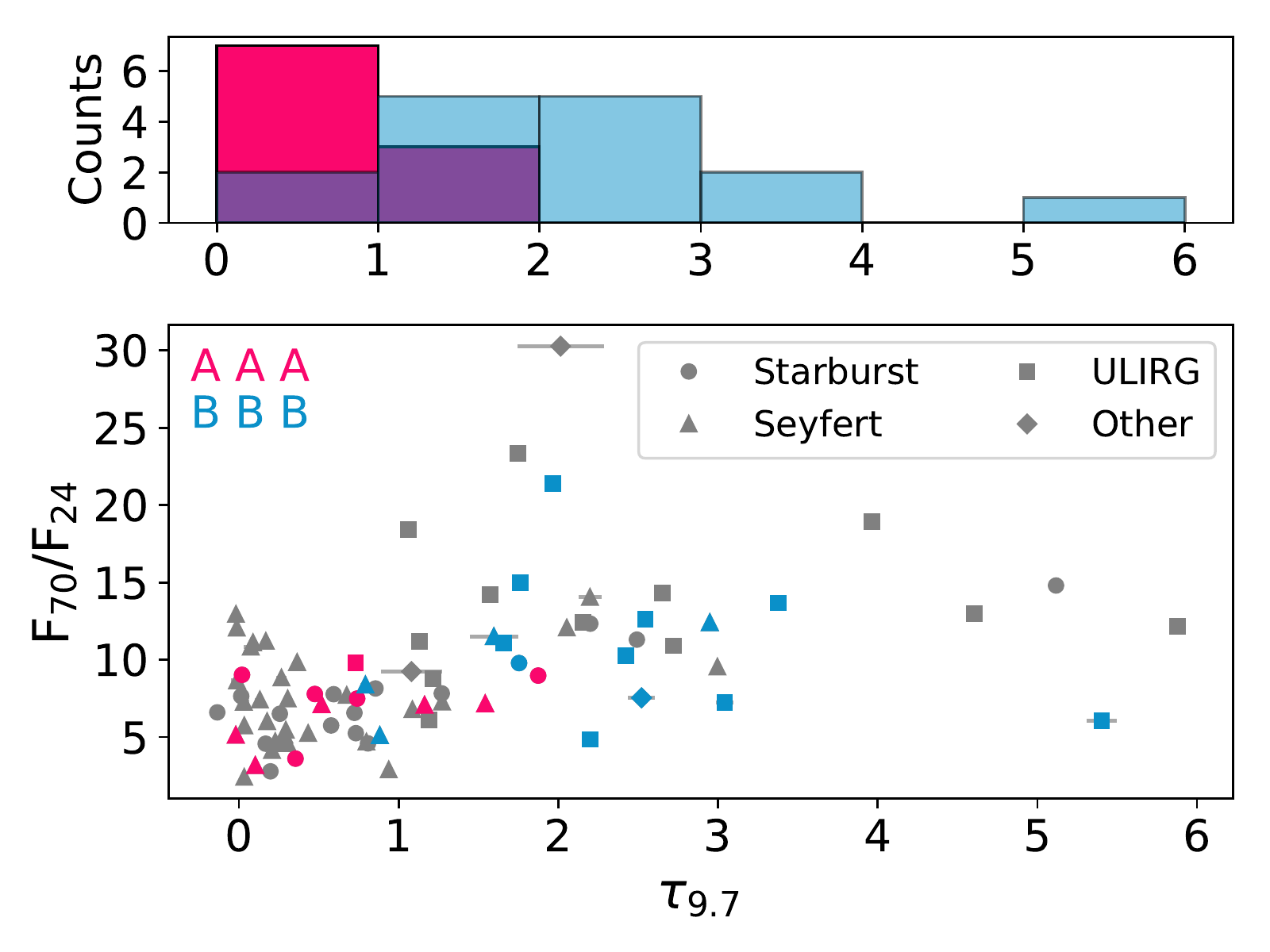}
\caption{Comparison between F$_{70}$/F$_{24}$ ratio and $\tau_{9.7}$. The types of the galaxies are shown in the label. Galaxies that received the same ``AAA'' or ``BBB'' classification are indicated by the colours pink and blue, respectively. The upper panel shows a histogram of the ``AAA'' and ``BBB'' galaxies according to the $\tau_{9.7}$ values. }
\label{fig:color70-24}
\end{figure*}

The Fig.\ref{fig:62x77} compares the Peeters' classification for the 6.2 and 7.7~$\mu$m bands and shows the possible combinations for the classifications -- ``A A'', ``A B'', ``B A'' and ``B B'', with the first letter always corresponding to the 6.2~$\mu$m band. As can be reinforced by the graph, class A objects are more correlated once one can see more ``A A'' objects than ``B B'', although most of the sources (41 per cent) received different classification for these bands. It is also interesting to notice that, in such cases, ``A B'' objects are more abundant than ``B A''. The majority of starbursts and Seyferts are classified as A objects for 6.2~$\mu$m band while the other galaxy types are more sparsely distributed in the plot.  Although the 7.7~$\mu$m band is not necessarily connected to a possible PANH emission such as the 6.2~$\mu$m band, galaxies equally classified as A objects for both bands have their classification reinforced.

The correlation between the 6.2 and 8.6~$\mu$m bands is shown in  Fig.\ref{fig:62x77}. ULIRGs are sparsely distributed along the plot but they dominated ``A A'' and ``B B'' sources. Starbursts and Seyferts are more represented by ``A B'' sources. We can note that fewer sources received the ``B A'' classification. Considering all three bands, the most common classification in our sample is ``A A B'' and ``A B B'', respectively to the 6.2, 7.7 and 8.6~$\mu$m bands.

\begin{figure*}
\centering
\includegraphics[scale=0.65]{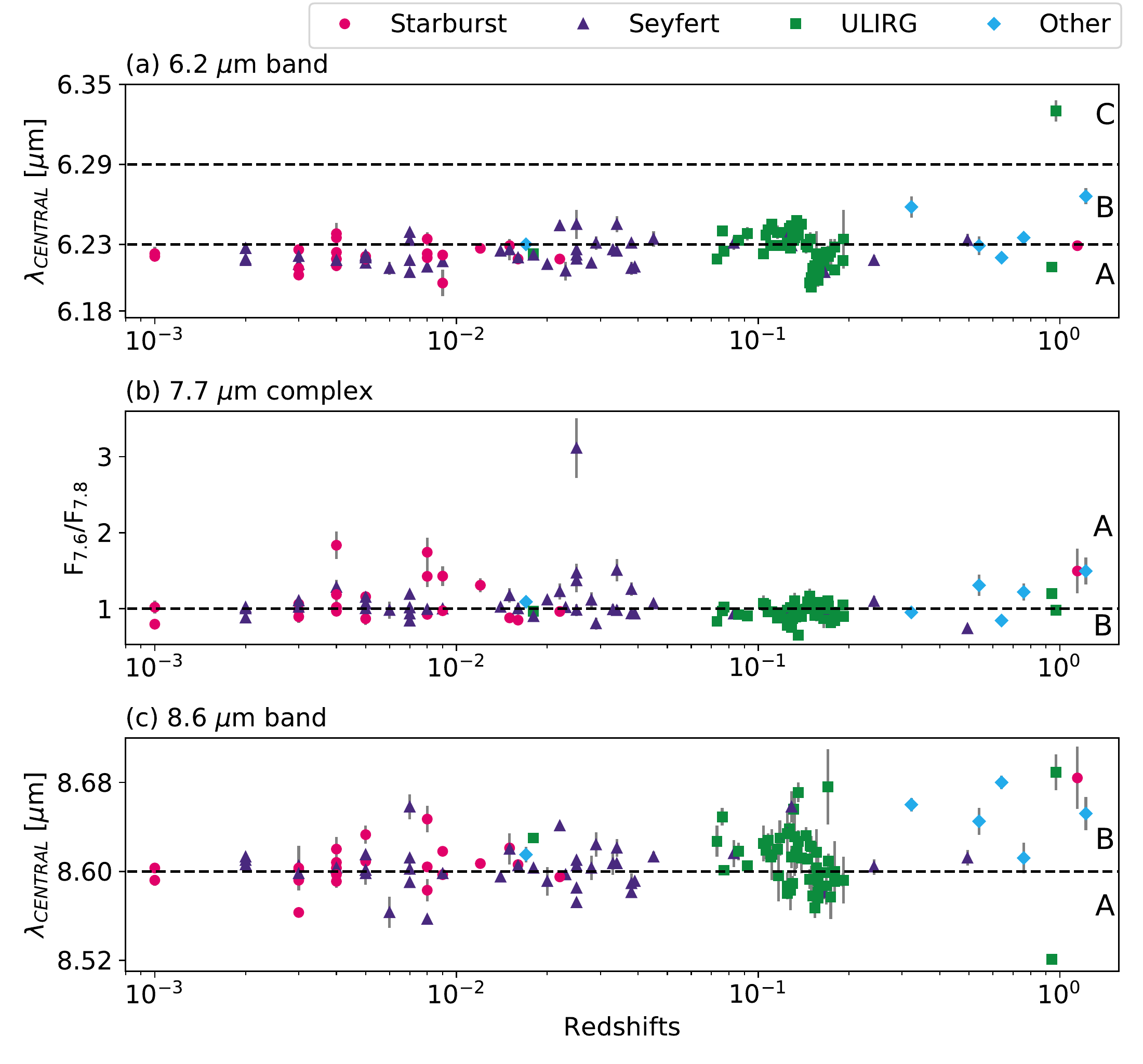}
\caption{Distribution of the 6.2, 7.7 and 8.6~$\mu$m bands, respectively, according to the galaxies redshift. The dashed lines are the limits among the Peeters' classes, indicated also by A, B or C letter. The redshift axis is in logarithmic scale. The type of the galaxies were divided into four main groups and their data points are represented in the plots by different symbols. The uncertainties are displayed as grey error bars.}
\label{fig:red-all}
\end{figure*}

Stronger 7.8~$\mu$m flux could arise from very large PAHs with multiple bay regions, irregular edges and modified duo CH groups (such as N substitution) in regions closest to the radiation sources \citep{Peeters17}. However, this kind of PANH with nitrogen in the bay regions does not contribute to the class A 6.2~$\mu$m band emission \citep{Hud05}. Roughly 20 per cent of our sources were doubly classified as B objects for 6.2 and 7.7~$\mu$m bands and they are better represented by ULIRGs. Class C does not have enough data for any interpretation. Although our sample is composed by galaxies, the gradual variation noticed by \citet{Peeters02}, in which lower 6.2~$\mu$m central wavelengths can present higher  F$_{7.6}$/F$_{7.8}$ ratios, is also recovered in our analyses. 

In order to analyse if the silicate absorption influences the fitting and the classification, the band comparisons were also performed using the $\tau_{9.7}$ values available in the ATLAS. They were divided into six different groups for better visualisation: $\tau_{9.7}$ < 1, which corresponds to 40 per cent of our sample; 1~$\leq \tau_{9.7}$~<~2 , representing 30 per cent of the objects; 2~$\leq \tau_{9.7}$~<~3, representing 13 per cent of the objects; 3~$\leq \tau_{9.7}$~<~4, representing 7 per cent of the objects; and 4~$\leq \tau_{9.7}$~<~5 and $\tau_{9.7} \geq$~5,  representing 10 per cent of our sample. Figure \ref{fig:colormapTau} shows the results for the three bands. Galaxies with $\tau_{9.7} \geq$~2 are mainly ULIRGs, and tend to be classified as ``B''. We can not identify a strong correlation between the optical depth and the PAH classification for the 8.6~$\mu$m band, although  the silicate absorption may redshift its peak position for $\tau_{9.7} \geq$~4, as already mentioned. On the other hand, the 7.7~$\mu$m band appeared to be more susceptible to $\tau_{9.7}$ values. Nevertheless, there is no clear indication in Appendix~\ref{sec:ext} that supports such trend because the decrease in the F$_{7.6}$/F$_{7.8}$ ratio  with high $\tau_{9.7}$ values is not enough to modify the Peeters' classification of this band. In fact, dusty environments with larger PAHs are typical of class B sources \citep[e.g. ][]{Shannon19}, which explains the dominance of the B class  in ULIRGs. The large majority of the ULIRGs and hyperluminous IRGs spectra studied by \citet{Spoon2007} are, indeed,  characterised by increasingly  apparent  silicate  absorption  and  less-pronounced  PAH emission features, with deeply obscured galactic nuclei.

Considering sources with $\tau_{9.7}$ < 1, the dominance of class A objects does increase to 80 and 66 per cent for the 6.2 and 7.7~$\mu$m bands, respectively. The 8.6~$\mu$m band classification remains the same throughout our sample. However, this is expected because mostly galaxies from our sample that have $\tau_{9.7}$ < 1 are Starbursts and Seyferts, which are widely dominated by class A sources as has already been demonstrated by \citet{Canelo18}. Therefore, the extinction and the  silicate absorption seemingly does not play an important role to the fitting and/or to the PAH classification results. The main influence could be an overestimation of class B sources for the 8.6~$\mu$m band in a maximum of 10 per cent of our sample ($\tau_{9.7} \geq$~4), however further investigation need to be address.

The optical depth was also compared with the F$_{70}$/F$_{24}$ ratio and can be seen in Figure \ref{fig:color70-24}. The fluxes at 24 and 70$\mu$m were extracted directly from the ATLAS. As expected, higher values of $\tau_{9.7}$ are followed by higher F$_{70}$/F$_{24}$ ratio, indicating that colder regions have more dust absorption. Considering the galaxies that obtained the same classification for the three bands, it is possible to see that class ``A A A'' sources are located mainly at lower values of F$_{70}$/F$_{24}$ and $\tau_{9.7}$, in which the temperature is higher and there is less dust in the ISM. This could suggest that dust grains have also been processed into PAH molecules and even in heteroatomic PAHs, such as PANHs. On the other hand, class ``B B B'' sources are shifted to higher values of F$_{70}$/F$_{24}$ and $\tau_{9.7}$, suggesting environments with greater abundance of grains. In general, our results point out that galaxies with  $\tau_{9.7}<$1 are Starbursts and Seyferts, as already mentioned, in which this kind of source usually has higher temperature than ULIRGs.

Regarding to the evolution of aromatic species according to the object type, \citet{Shannon19} claim that exposed interstellar environments show class A profiles while class B profiles are observed from circumstellar environments. We also find an spread in the classification of the 6.2, 7.7 and 8.6~$\mu$m band profiles through the ionisation source of our sample.  In addition, Starbursts and Seyferts were the type of galaxies that most varied the classification along the bands and they are concentrated in the lower region of the plot. This could also indicate different emitting PAH populations due to the physical conditions of exposed interstellar environments.

An overview of the results is shown in Fig.~\ref{fig:red-all}, which displays the distribution of the central wavelengths (for the 6.2 and 8.6~$\mu$m bands) and the flux ratio (for the 7.7~$\mu$m complex) in comparison of redshifts of the galaxies. It is possible to perceive the predominance of class A objects over class B objects for the 6.2~$\mu$m band. The opposite occurs for the 8.6~$\mu$m band. Again, the type of galaxies does not seem to interfere with the classification. Higher $\tau_{9.7}$ values occur mainly for ULIRGs but all types of galaxies have a dominance of B objects for the 8.6~$\mu$m band. Instead, the redshift may play an important role in the results. 
Although there is a luminosity selection such that only very high LIR sources will have IRS spectroscopy above z~=~1 and the bands of interest can be shifted out of the Spitzer/IRS SL2 band, across a break in the spectral coverage, into the lower spectral resolution IRS SL1 band, the comparison of galaxies at different redshifts could expand the aromatic evolution in stellar lifecycle of \citet{Shannon19} to an extragalactic point of view. Actually, for all three bands, the quantity of class B sources seems to increase to galaxies at higher redshifts, which may suggest a evolutionary timescale of the PAH population through the galaxies evolution in the Universe.

From the light of Astrochemistry, chemically young astrophysical sources might have reduced PAH abundances and PAH molecules are not as efficiently produced in low-metallicity environments because fewer carbon atoms are available in the ISM \citep{Shivaei17}. As can be seen in Fig. \ref{fig:red-all}, lower F$_{7.6}$/F$_{7.8}$ ratios are often for galaxies at redshifts higher than 0.01 indicating a greater predominance of the 7.8~$\mu$m component respect to the 7.7~$\mu$m complex. This feature has also been attributed to evaporating very small grains \citep[eVSGs, ][]{Rapacioli05,Berne07}, which may suggest a greater contribution of these molecular material to galaxies at higher redshifts. In this sense, these findings support future research
that should occur, for instance, with the James Webb Space Telescope (JWST). The JWST will allow the observation of galaxies at higher redshifts and with greater resolution in the MIR, which will be fundamental for a  complementary analysis of these
questions \citep{Stiavelli09}.

\section{Conclusions}
\label{sec:conclusion}
We have analysed the MIR spectra of 126 starburst-dominated galaxies, searching for the contribution of the Peeters' classes to the 6.2, 7.7 and 8.6~$\mu$m PAH bands. It is the first time that such statistical analysis is performed to a sample of galaxies. To date, observed 6.2~$\mu$m peak position for class A systems can only be attributed to PANHs, PAHs containing N atoms. Thus, this exemplifies how a detailed analysis of PAH feature profiles can help us identify and uncover different PAH populations including the presence of nitrogen incorporated to the rings. 

The class A of the 6.2~$\mu$m PAH emission band seems to dominate this spectral feature in starburst-dominated galaxies, suggesting a significant presence of these molecules in extragalactic environments \citep{Canelo18}. We extend this analysis to the  7.7 and 8.6~$\mu$m bands, classifying them into classes A, B and C following \citet{Peeters02}. The 7.7~$\mu$m complex presents 50 per cent of class A objects while 8.6~$\mu$m band presents 63 per cent of class B sources. Only the 6.2~$\mu$m band has a class C profile with 1 per cent of our sample. The  extinction and silicate feature at 9.7~$\mu$m, for which we have not made corrections as part of our analysis, do not seem to have a significant impact on our analysis (see Appendix~\ref{sec:ext}). Most objects of our sample with $\tau_{9.7}$~>~2 are ULIRGs, which present dusty environments typically classified as B sources, consistent with our findings.  Nevertheless, a non correction of the extinction could lead to an overestimation of class B sources for the 8.6~$\mu$m band in 10 per cent of our sample, for the objects with $\tau_{9.7} \geq$~4.

Considering the class A objects, 39 per cent of the galaxies were distributed into class A objects for both 6.2 and 7.7~$\mu$m bands and only 18 per cent received the same A classification for all three bands. Although this result may not allow us to indirectly study the PANHs emission at 6.2~$\mu$m with the other bands correlation, it supports the complexity of PAH emission. The differences in PAH profile classes, especially when these differences are present for one same source, arise from the astrophysical and chemical conditions of the environments, including all the molecular species that contribute differently for each band emission. For instance, according to \citet{Monfredini19}, a more sophisticated interplay between PAHs and dust grains should be present in order to circumvent molecular destruction in AGN circumnuclear medium.

Analysis of other types of objects also available in the ATLAS project, such as AGNs, could shed light on how the starburst-dominated emission of the sources is responsible for the majority of class A objects and could provide a broader overview of the PAH band behaviour in astrophysical environments. It would be also important to compare extended intervals of redshift, specially the highest, in order to probe for signs of a evolution in the PAH molecular composition. Furthermore, our results could also indicate an aromatic evolution in the ISM of galaxies along to redshifts lower than 1.5 similar to the aromatic evolution  in stellar lifecycle proposed by \citet{Shannon19}.

The PAH profile variations that reproduce different Peeters' classification for the same source could be addressed with chemical evolution models taking into account differences in metallicity, star formation history and the nature of molecular clouds in the host galaxy, e.g. the chemodynamical model in \citet{Friaca17}. In addition, further computational calculations taken together with laboratory and observational measurements are needed to address issues about emitting PAH population, mainly in the conditions prevailing in active galaxies with high star formation and/or super-massive black hole. Finally, new observations with JWST at higher redshifts and with greater resolution in the MIR will allow a more detailed  analysis on the PAH bands and the PAH evolutionary timescale in galaxies.

\section*{Acknowledgements}

We thank the anonymous referee for the useful comments that improved the article. We also would like to thank Dr. Els Peeters (Dept. of Physics \& Astronomy, University of Western Ontario, SETI Institute) for the comments and discussions. CMC acknowledges the support of CNPq, Conselho Nacional de Desenvolvimento Cient\'ifico e Tecnol\'ogico - Brazil, process number 141714/2016-6. This study was financed in part by the Coordena\c{c}\~ao de Aperfei\c{c}oamento de Pessoal de N\'ivel Superior - Brasil (CAPES) - Finance Code 001. DASales  acknowledges the support of CNPq and of the Funda\c{c}\~ao de Amparo \`a Pesquisa do Estado do Rio Grande do Sul (FAPERGS), Brazil. KMD thanks the support of the Serrapilheira Institute as well as that of the CNPq and of the Funda\c{c}\~ao de Amparo \`a Pesquisa do Estado do Rio de Janeiro (FAPERJ), Brazil.

\section*{Data Availability}

The data underlying this article are available in the Spitzer/IRS ATLAS project\footnote{http://www.denebola.org/atlas/} \citep{caballero} and in the online supplementary material of this article. Any additional data will be shared on reasonable request to the corresponding author.




\bibliographystyle{mnras}
\bibliography{mnras_canelo_20} 

\begin{thebibliography}{}
\makeatletter
\relax
\def\mn@urlcharsother{\let\do\@makeother \do\$\do\&\do\#\do\^\do\_\do\%\do\~}
\def\mn@doi{\begingroup\mn@urlcharsother \@ifnextchar [ {\mn@doi@}
  {\mn@doi@[]}}
\def\mn@doi@[#1]#2{\def\@tempa{#1}\ifx\@tempa\@empty \href
  {http://dx.doi.org/#2} {doi:#2}\else \href {http://dx.doi.org/#2} {#1}\fi
  \endgroup}
\def\mn@eprint#1#2{\mn@eprint@#1:#2::\@nil}
\def\mn@eprint@arXiv#1{\href {http://arxiv.org/abs/#1} {{\tt arXiv:#1}}}
\def\mn@eprint@dblp#1{\href {http://dblp.uni-trier.de/rec/bibtex/#1.xml}
  {dblp:#1}}
\def\mn@eprint@#1:#2:#3:#4\@nil{\def\@tempa {#1}\def\@tempb {#2}\def\@tempc
  {#3}\ifx \@tempc \@empty \let \@tempc \@tempb \let \@tempb \@tempa \fi \ifx
  \@tempb \@empty \def\@tempb {arXiv}\fi \@ifundefined
  {mn@eprint@\@tempb}{\@tempb:\@tempc}{\expandafter \expandafter \csname
  mn@eprint@\@tempb\endcsname \expandafter{\@tempc}}}

\bibitem[\protect\citeauthoryear{{Allamandola}, {Hudgins}  \&
  {Sandford}}{{Allamandola} et~al.}{1999}]{allamandola99}
{Allamandola} L.~J.,  {Hudgins} D.~M.,   {Sandford} S.~A.,  1999, \mn@doi
  [\apjl] {10.1086/311843}, \href
  {http://adsabs.harvard.edu/abs/1999ApJ...511L.115A} {511, L115}

\bibitem[\protect\citeauthoryear{{Alonso-Herrero} et~al.,}{{Alonso-Herrero}
  et~al.}{2014}]{Alonso-Herrero14}
{Alonso-Herrero} A.,  et~al., 2014, \mn@doi [\mnras] {10.1093/mnras/stu1293},
  \href {https://ui.adsabs.harvard.edu/abs/2014MNRAS.443.2766A} {443, 2766}

\bibitem[\protect\citeauthoryear{{Alonso-Herrero} et~al.,}{{Alonso-Herrero}
  et~al.}{2016}]{Alonso-Herrero16}
{Alonso-Herrero} A.,  et~al., 2016, \mn@doi [\mnras] {10.1093/mnras/stv2342},
  \href {https://ui.adsabs.harvard.edu/abs/2016MNRAS.455..563A} {455, 563}

\bibitem[\protect\citeauthoryear{{Bern{\'e}} et~al.,}{{Bern{\'e}}
  et~al.}{2007}]{Berne07}
{Bern{\'e}} O.,  et~al., 2007, \mn@doi [\aap] {10.1051/0004-6361:20066282},
  \href {http://adsabs.harvard.edu/abs/2007A%26A...469..575B} {469, 575}

\bibitem[\protect\citeauthoryear{{Bern{\'e}}, {Mulas}  \& {Joblin}}{{Bern{\'e}}
  et~al.}{2013}]{Berne13}
{Bern{\'e}} O.,  {Mulas} G.,   {Joblin} C.,  2013, \mn@doi [\aap]
  {10.1051/0004-6361/201220730}, \href
  {http://adsabs.harvard.edu/abs/2013A%26A...550L...4B} {550, L4}

\bibitem[\protect\citeauthoryear{{Brandl} et~al.,}{{Brandl}
  et~al.}{2006}]{Brandl06}
{Brandl} B.~R.,  et~al., 2006, \mn@doi [\apj] {10.1086/508849}, \href
  {http://adsabs.harvard.edu/abs/2006ApJ...653.1129B} {653, 1129}

\bibitem[\protect\citeauthoryear{{Candian} \& {Sarre}}{{Candian} \&
  {Sarre}}{2015}]{Candian15}
{Candian} A.,  {Sarre} P.~J.,  2015, \mn@doi [\mnras] {10.1093/mnras/stv192},
  \href {http://adsabs.harvard.edu/abs/2015MNRAS.448.2960C} {448, 2960}

\bibitem[\protect\citeauthoryear{{Canelo}, {Fria{\c c}a}, {Sales}, {Pastoriza}
  \& {Ruschel-Dutra}}{{Canelo} et~al.}{2018}]{Canelo18}
{Canelo} C.~M.,  {Fria{\c c}a} A.~C.~S.,  {Sales} D.~A.,  {Pastoriza} M.~G.,
  {Ruschel-Dutra} D.,  2018, \mn@doi [\mnras] {10.1093/mnras/stx3351}, \href
  {http://adsabs.harvard.edu/abs/2018MNRAS.475.3746C} {475, 3746}

\bibitem[\protect\citeauthoryear{{Chiar} \& {Tielens}}{{Chiar} \&
  {Tielens}}{2006}]{Chiar06}
{Chiar} J.~E.,  {Tielens} A.~G.~G.~M.,  2006, \mn@doi [\apj] {10.1086/498406},
  \href {https://ui.adsabs.harvard.edu/abs/2006ApJ...637..774C} {637, 774}

\bibitem[\protect\citeauthoryear{{Deo}, {Crenshaw}, {Kraemer}, {Dietrich},
  {Elitzur}, {Teplitz}  \& {Turner}}{{Deo} et~al.}{2007}]{Deo07}
{Deo} R.~P.,  {Crenshaw} D.~M.,  {Kraemer} S.~B.,  {Dietrich} M.,  {Elitzur}
  M.,  {Teplitz} H.,   {Turner} T.~J.,  2007, \mn@doi [\apj] {10.1086/522823},
  \href {http://adsabs.harvard.edu/abs/2007ApJ...671..124D} {671, 124}

\bibitem[\protect\citeauthoryear{{Draine} \& {Li}}{{Draine} \&
  {Li}}{2001}]{Draine01}
{Draine} B.~T.,  {Li} A.,  2001, in American Astronomical Society Meeting
  Abstracts. p.~1451

\bibitem[\protect\citeauthoryear{{Draine} \& {Li}}{{Draine} \&
  {Li}}{2007}]{Draine07}
{Draine} B.~T.,  {Li} A.,  2007, \mn@doi [\apj] {10.1086/511055}, \href
  {http://adsabs.harvard.edu/abs/2007ApJ...657..810D} {657, 810}

\bibitem[\protect\citeauthoryear{{Ehrenfreund} et~al.,}{{Ehrenfreund}
  et~al.}{2002}]{Eh02}
{Ehrenfreund} P.,  et~al., 2002, \mn@doi [Reports on Progress in Physics]
  {10.1088/0034-4885/65/10/202}, \href
  {http://adsabs.harvard.edu/abs/2002RPPh...65.1427E} {65, 1427}

\bibitem[\protect\citeauthoryear{{Ehrenfreund}, {Rasmussen}, {Cleaves}  \&
  {Chen}}{{Ehrenfreund} et~al.}{2006}]{Eh06}
{Ehrenfreund} P.,  {Rasmussen} S.,  {Cleaves} J.,   {Chen} L.,  2006, \mn@doi
  [Astrobiology] {10.1089/ast.2006.6.490}, \href
  {http://adsabs.harvard.edu/abs/2006AsBio...6..490E} {6, 490}

\bibitem[\protect\citeauthoryear{{Fria{\c c}a} \& {Barbuy}}{{Fria{\c c}a} \&
  {Barbuy}}{2017}]{Friaca17}
{Fria{\c c}a} A.~C.~S.,  {Barbuy} B.,  2017, \mn@doi [\aap]
  {10.1051/0004-6361/201629941}, \href
  {http://adsabs.harvard.edu/abs/2017A%26A...598A.121F} {598, A121}

\bibitem[\protect\citeauthoryear{{Galliano}, {Madden}, {Tielens}, {Peeters}  \&
  {Jones}}{{Galliano} et~al.}{2008}]{Galliano08}
{Galliano} F.,  {Madden} S.~C.,  {Tielens} A.~G.~G.~M.,  {Peeters} E.,
  {Jones} A.~P.,  2008, \mn@doi [\apj] {10.1086/587051}, \href
  {http://adsabs.harvard.edu/abs/2008ApJ...679..310G} {679, 310}

\bibitem[\protect\citeauthoryear{{Genzel} \& {Cesarsky}}{{Genzel} \&
  {Cesarsky}}{2000}]{Genzel00}
{Genzel} R.,  {Cesarsky} C.~J.,  2000, \mn@doi [\araa]
  {10.1146/annurev.astro.38.1.761}, \href
  {http://adsabs.harvard.edu/abs/2000ARA%26A..38..761G} {38, 761}

\bibitem[\protect\citeauthoryear{{Hensley} \& {Draine}}{{Hensley} \&
  {Draine}}{2020}]{Hensley2020}
{Hensley} B.~S.,  {Draine} B.~T.,  2020, \mn@doi [\apj]
  {10.3847/1538-4357/ab8cc3}, \href
  {https://ui.adsabs.harvard.edu/abs/2020ApJ...895...38H} {895, 38}

\bibitem[\protect\citeauthoryear{{Hern{\'a}n-Caballero} \&
  {Hatziminaoglou}}{{Hern{\'a}n-Caballero} \&
  {Hatziminaoglou}}{2011}]{caballero}
{Hern{\'a}n-Caballero} A.,  {Hatziminaoglou} E.,  2011, \mn@doi [\mnras]
  {10.1111/j.1365-2966.2011.18413.x}, \href
  {http://adsabs.harvard.edu/abs/2011MNRAS.414..500H} {414, 500}

\bibitem[\protect\citeauthoryear{{Hern{\'a}n-Caballero}
  et~al.,}{{Hern{\'a}n-Caballero} et~al.}{2009}]{Hernot-Caballero09}
{Hern{\'a}n-Caballero} A.,  et~al., 2009, \mn@doi [\mnras]
  {10.1111/j.1365-2966.2009.14660.x}, \href
  {http://adsabs.harvard.edu/abs/2009MNRAS.395.1695H} {395, 1695}

\bibitem[\protect\citeauthoryear{{Hirashita}, {Deng}  \& {Murga}}{{Hirashita}
  et~al.}{2020}]{Hirashita2020}
{Hirashita} H.,  {Deng} W.,   {Murga} M.~S.,  2020, \mn@doi [\mnras]
  {10.1093/mnras/staa3101}, \href
  {https://ui.adsabs.harvard.edu/abs/2020MNRAS.499.3046H} {499, 3046}

\bibitem[\protect\citeauthoryear{{Houck} et~al.,}{{Houck}
  et~al.}{2004}]{Houck04}
{Houck} J.~R.,  et~al., 2004, \mn@doi [\apjs] {10.1086/423134}, \href
  {http://adsabs.harvard.edu/abs/2004ApJS..154...18H} {154, 18}

\bibitem[\protect\citeauthoryear{{Hudgins}, {Bauschlicher}  \&
  {Allamandola}}{{Hudgins} et~al.}{2005}]{Hud05}
{Hudgins} D.~M.,  {Bauschlicher} Jr. C.~W.,   {Allamandola} L.~J.,  2005,
  \mn@doi [\apj] {10.1086/432495}, \href
  {http://adsabs.harvard.edu/abs/2005ApJ...632..316H} {632, 316}

\bibitem[\protect\citeauthoryear{{Imanishi}, {Dudley}, {Maiolino}, {Maloney},
  {Nakagawa}  \& {Risaliti}}{{Imanishi} et~al.}{2007}]{Imanishi07}
{Imanishi} M.,  {Dudley} C.~C.,  {Maiolino} R.,  {Maloney} P.~R.,  {Nakagawa}
  T.,   {Risaliti} G.,  2007, \mn@doi [\apjs] {10.1086/513715}, \href
  {http://adsabs.harvard.edu/abs/2007ApJS..171...72I} {171, 72}

\bibitem[\protect\citeauthoryear{{Imanishi}, {Maiolino}  \&
  {Nakagawa}}{{Imanishi} et~al.}{2010}]{Imanishi10}
{Imanishi} M.,  {Maiolino} R.,   {Nakagawa} T.,  2010, \mn@doi [\apj]
  {10.1088/0004-637X/709/2/801}, \href
  {http://adsabs.harvard.edu/abs/2010ApJ...709..801I} {709, 801}

\bibitem[\protect\citeauthoryear{{Joblin}, {Leger}  \& {Martin}}{{Joblin}
  et~al.}{1992}]{job92}
{Joblin} C.,  {Leger} A.,   {Martin} P.,  1992, \mn@doi [\apjl]
  {10.1086/186456}, \href {http://adsabs.harvard.edu/abs/1992ApJ...393L..79J}
  {393, L79}

\bibitem[\protect\citeauthoryear{{Joblin}, {Szczerba}, {Bern{\'e}}  \&
  {Szyszka}}{{Joblin} et~al.}{2008}]{Joblin08}
{Joblin} C.,  {Szczerba} R.,  {Bern{\'e}} O.,   {Szyszka} C.,  2008, \mn@doi
  [\aap] {10.1051/0004-6361:20079061}, \href
  {http://adsabs.harvard.edu/abs/2008A%26A...490..189J} {490, 189}

\bibitem[\protect\citeauthoryear{{Lebouteiller}, {Barry}, {Spoon},
  {Bernard-Salas}, {Sloan}, {Houck}  \& {Weedman}}{{Lebouteiller}
  et~al.}{2011}]{cassis}
{Lebouteiller} V.,  {Barry} D.~J.,  {Spoon} H.~W.~W.,  {Bernard-Salas} J.,
  {Sloan} G.~C.,  {Houck} J.~R.,   {Weedman} D.~W.,  2011, \mn@doi [\apjs]
  {10.1088/0067-0049/196/1/8}, \href
  {https://ui.adsabs.harvard.edu/abs/2011ApJS..196....8L} {196, 8}

\bibitem[\protect\citeauthoryear{{Leipski}, {Antonucci}, {Ogle}  \&
  {Whysong}}{{Leipski} et~al.}{2009}]{Leipski09}
{Leipski} C.,  {Antonucci} R.,  {Ogle} P.,   {Whysong} D.,  2009, \mn@doi
  [\apj] {10.1088/0004-637X/701/2/891}, \href
  {http://adsabs.harvard.edu/abs/2009ApJ...701..891L} {701, 891}

\bibitem[\protect\citeauthoryear{{Li}}{{Li}}{2004}]{Li04}
{Li} A.,  2004, in {Witt} A.~N.,  {Clayton} G.~C.,   {Draine} B.~T.,  eds,
  Astronomical Society of the Pacific Conference Series Vol. 309, Astrophysics
  of Dust. p.~417 (\mn@eprint {} {astro-ph/0311066})

\bibitem[\protect\citeauthoryear{{Maragkoudakis}, {Ivkovich}, {Peeters},
  {Stock}, {Hemachandra}  \& {Tielens}}{{Maragkoudakis} et~al.}{2018}]{Marag18}
{Maragkoudakis} A.,  {Ivkovich} N.,  {Peeters} E.,  {Stock} D.~J.,
  {Hemachandra} D.,   {Tielens} A.~G.~G.~M.,  2018, \mn@doi [\mnras]
  {10.1093/mnras/sty2658}, \href
  {http://adsabs.harvard.edu/abs/2018MNRAS.481.5370M} {481, 5370}

\bibitem[\protect\citeauthoryear{{Monfredini} et~al.,}{{Monfredini}
  et~al.}{2019}]{Monfredini19}
{Monfredini} T.,  et~al., 2019, \mn@doi [\mnras] {10.1093/mnras/stz1021}, \href
  {https://ui.adsabs.harvard.edu/abs/2019MNRAS.488..451M} {488, 451}

\bibitem[\protect\citeauthoryear{{Papovich} et~al.,}{{Papovich}
  et~al.}{2006}]{Papovich06}
{Papovich} C.,  et~al., 2006, \mn@doi [\apj] {10.1086/499915}, \href
  {http://adsabs.harvard.edu/abs/2006ApJ...640...92P} {640, 92}

\bibitem[\protect\citeauthoryear{{Peeters}, {Hony}, {Van Kerckhoven},
  {Tielens}, {Allamandola}, {Hudgins}  \& {Bauschlicher}}{{Peeters}
  et~al.}{2002}]{Peeters02}
{Peeters} E.,  {Hony} S.,  {Van Kerckhoven} C.,  {Tielens} A.~G.~G.~M.,
  {Allamandola} L.~J.,  {Hudgins} D.~M.,   {Bauschlicher} C.~W.,  2002, \mn@doi
  [\aap] {10.1051/0004-6361:20020773}, \href
  {http://adsabs.harvard.edu/abs/2002A%26A...390.1089P} {390, 1089}

\bibitem[\protect\citeauthoryear{{Peeters}, {Bauschlicher}, {Allamandola},
  {Tielens}, {Ricca}  \& {Wolfire}}{{Peeters} et~al.}{2017}]{Peeters17}
{Peeters} E.,  {Bauschlicher} Jr. C.~W.,  {Allamandola} L.~J.,  {Tielens}
  A.~G.~G.~M.,  {Ricca} A.,   {Wolfire} M.~G.,  2017, \mn@doi [\apj]
  {10.3847/1538-4357/836/2/198}, \href
  {http://adsabs.harvard.edu/abs/2017ApJ...836..198P} {836, 198}

\bibitem[\protect\citeauthoryear{{Pino} et~al.,}{{Pino} et~al.}{2008}]{pino08}
{Pino} T.,  et~al., 2008, \mn@doi [\aap] {10.1051/0004-6361:200809927}, \href
  {http://adsabs.harvard.edu/abs/2008A\%26A...490..665P} {490, 665}

\bibitem[\protect\citeauthoryear{{Pope} et~al.,}{{Pope} et~al.}{2008}]{Pope08}
{Pope} A.,  et~al., 2008, \mn@doi [\apj] {10.1086/527030}, \href
  {http://adsabs.harvard.edu/abs/2008ApJ...675.1171P} {675, 1171}

\bibitem[\protect\citeauthoryear{{Rapacioli}, {Joblin}  \&
  {Boissel}}{{Rapacioli} et~al.}{2005}]{Rapacioli05}
{Rapacioli} M.,  {Joblin} C.,   {Boissel} P.,  2005, \mn@doi [\aap]
  {10.1051/0004-6361:20041247}, \href
  {http://adsabs.harvard.edu/abs/2005A%26A...429..193R} {429, 193}

\bibitem[\protect\citeauthoryear{{Ricca}, {Bauschlicher}, {Boersma}, {Tielens}
  \& {Allamandola}}{{Ricca} et~al.}{2012}]{Ricca12}
{Ricca} A.,  {Bauschlicher} Jr. C.~W.,  {Boersma} C.,  {Tielens} A.~G.~G.~M.,
  {Allamandola} L.~J.,  2012, \mn@doi [\apj] {10.1088/0004-637X/754/1/75},
  \href {http://adsabs.harvard.edu/abs/2012ApJ...754...75R} {754, 75}

\bibitem[\protect\citeauthoryear{{Ricca}, {Bauschlicher}, {Roser}  \&
  {Peeters}}{{Ricca} et~al.}{2018}]{Ricca18}
{Ricca} A.,  {Bauschlicher} Jr. C.~W.,  {Roser} J.~E.,   {Peeters} E.,  2018,
  \mn@doi [\apj] {10.3847/1538-4357/aaa757}, \href
  {http://adsabs.harvard.edu/abs/2018ApJ...854..115R} {854, 115}

\bibitem[\protect\citeauthoryear{{Ruschel-Dutra}, {Pastoriza}, {Riffel},
  {Sales}  \& {Winge}}{{Ruschel-Dutra} et~al.}{2014}]{ruschel-dutra14}
{Ruschel-Dutra} D.,  {Pastoriza} M.,  {Riffel} R.,  {Sales} D.~A.,   {Winge}
  C.,  2014, \mn@doi [\mnras] {10.1093/mnras/stt2448}, \href
  {https://ui.adsabs.harvard.edu/abs/2014MNRAS.438.3434R} {438, 3434}

\bibitem[\protect\citeauthoryear{{Sales}, {Pastoriza}  \& {Riffel}}{{Sales}
  et~al.}{2010}]{sales10}
{Sales} D.~A.,  {Pastoriza} M.~G.,   {Riffel} R.,  2010, \mn@doi [\apj]
  {10.1088/0004-637X/725/1/605}, \href
  {https://ui.adsabs.harvard.edu/abs/2010ApJ...725..605S} {725, 605}

\bibitem[\protect\citeauthoryear{{Sales}, {Pastoriza}, {Riffel}  \&
  {Winge}}{{Sales} et~al.}{2013}]{sales13}
{Sales} D.~A.,  {Pastoriza} M.~G.,  {Riffel} R.,   {Winge} C.,  2013, \mn@doi
  [\mnras] {10.1093/mnras/sts542}, \href
  {https://ui.adsabs.harvard.edu/abs/2013MNRAS.429.2634S} {429, 2634}

\bibitem[\protect\citeauthoryear{{Schlafly} et~al.,}{{Schlafly}
  et~al.}{2016}]{Schlafly2016}
{Schlafly} E.~F.,  et~al., 2016, \mn@doi [\apj] {10.3847/0004-637X/821/2/78},
  \href {https://ui.adsabs.harvard.edu/abs/2016ApJ...821...78S} {821, 78}

\bibitem[\protect\citeauthoryear{{Shannon} \& {Boersma}}{{Shannon} \&
  {Boersma}}{2019}]{Shannon19}
{Shannon} M.~J.,  {Boersma} C.,  2019, \mn@doi [\apj]
  {10.3847/1538-4357/aaf562}, \href
  {https://ui.adsabs.harvard.edu/abs/2019ApJ...871..124S} {871, 124}

\bibitem[\protect\citeauthoryear{{Shivaei} et~al.,}{{Shivaei}
  et~al.}{2017}]{Shivaei17}
{Shivaei} I.,  et~al., 2017, \mn@doi [\apj] {10.3847/1538-4357/aa619c}, \href
  {http://adsabs.harvard.edu/abs/2017ApJ...837..157S} {837, 157}

\bibitem[\protect\citeauthoryear{{Sloan} et~al.,}{{Sloan}
  et~al.}{2007}]{Sloan07}
{Sloan} G.~C.,  et~al., 2007, \mn@doi [\apj] {10.1086/519236}, \href
  {https://ui.adsabs.harvard.edu/abs/2007ApJ...664.1144S} {664, 1144}

\bibitem[\protect\citeauthoryear{{Smith} et~al.,}{{Smith}
  et~al.}{2007}]{Smith07}
{Smith} J.~D.~T.,  et~al., 2007, \mn@doi [\apj] {10.1086/510549}, \href
  {http://adsabs.harvard.edu/abs/2007ApJ...656..770S} {656, 770}

\bibitem[\protect\citeauthoryear{{Spoon}, {Marshall}, {Houck}, {Elitzur},
  {Hao}, {Armus}, {Brandl}  \& {Charmandaris}}{{Spoon}
  et~al.}{2007}]{Spoon2007}
{Spoon} H.~W.~W.,  {Marshall} J.~A.,  {Houck} J.~R.,  {Elitzur} M.,  {Hao} L.,
  {Armus} L.,  {Brandl} B.~R.,   {Charmandaris} V.,  2007, \mn@doi [\apjl]
  {10.1086/511268}, \href
  {https://ui.adsabs.harvard.edu/abs/2007ApJ...654L..49S} {654, L49}

\bibitem[\protect\citeauthoryear{{Stiavelli} et~al.,}{{Stiavelli}
  et~al.}{2009}]{Stiavelli09}
{Stiavelli} M.,  et~al., 2009, in astro2010: The Astronomy and Astrophysics
  Decadal Survey. p.~287

\bibitem[\protect\citeauthoryear{{Stock} \& {Peeters}}{{Stock} \&
  {Peeters}}{2017}]{Stoch17}
{Stock} D.~J.,  {Peeters} E.,  2017, \mn@doi [\apj] {10.3847/1538-4357/aa5f54},
  \href {http://adsabs.harvard.edu/abs/2017ApJ...837..129S} {837, 129}

\bibitem[\protect\citeauthoryear{{Teplitz} et~al.,}{{Teplitz}
  et~al.}{2007}]{Teplitz07}
{Teplitz} H.~I.,  et~al., 2007, \mn@doi [\apj] {10.1086/512802}, \href
  {http://adsabs.harvard.edu/abs/2007ApJ...659..941T} {659, 941}

\bibitem[\protect\citeauthoryear{{Tielens}}{{Tielens}}{2008}]{Tielens08}
{Tielens} A.~G.~G.~M.,  2008, \mn@doi [\araa]
  {10.1146/annurev.astro.46.060407.145211}, \href
  {http://adsabs.harvard.edu/abs/2008ARA%26A..46..289T} {46, 289}

\bibitem[\protect\citeauthoryear{{Vega} et~al.,}{{Vega} et~al.}{2010}]{Vega10}
{Vega} O.,  et~al., 2010, \mn@doi [\apj] {10.1088/0004-637X/721/2/1090}, \href
  {http://adsabs.harvard.edu/abs/2010ApJ...721.1090V} {721, 1090}

\bibitem[\protect\citeauthoryear{{Weedman} \& {Houck}}{{Weedman} \&
  {Houck}}{2009}]{Weedman09}
{Weedman} D.~W.,  {Houck} J.~R.,  2009, \mn@doi [\apj]
  {10.1088/0004-637X/693/1/370}, \href
  {http://adsabs.harvard.edu/abs/2009ApJ...693..370W} {693, 370}

\bibitem[\protect\citeauthoryear{{Werner} et~al.,}{{Werner}
  et~al.}{2004}]{Werner04}
{Werner} M.~W.,  et~al., 2004, \mn@doi [\apjs] {10.1086/422992}, \href
  {http://adsabs.harvard.edu/abs/2004ApJS..154....1W} {154, 1}

\bibitem[\protect\citeauthoryear{{Willett}, {Stocke}, {Darling}  \&
  {Perlman}}{{Willett} et~al.}{2010}]{Willett2010}
{Willett} K.~W.,  {Stocke} J.~T.,  {Darling} J.,   {Perlman} E.~S.,  2010,
  \mn@doi [\apj] {10.1088/0004-637X/713/2/1393}, \href
  {https://ui.adsabs.harvard.edu/abs/2010ApJ...713.1393W} {713, 1393}

\bibitem[\protect\citeauthoryear{{Wu}, {Charmandaris}, {Huang}, {Spinoglio}  \&
  {Tommasin}}{{Wu} et~al.}{2009}]{Wu09}
{Wu} Y.,  {Charmandaris} V.,  {Huang} J.,  {Spinoglio} L.,   {Tommasin} S.,
  2009, \mn@doi [\apj] {10.1088/0004-637X/701/1/658}, \href
  {http://adsabs.harvard.edu/abs/2009ApJ...701..658W} {701, 658}

\bibitem[\protect\citeauthoryear{{Yan} et~al.,}{{Yan} et~al.}{2005}]{yan05}
{Yan} L.,  et~al., 2005, \mn@doi [\apj] {10.1086/431205}, \href
  {http://adsabs.harvard.edu/abs/2005ApJ...628..604Y} {628, 604}

\bibitem[\protect\citeauthoryear{{Yan} et~al.,}{{Yan} et~al.}{2007}]{yan07}
{Yan} L.,  et~al., 2007, \mn@doi [\apj] {10.1086/511516}, \href
  {http://adsabs.harvard.edu/abs/2007ApJ...658..778Y} {658, 778}

\bibitem[\protect\citeauthoryear{{van Diedenhoven}, {Peeters}, {Van
  Kerckhoven}, {Hony}, {Hudgins}, {Allamandola}  \& {Tielens}}{{van
  Diedenhoven} et~al.}{2004}]{died04}
{van Diedenhoven} B.,  {Peeters} E.,  {Van Kerckhoven} C.,  {Hony} S.,
  {Hudgins} D.~M.,  {Allamandola} L.~J.,   {Tielens} A.~G.~G.~M.,  2004,
  \mn@doi [\apj] {10.1086/422404}, \href
  {http://adsabs.harvard.edu/abs/2004ApJ...611..928V} {611, 928}

\makeatother
\end{thebibliography}




\appendix

\section{Sources -- identification and derived properties}


\begin{table*}
    \centering
    \caption{Sources and their respective information extracted from the MIR\_SB sample (Spitzer/IRS ATLAS, version 1.0) and \citet{yan07}, including their ID, type, source reference, right ascension, declination and redshift. Acronyms: AGN -- Active Galactic Nucleus, FR -- Fanaroff-Riley galaxy, HII --  HII region, IRgal -- Infrared galaxy, LINER -- Low-Ionization Nuclear Emission-line Region, QSO -- Quasi-Stellar Object, SB -- Starburst galaxy, SMG -- Submillimeter Galaxy, Sy -- Seyfert galaxy, ULIRG -- Ultra-Luminous Infrared Galaxy. The full table is available online.}
    \label{tab:sources}    
    \begin{tabular}{lccccc}
    \hline
    ID & Type & Reference & RA (hms) & Dec (dms) & z \\
    \hline
    3C293 & Sy3 & \citet{Leipski09} & 13:52:17.80 & 31:26:46.50 & 0.045 \\
    3C31 & FR-1 & \citet{Leipski09} & 01:07:24.90 & 32:24:45.20 & 0.017 \\
    AGN15* & LINER & \citet{Weedman09} & 17:18:52.71 & 59:14:32.00 & 0.322 \\
    Arp220 & ULIRG & \citet{Imanishi07} & 15:34:57.10 & 23:30:11.00 & 0.018 \\
    E12-G21 & Sy1 & \citet{Wu09} & 00:40:47.80 & -79:14:27.00 & 0.033 \\
    EIRS-2* & SB & \citet{Hernot-Caballero09} & 16:13:49.94 & 54:26:28.40 & 1.143 \\
    GN26 & SMG & \citet{Pope08} & 12:36:34.51 & 62:12:40.90 & 1.219 \\
    IC342 & SB & \citet{Brandl06} & 03:46:48.51 & 68:05:46.00 & 0.001 \\
    IRAS02021-2103 & ULIRG & \citet{Imanishi10} & 02:04:27.30 & -20:49:41 & 0.116 \\
    \vdots & \vdots & \vdots & \vdots & \vdots & \vdots \\
    UGC12138 & Sy1.8 & \citet{Deo07} & 22:40:17.00 & 08:03:14.00 & 0.025 \\
    \hline
    \end{tabular}
    
    *Objects with redshift obtained through the IRS spectrum.
    
\end{table*}


\section{Analysis of the extinction effect in object Mrk 52}
\label{sec:ext}
\begin{figure}
    \centering
    \includegraphics[width=\columnwidth,keepaspectratio]{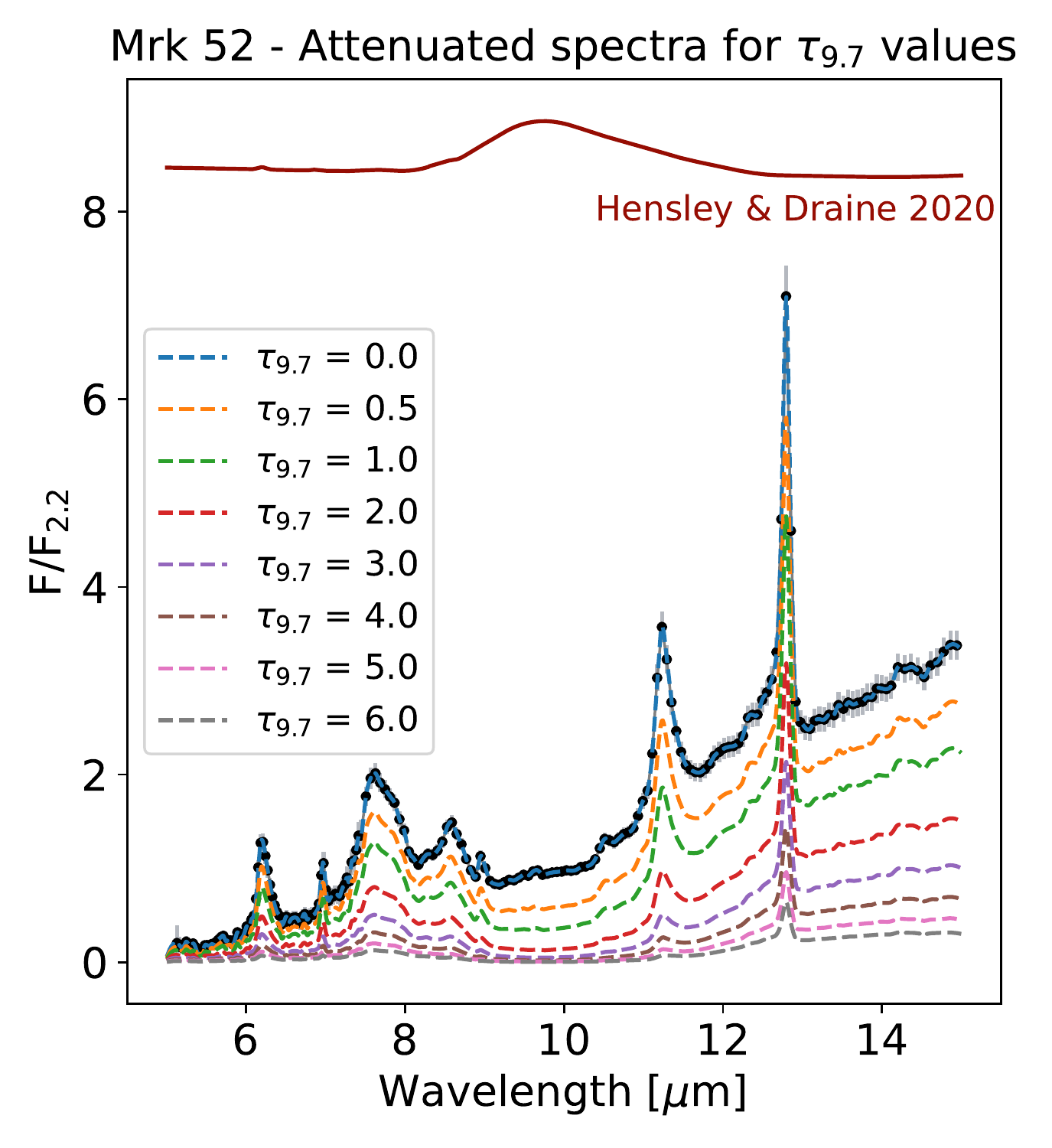}
    \caption{Attenuated spectra of the galaxy Mrk 52 for different values of $\tau_{9.7}$. The spectrum and the error bars are shown in black and grey, respectively. The extinction curve of \citet{Hensley2020} is also show, offset for better visualisation. }
    \label{fig:ext-cont}
\end{figure}

In order to analysed the effects of the extinction in our sample, we selected the starburst galaxy Mrk~52 as a general template of low silicate absorption spectra and we adopted the extinction curve from \citet{Hensley2020}\footnote{``Data behind the figure'', \url{https://iopscience.iop.org/article/10.3847/1538-4357/ab8cc3}}, normalised to the K band (at 2.2~$\mu$m). The flux at 2.2~$\mu$m for Mrk~52 was extracted from the ATLAS (F$_{2.2}$ = 82 $\pm$ 3.55 mJy). After the normalisation of the spectrum, we multiplied it by the following function:

\begin{equation}
    F_{ext} (\lambda) = \frac{F}{F_{2.2}}(\lambda) \times e^{ (-\tau_{norm}(\lambda)\times\tau_{9.7})}  
\end{equation}

\begin{equation}
    \tau_{norm}(\lambda) = \frac{\tau (\lambda)}{\tau (9.7)}
\end{equation}

\noindent where F$_{ext}$ is the spectrum of Mrk~52 with an extinction component added, F/F$_{2.2}$ is the normalised Mrk~52 spectrum, $\tau_{norm}(\lambda)$ is the normalised extinction curve \citep[$\tau (\lambda)$,][]{Hensley2020} to its respective value at 9.7~$\mu$m ($\tau(9.7)$), and $\tau_{9.7}$ is a scaling value that we varied from 0 to 6, as in the previous sections. The resulting spectra are displayed in Figure~\ref{fig:ext-cont}, together with the extinction curve, in the spectral interval of 5~--~15~$\mu$m. To perform this analysis, the original Mrk~52 spectrum needed to be interpolated and, therefore, the resulting spectra are just models and the original flux uncertainties were not considered.

The eight attenuated spectra (for $\tau_{9.7}$ = 0, 0.5, 1, 2, 3, 4, 5, and 6) were firstly submitted to the local spline decomposition, and then the 6.2, 7.7 and 8.6~$\mu$m PAH bands were fitted. The results can be seen in Figure~\ref{fig:ext-fits} and Table~\ref{tab:fits-ext}. The uncertainties shown in the table are derived from the \textit{scipy.optmize.curve\_fit} tool and do not take into account the observational errors. From the splines, it is possible to see that the silicate absorption at 9.7~$\mu$m gets broader while the entire spectral flux decreases for higher $\tau_{9.7}$ values. In particular, the 8.6~$\mu$m band is clearly the most affect by the extinction and almost disappears for $\tau_{9.7}$~=~6. The 6.2 and 7.7~$\mu$m bands also present a decrease in intensity, which seems to be more evident for the 6.2~$\mu$m band.  

The central wavelength of the  6.2~$\mu$m band is practically the same despite of the $\tau_{9.7}$ values and, therefore, the Peeters' classification is not affected by the extinction for this band. On the other hand, this parameter varies significantly for the 8.6~$\mu$m band, and the Peeters' classes changes from A to B with $\tau_{9.7}$ values higher than 4. This could lead to an overestimation of class B objects for this band. Nevertheless, sources with high extinction and silicate absorption are expected to present more dusty environments, which are consistent with the B classification. Finally, considering the 7.7~$\mu$m complex, F$_{7.6}$/F$_{7.8}$ ratio varied from 1.182~$\pm$~0.031 for $\tau_{9.7}$~=~0 to 1.135~$\pm$~0.028  for $\tau_{9.7}$~=~6.  Even though the ratio decreases, the extinction is not able to change the  Peeters' classification of this complex. These results suggest that the Peeters' classes are little or quite not influenced by the extinction, in the case of the 6.2 and 7.7~$\mu$m bands. However, the extinction could shift the classes of the  8.6~$\mu$m band from A to B for $\tau_{9.7}$~>~4.

\begin{table}
\centering
\caption{ Best-fit results for the 6.2, 7.7 and 8.6~$\mu$m bands. A is the amplitude in mJy/sr, $\lambda\_c$ is the central wavelength in $\mu$m and FWHM is the full width at half maximum. The respective $\tau_{9.7}$ values are also shown.}
\label{tab:fits-ext}
\begin{tabular}{ccccccc}
\hline
$\tau_{9.7}$ & $\lambda\_c$ & Err & A & Err & FWHM & Err \\
\hline
0.0 & 6.203 & 0.001 & 0.141 & 0.003 & 0.136 & 0.003 \\
 & 7.561 & 0.003 & 0.280 & 0.005 & 0.280 & --- \\
 & 7.821 & 0.004 & 0.238 & 0.005 & 0.320 & --- \\
 & 8.589 & 0.001 & 0.148 & 0.002 & 0.283 & 0.004 \\
\hline
0.5 & 6.202 & 0.001 & 0.109 & 0.002 & 0.136 & 0.003 \\
 & 7.562 & 0.003 & 0.219 & 0.004 & 0.280 & --- \\
 & 7.823 & 0.004 & 0.185 & 0.004 & 0.320 & --- \\
 & 8.590 & 0.001 & 0.109 & 0.001 & 0.280 & 0.004 \\
\hline 
1.0 & 6.203 & 0.001 & 0.085 & 0.002 & 0.136 & 0.003 \\
 & 7.562 & 0.003 & 0.172 & 0.003 & 0.280 & --- \\
 & 7.824 & 0.004 & 0.146 & 0.003 & 0.320 & --- \\
 & 8.592 & 0.001 & 0.082 & 0.001 & 0.278 & 0.004 \\
\hline
2.0 & 6.203 & 0.001 & 0.051 & 0.001 & 0.137 & 0.003 \\
 & 7.561 & 0.003 & 0.106 & 0.002 & 0.280 & --- \\
 & 7.827 & 0.004 & 0.090 & 0.002 & 0.320 & --- \\
 & 8.596 & 0.001 & 0.046 & 0.000 & 0.272 & 0.003 \\
\hline
3.0 & 6.203 & 0.001 & 0.031 & 0.001 & 0.138 & 0.003 \\
 & 7.561 & 0.003 & 0.066 & 0.001 & 0.280 & --- \\
 & 7.830 & 0.004 & 0.056 & 0.001 & 0.320 & --- \\
 & 8.598 & 0.001 & 0.024 & 0.000 & 0.262 & 0.003 \\
\hline
4.0 & 6.203 & 0.001 & 0.019 & 0.000 & 0.140 & 0.003 \\
 & 7.560 & 0.003 & 0.040 & 0.001 & 0.280 & --- \\
 & 7.830 & 0.004 & 0.035 & 0.001 & 0.320 & --- \\
 & 8.605 & 0.001 & 0.013 & 0.000 & 0.251 & 0.003 \\
\hline
5.0 & 6.203 & 0.001 & 0.012 & 0.000 & 0.140 & 0.003 \\
 & 7.560 & 0.003 & 0.025 & 0.000 & 0.280 & --- \\
 & 7.834 & 0.004 & 0.022 & 0.000 & 0.320 & --- \\
 & 8.608 & 0.001 & 0.006 & 0.000 & 0.229 & 0.003 \\
\hline
6.0 & 6.204 & 0.002 & 0.007 & 0.000 & 0.141 & 0.004 \\
 & 7.560 & 0.003 & 0.016 & 0.000 & 0.280 & --- \\
 & 7.838 & 0.004 & 0.014 & 0.000 & 0.320 & --- \\
 & 8.615 & 0.002 & 0.003 & 0.000 & 0.221 & 0.004 \\

 \hline
\end{tabular}
\end{table}

\begin{figure*}
    \centering
    \includegraphics[scale=0.38]{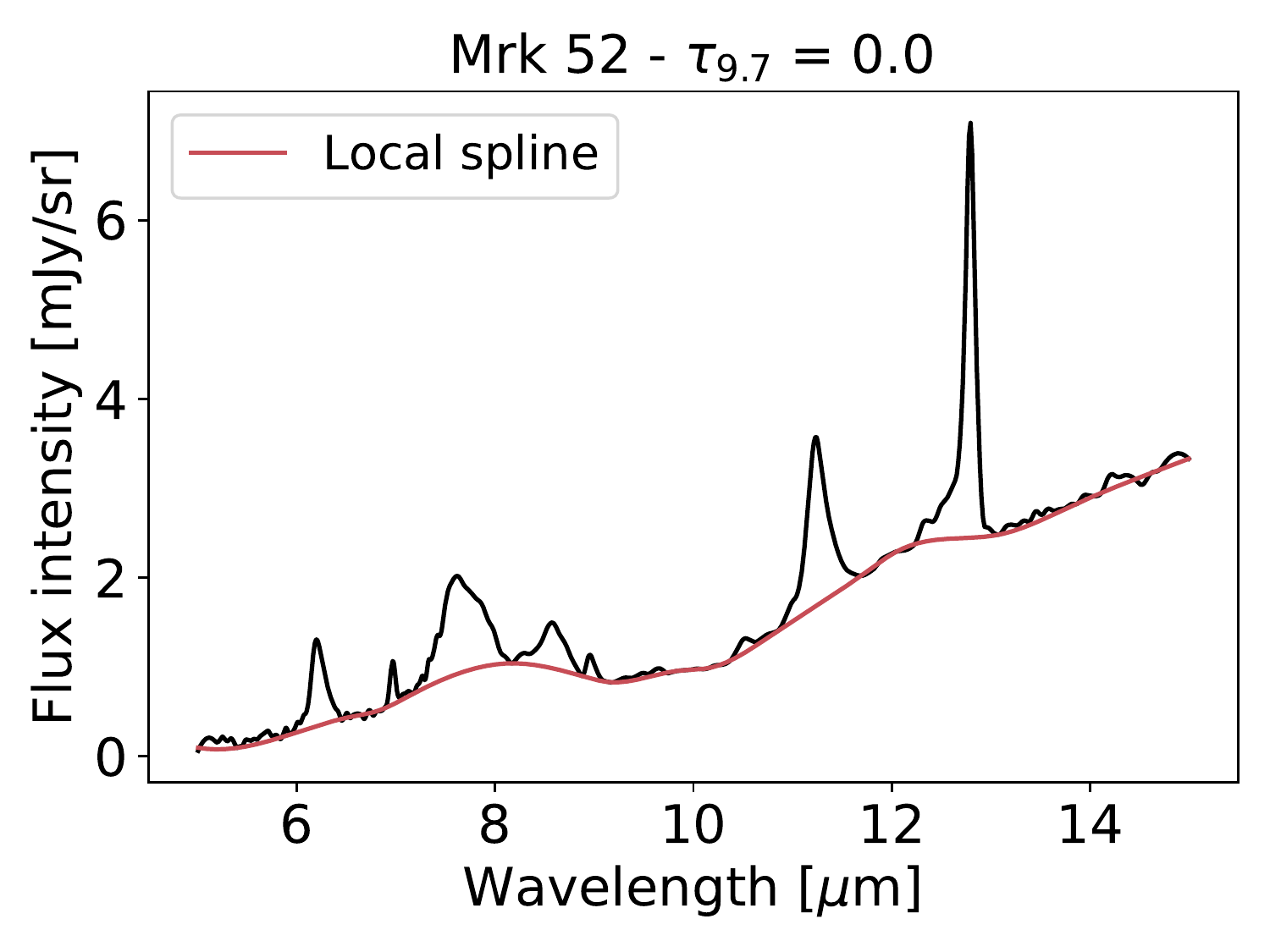}
    \includegraphics[scale=0.38]{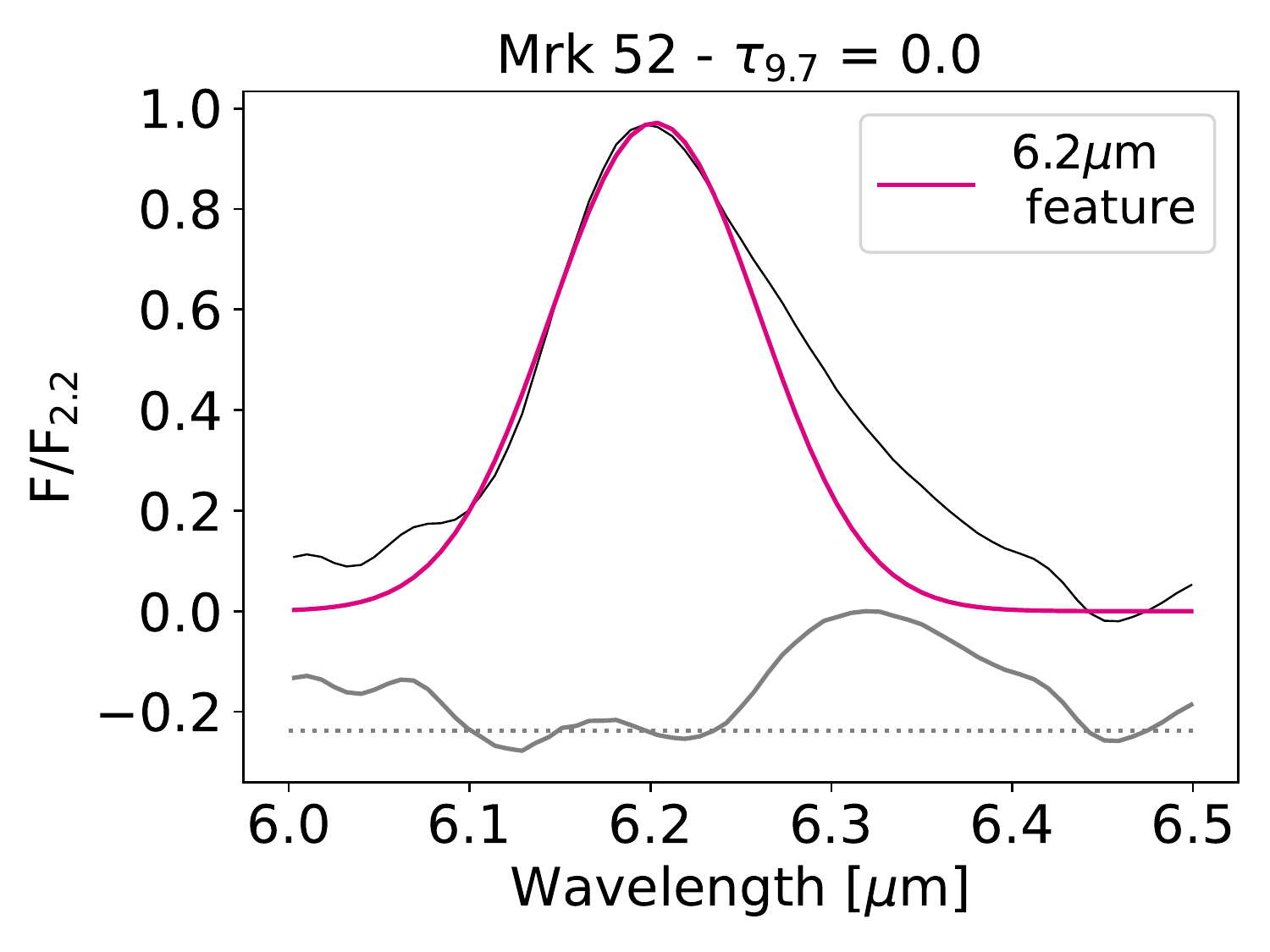}
    \includegraphics[scale=0.38]{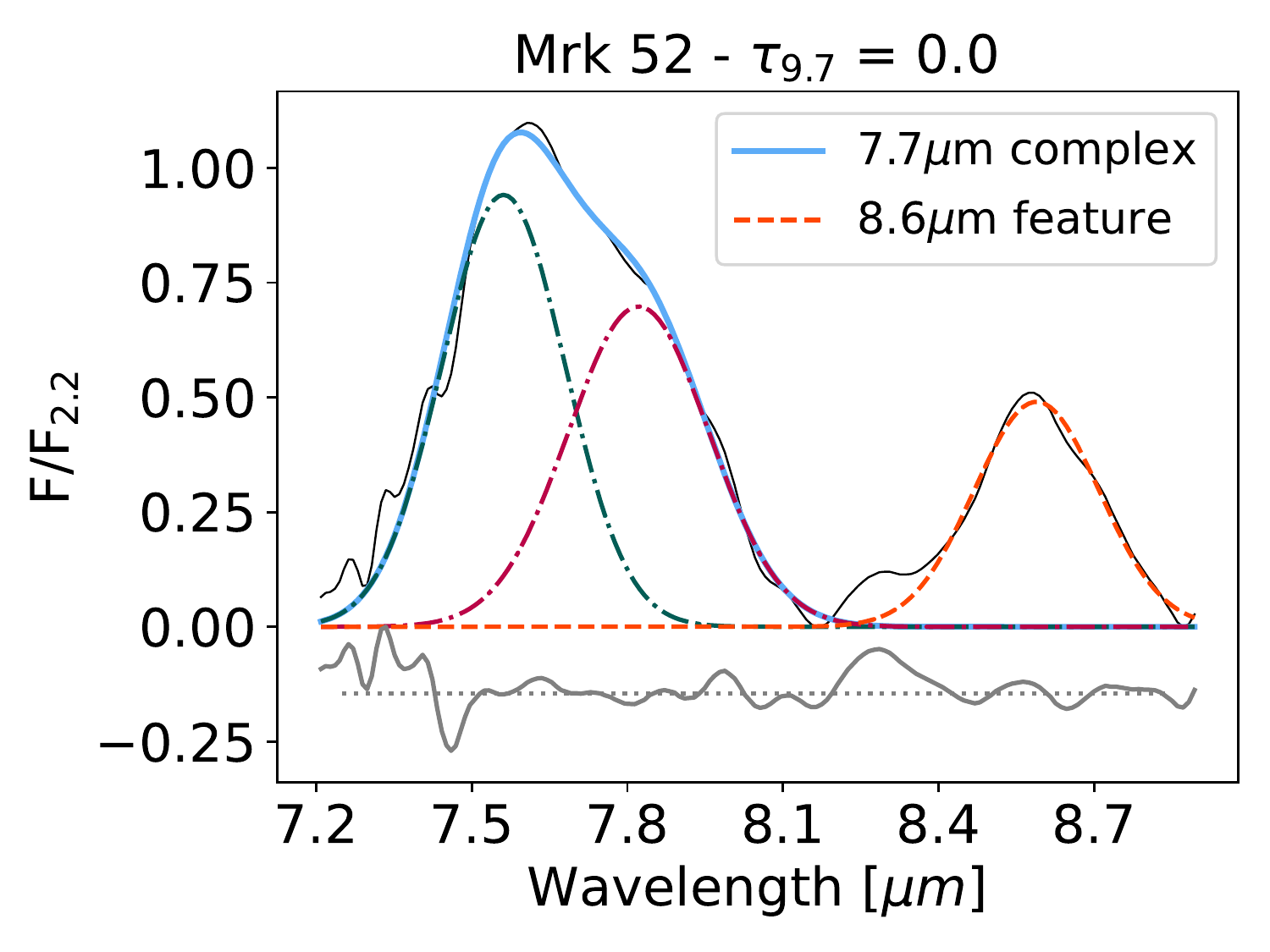}
    
    \includegraphics[scale=0.38]{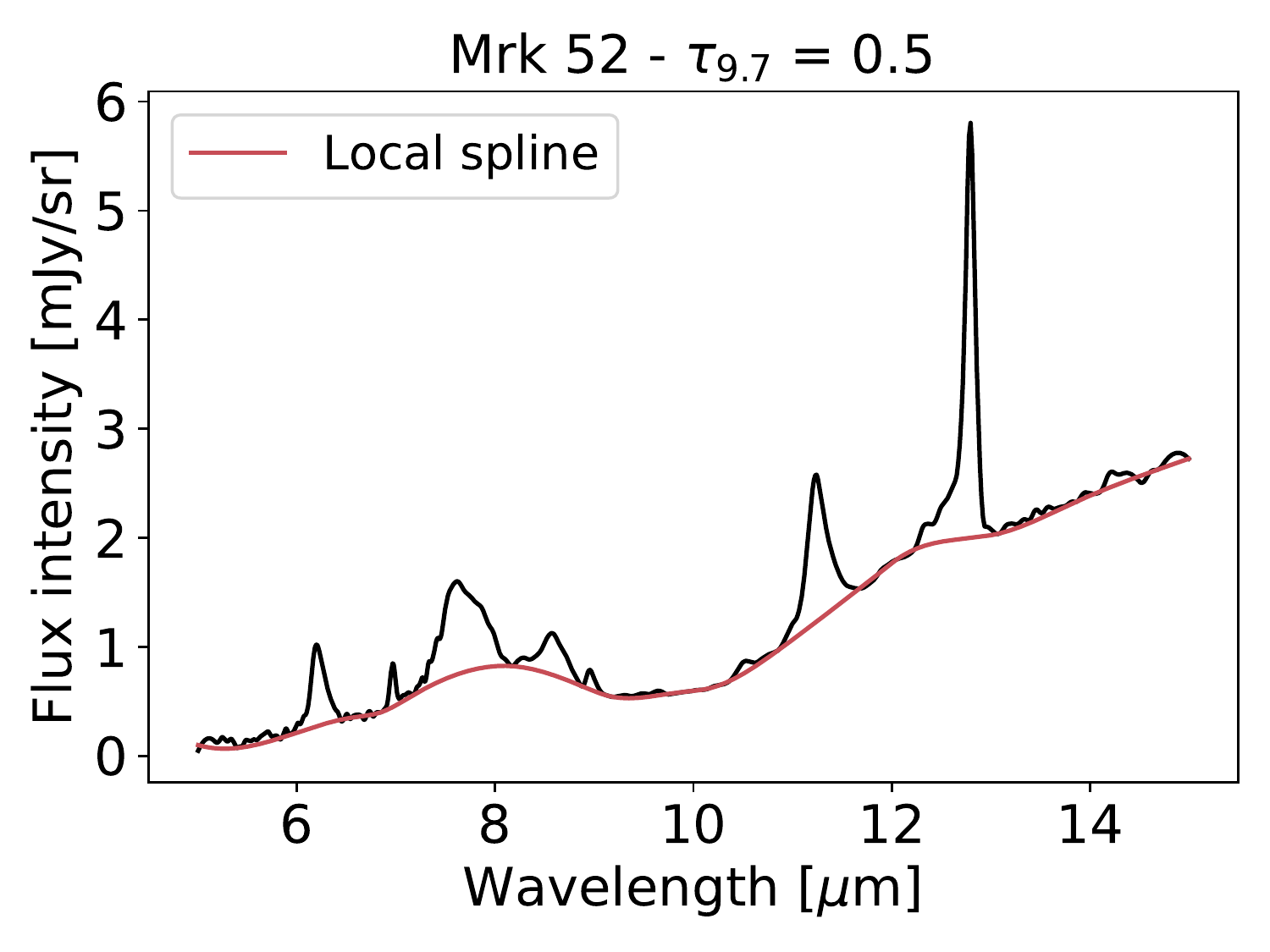}
    \includegraphics[scale=0.38]{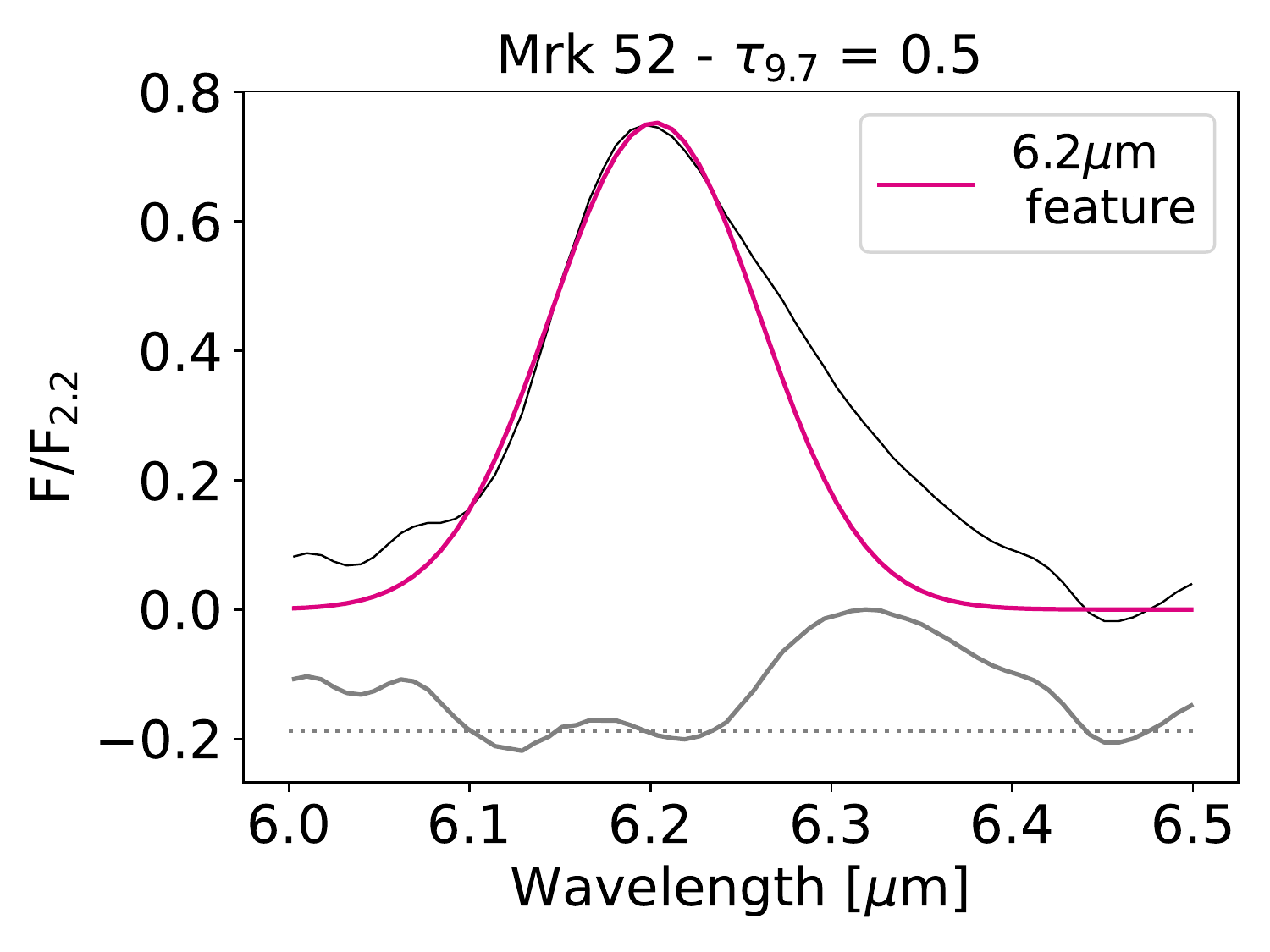}
    \includegraphics[scale=0.38]{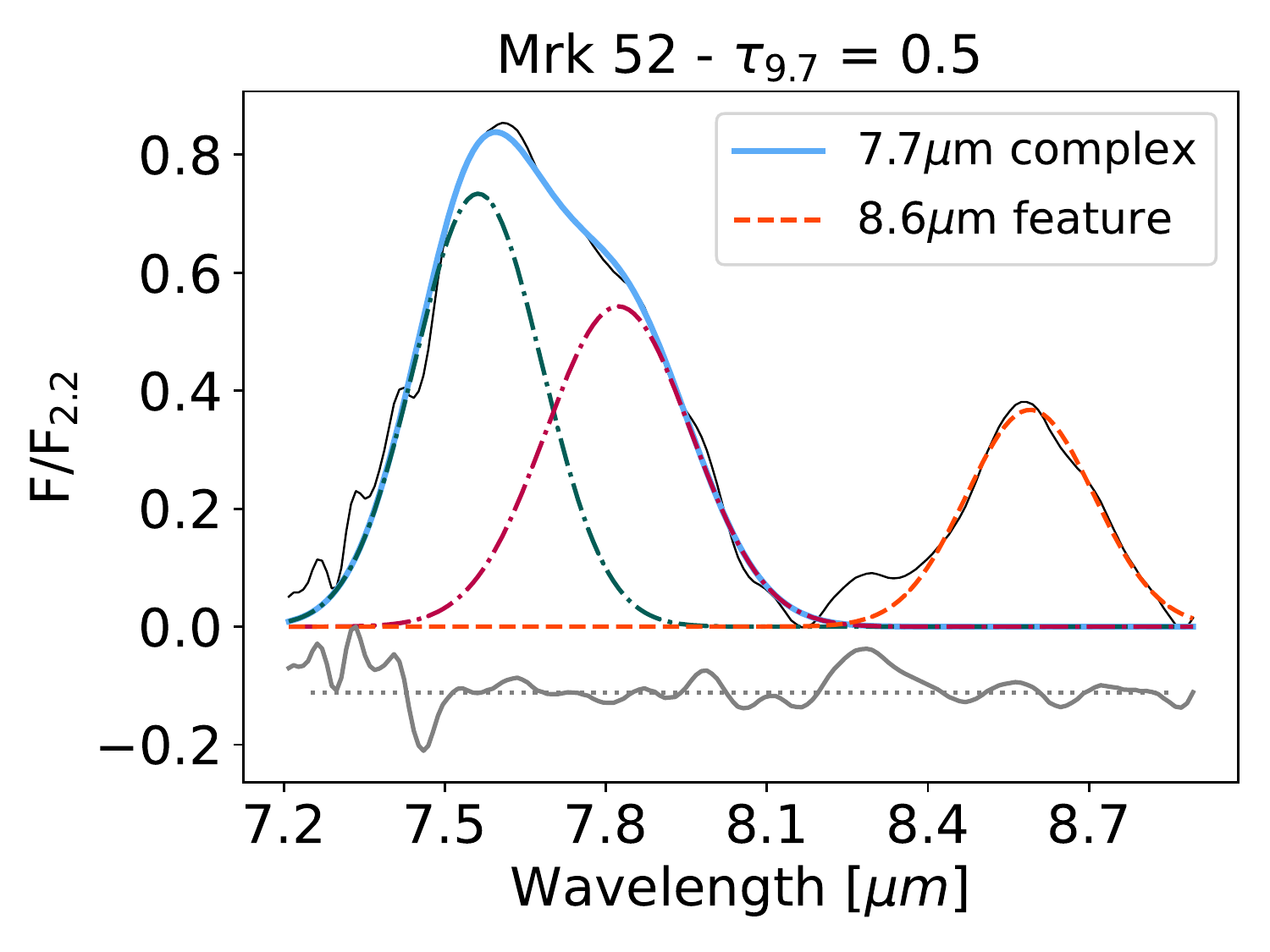}
    
    \includegraphics[scale=0.38]{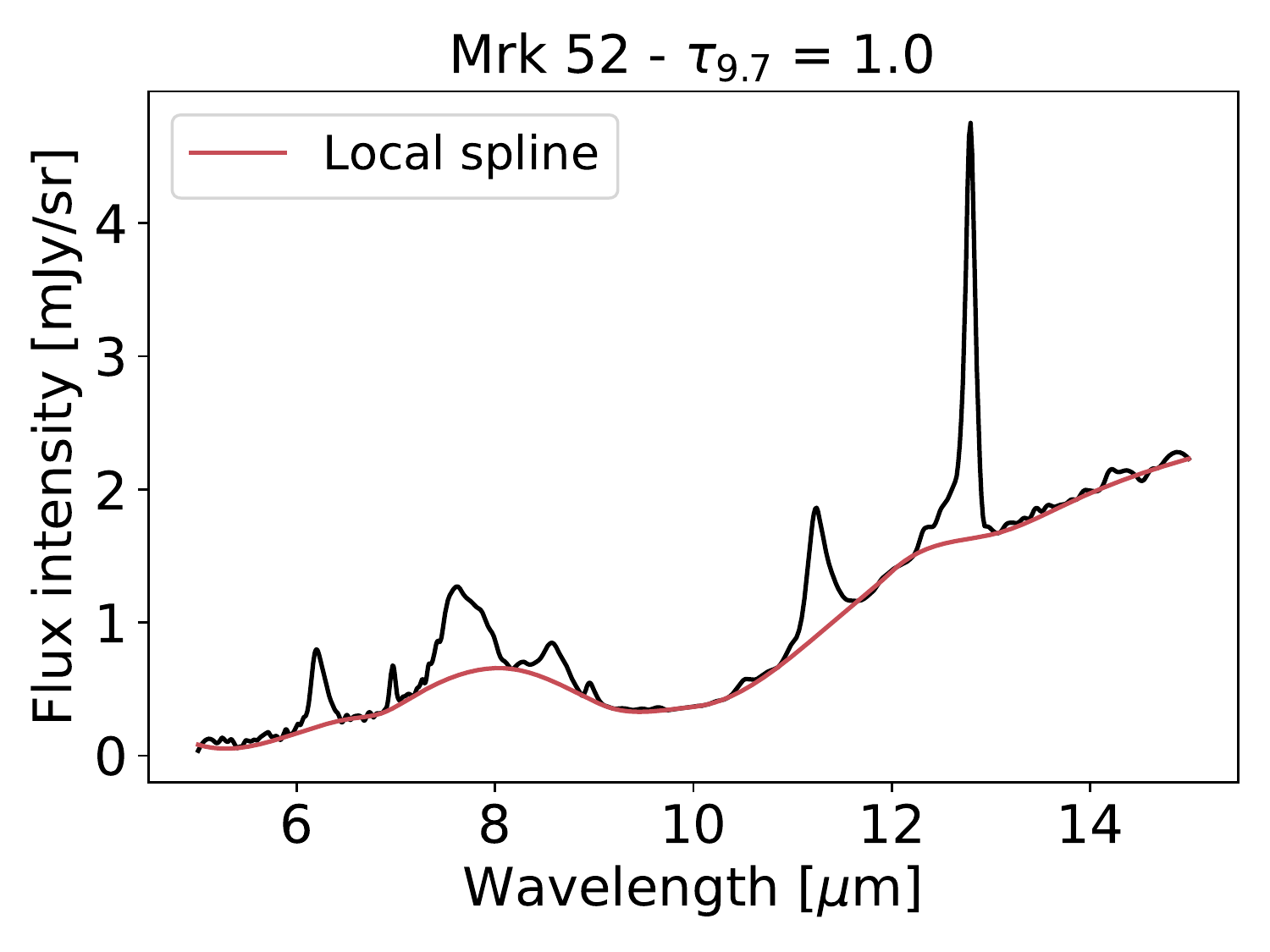}
    \includegraphics[scale=0.38]{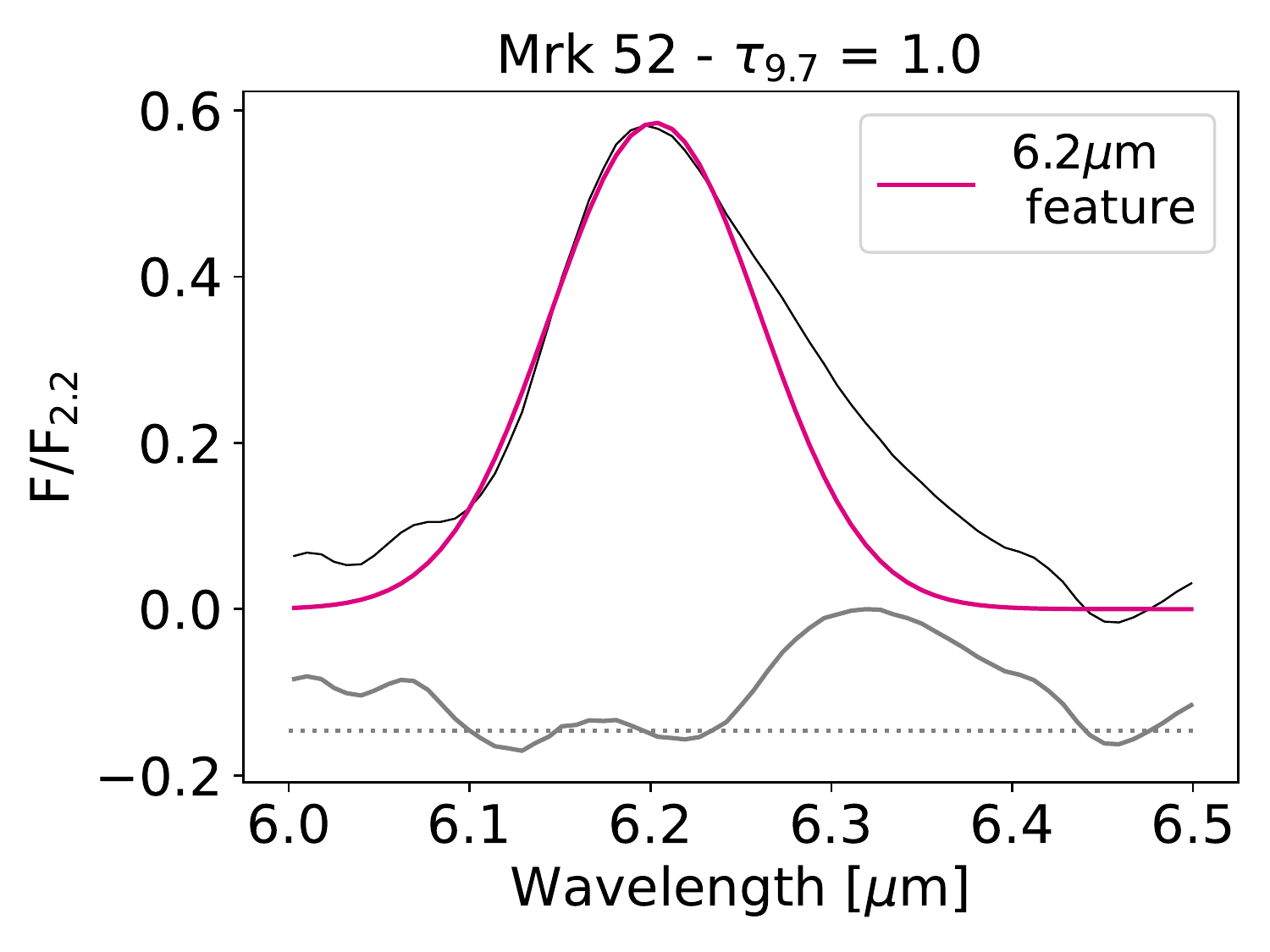}
    \includegraphics[scale=0.38]{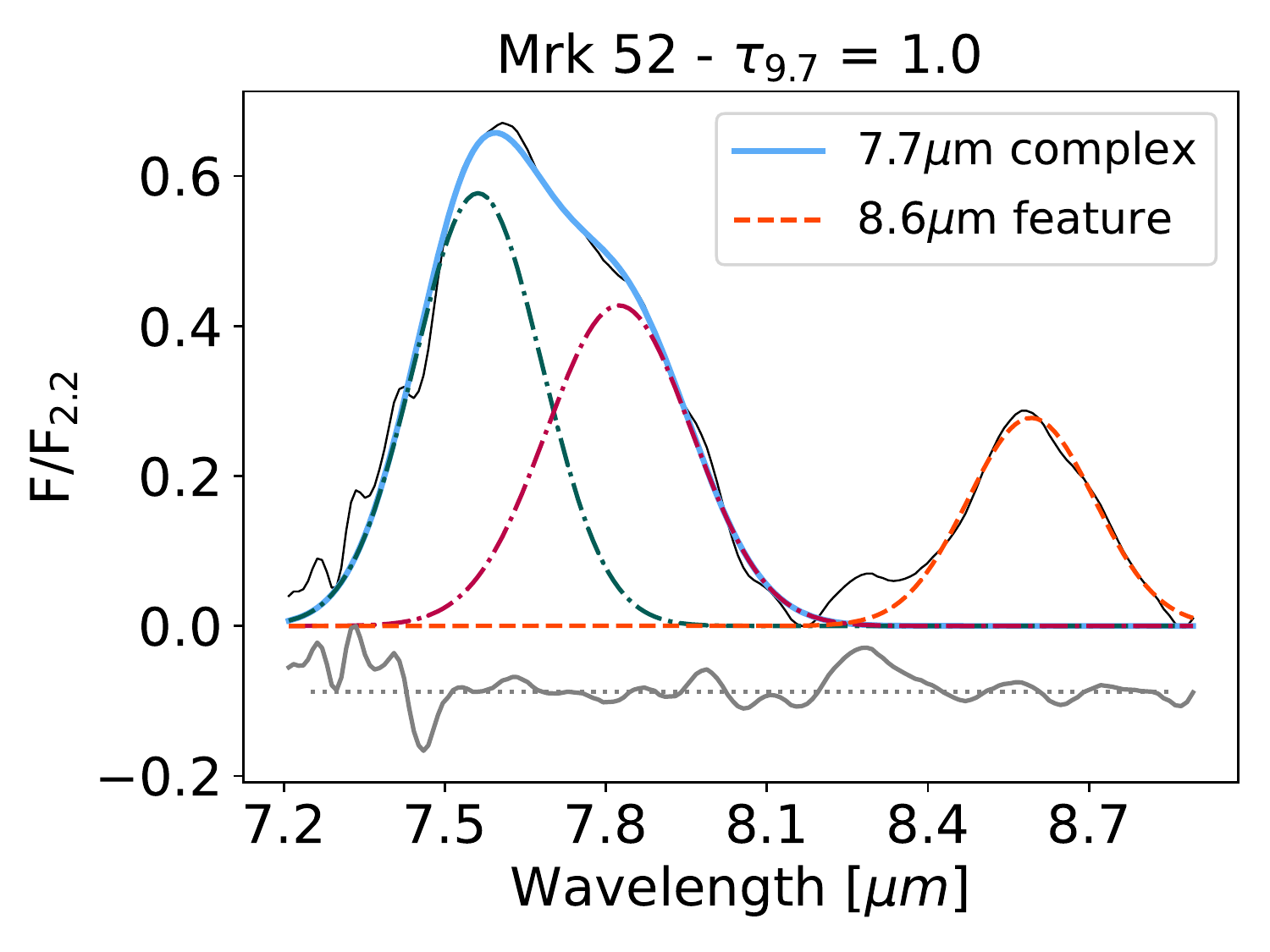}
    
    \includegraphics[scale=0.38]{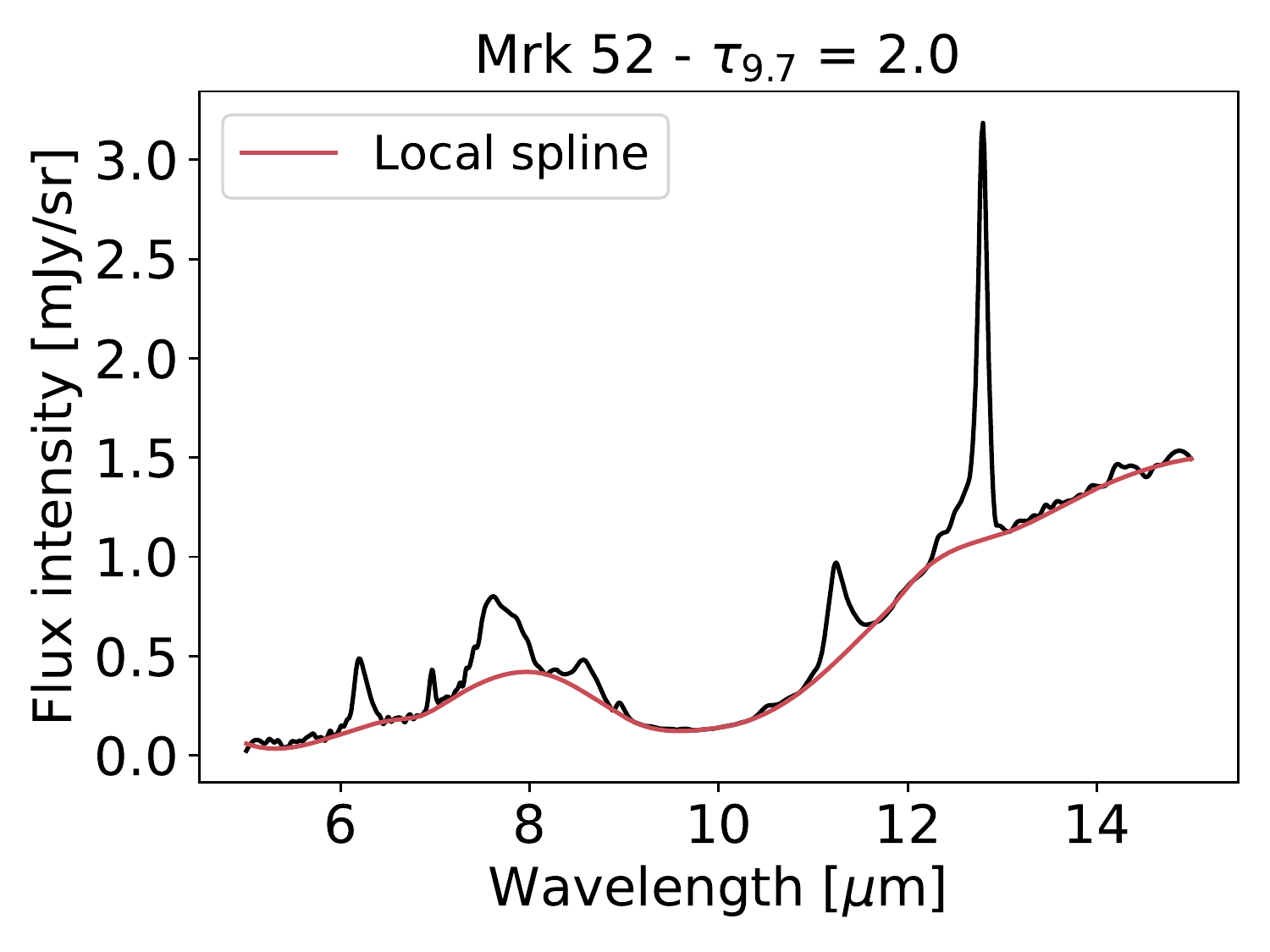}
    \includegraphics[scale=0.38]{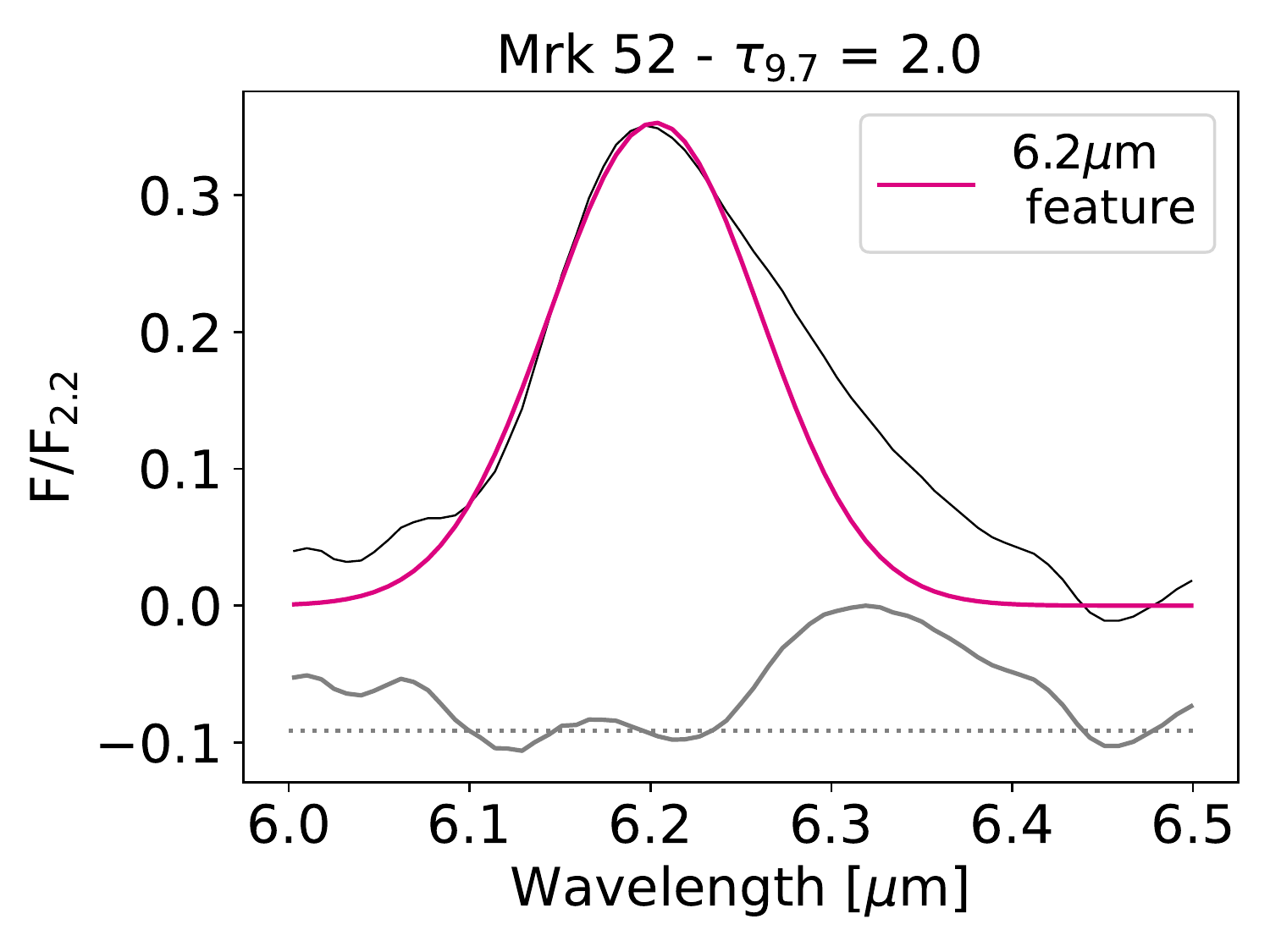}
    \includegraphics[scale=0.38]{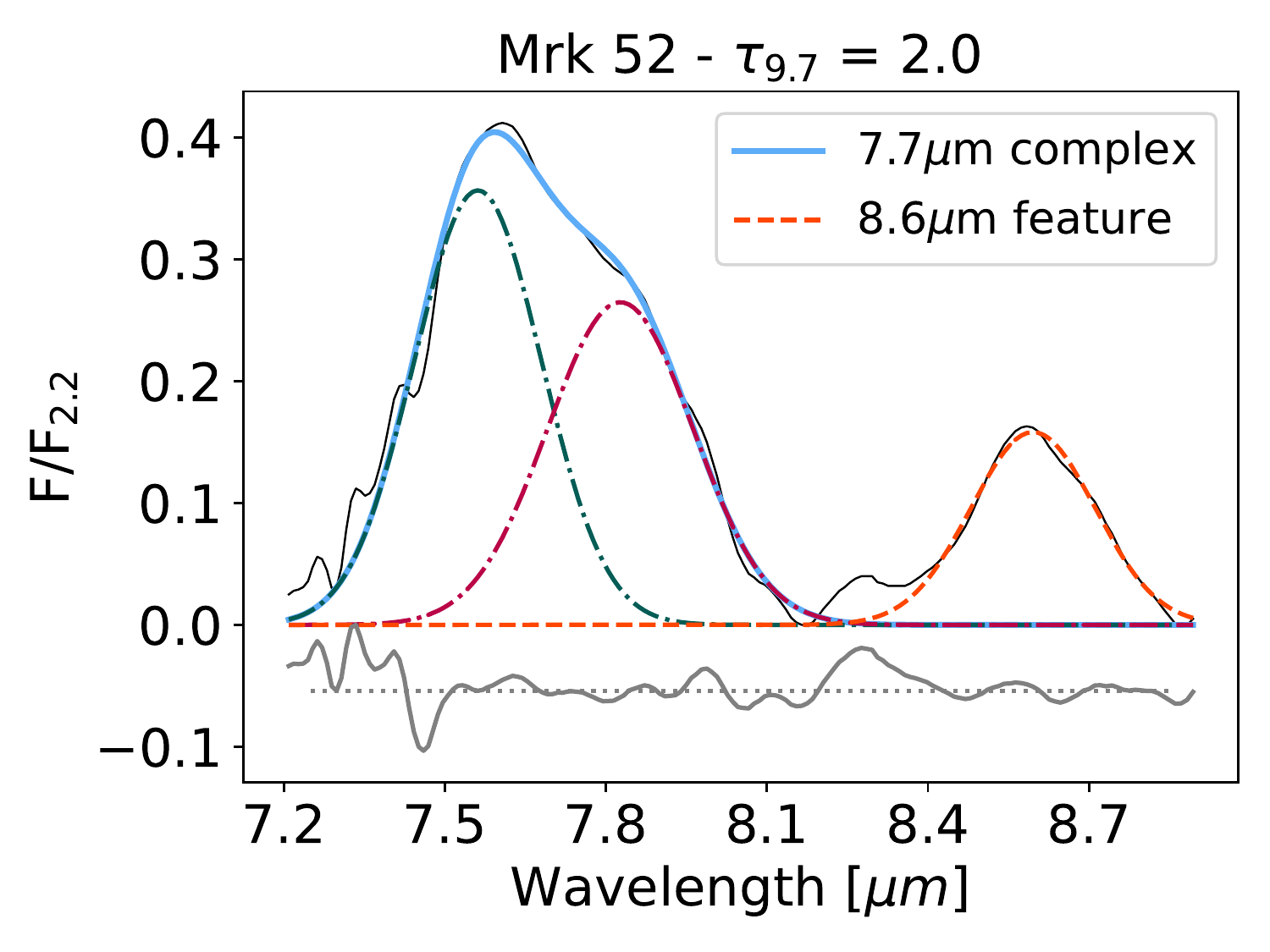}
    
    \includegraphics[scale=0.38]{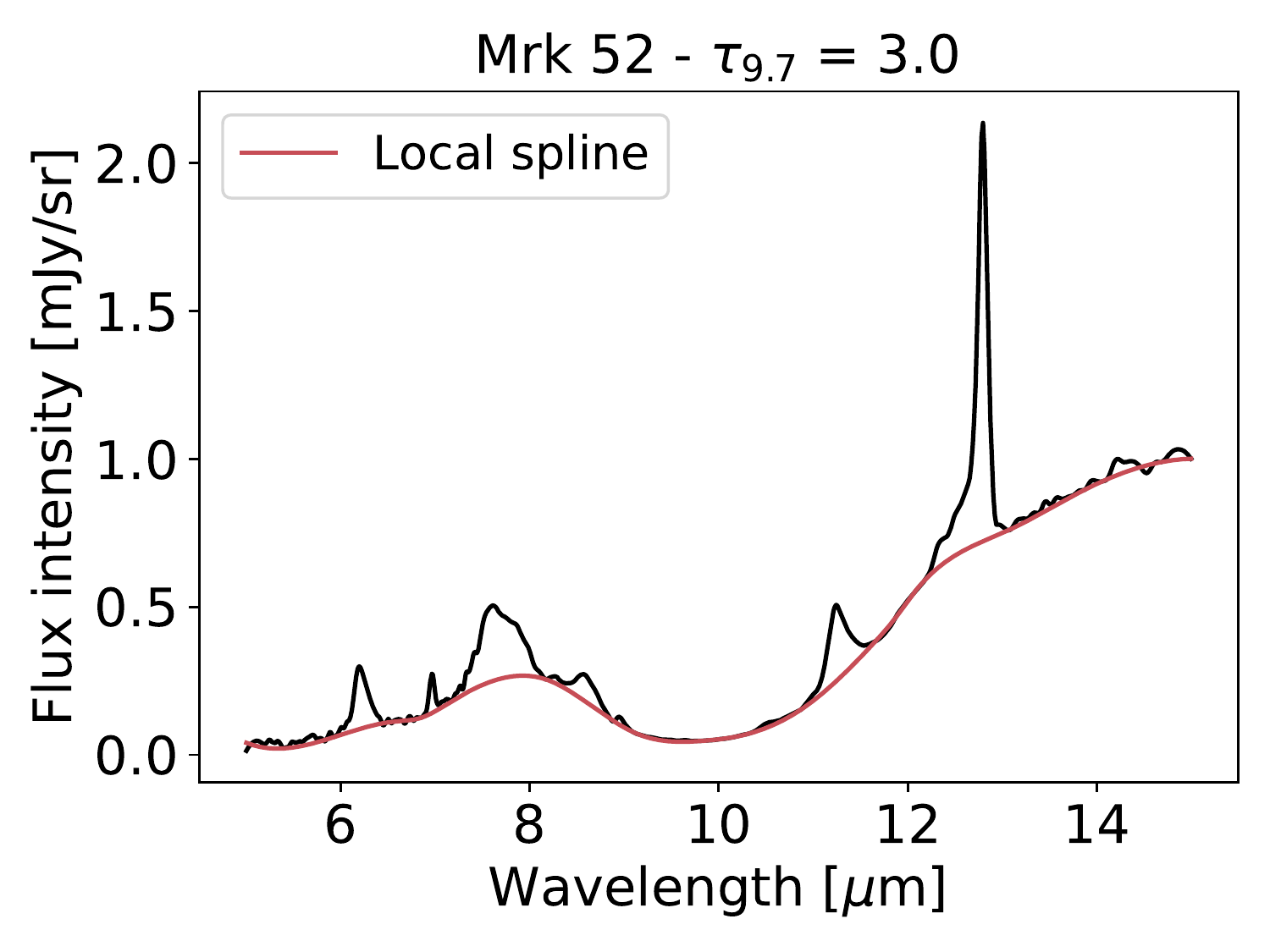}
    \includegraphics[scale=0.38]{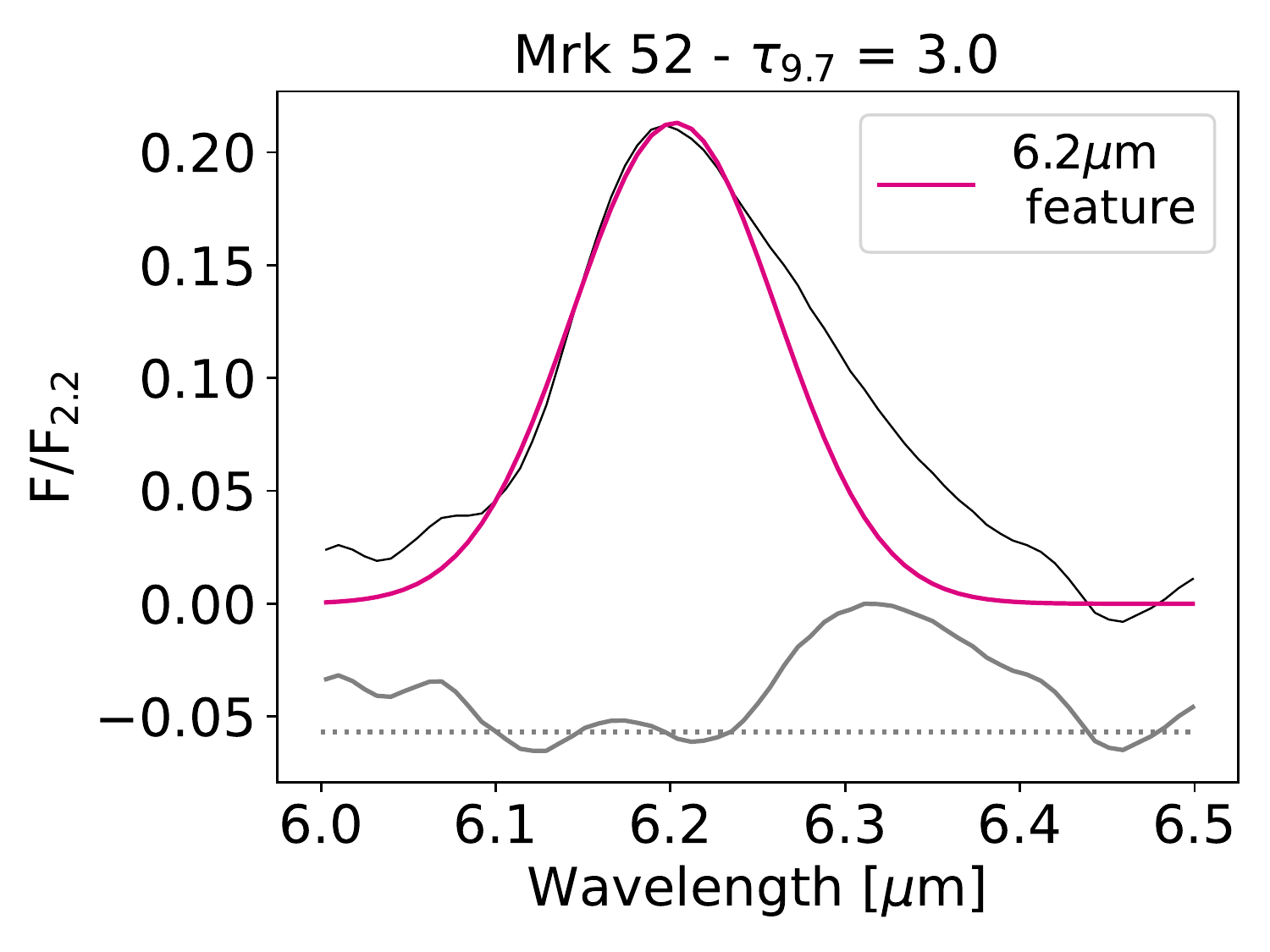}
    \includegraphics[scale=0.38]{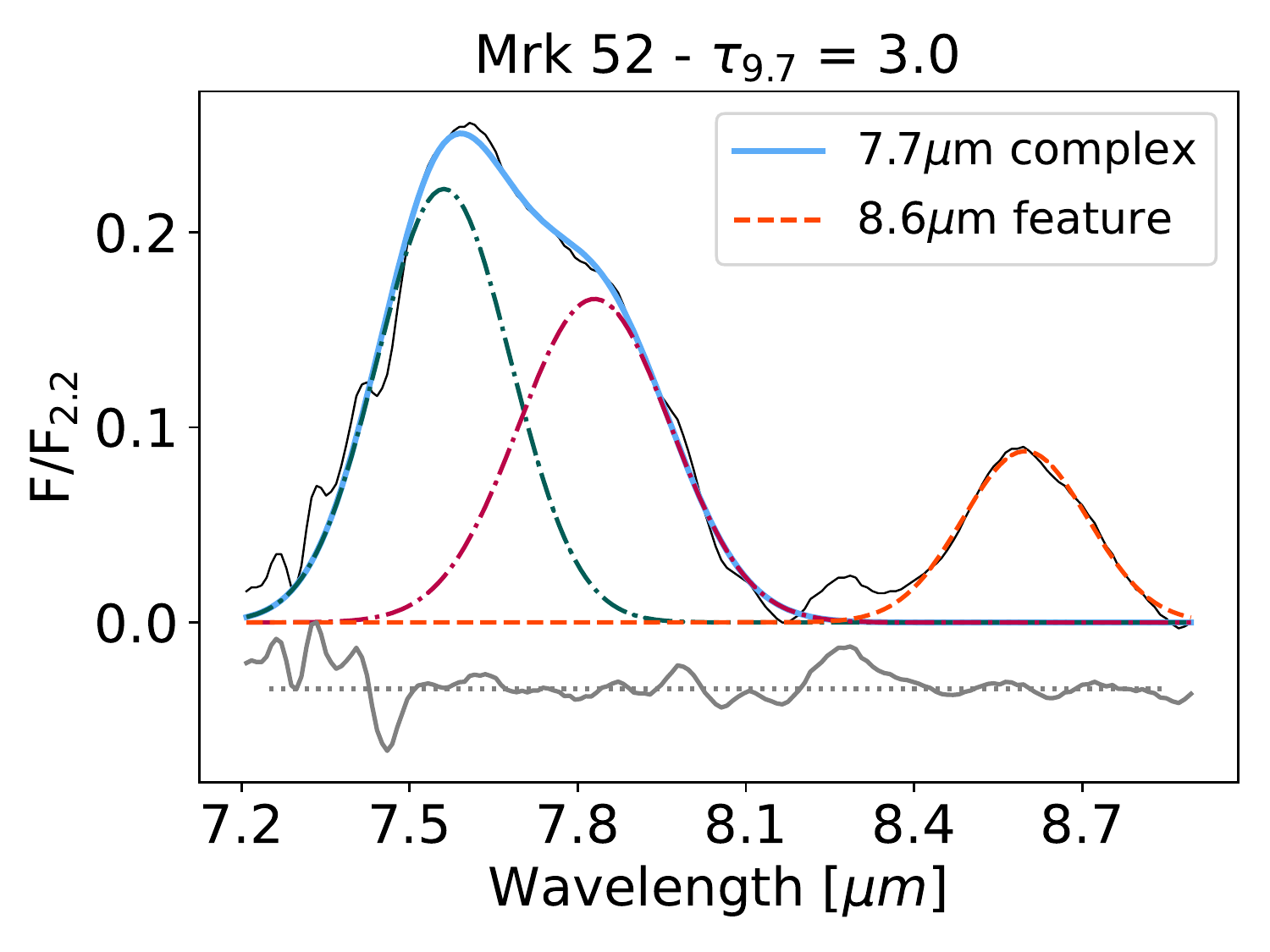}
    
    \caption{Local spline decomposition and fit results of the 6.2, 7.7 and 8.6 bands of the attenuated spectra of Mrk~52 for $\tau_{9.7}$ values of 0, 0.5, 1, 2, 3, 4, 5, and 6, from top to down.} 
    \label{fig:ext-fits}
\end{figure*}

\begin{figure*}
    \centering
    
    \includegraphics[scale=0.38]{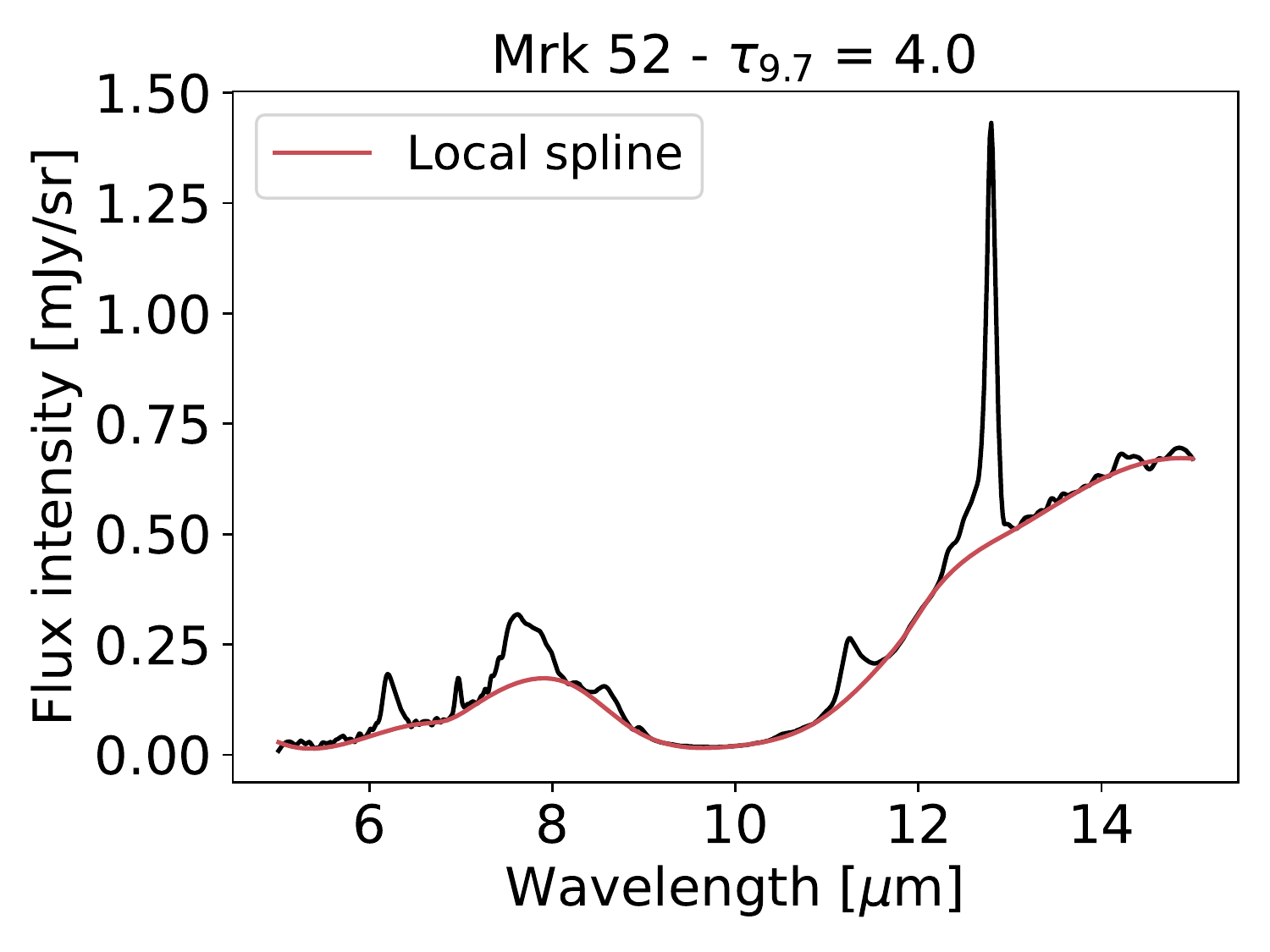}
    \includegraphics[scale=0.38]{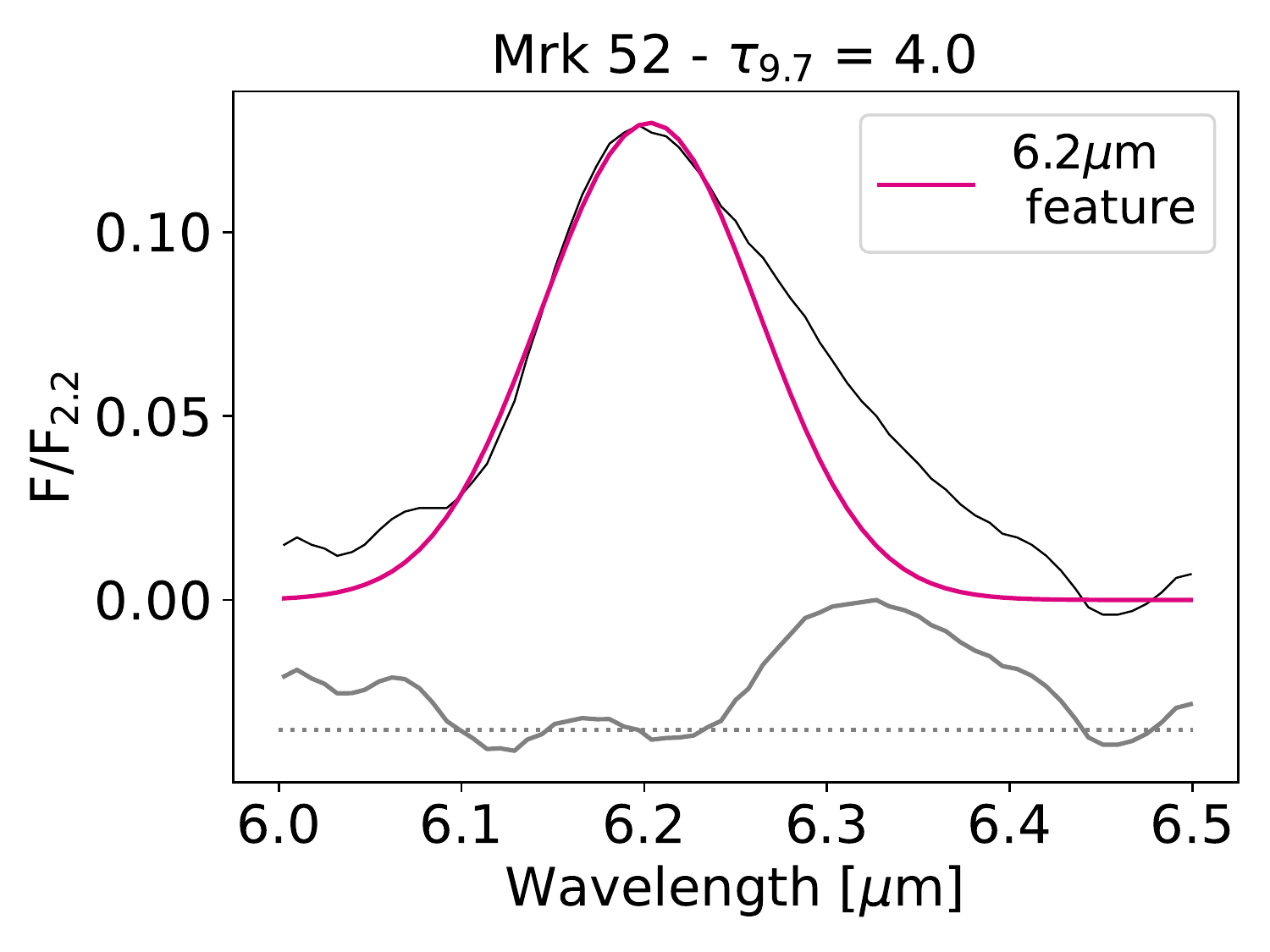}
    \includegraphics[scale=0.38]{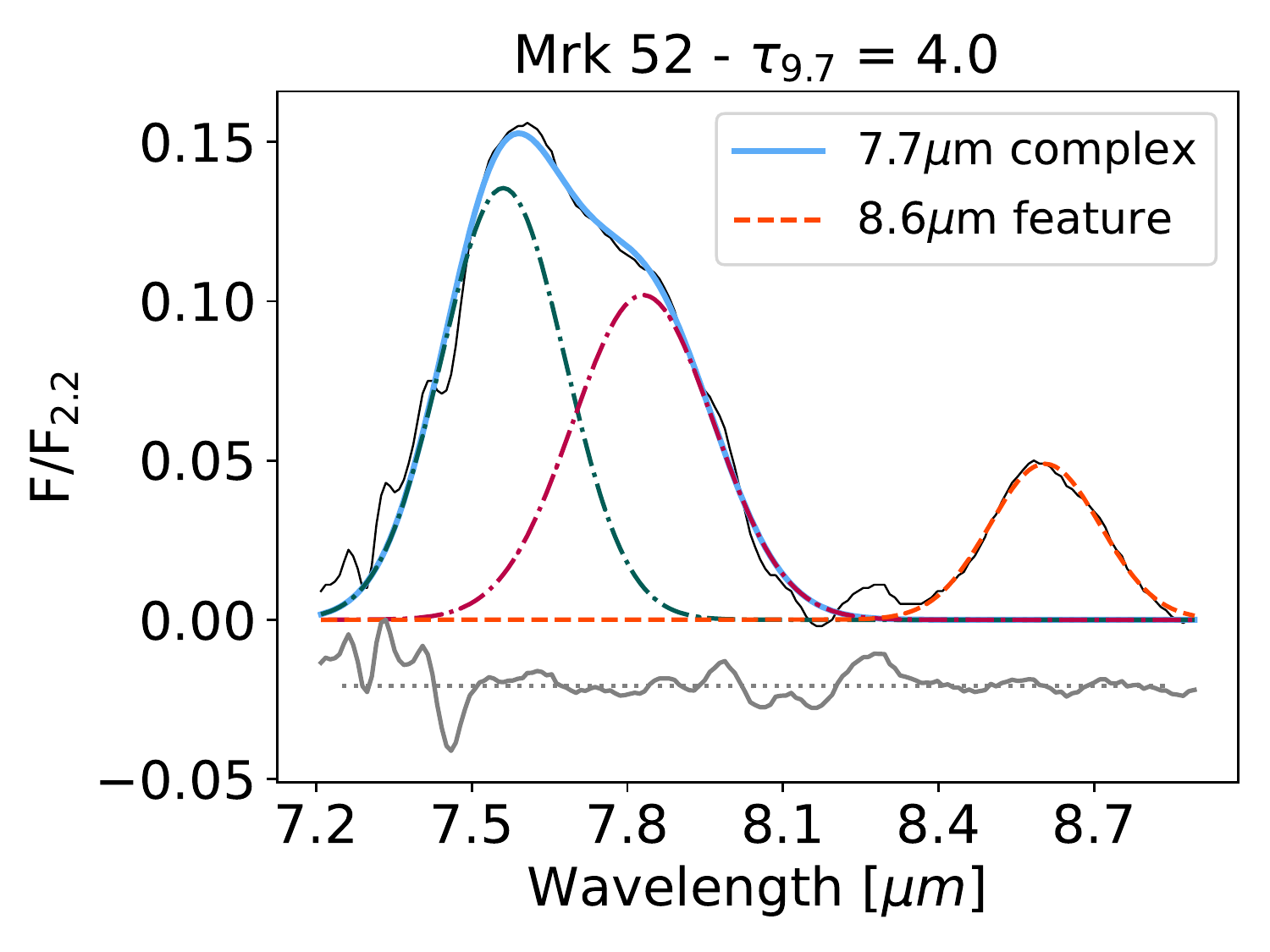}
    
    \includegraphics[scale=0.38]{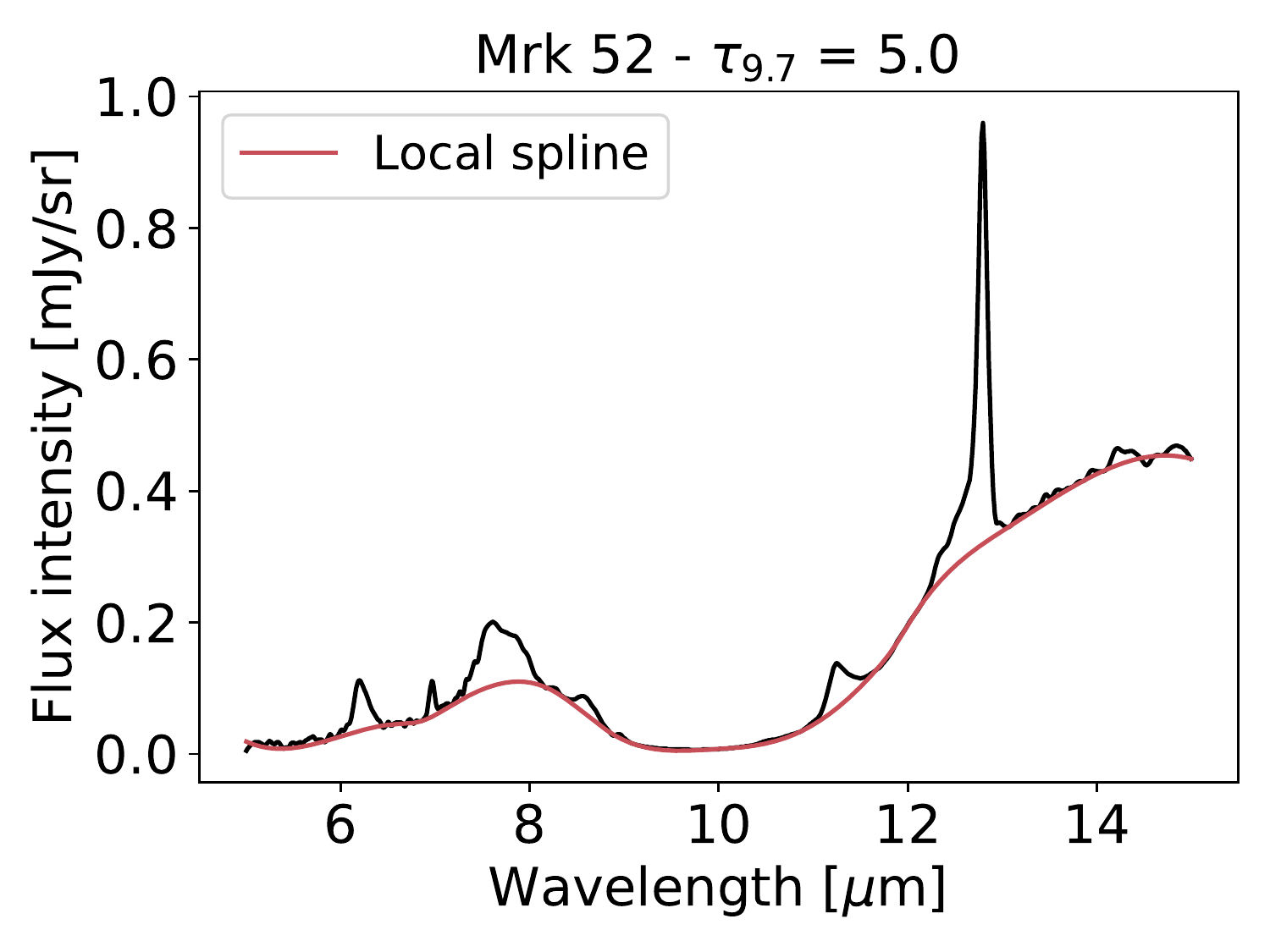}
    \includegraphics[scale=0.38]{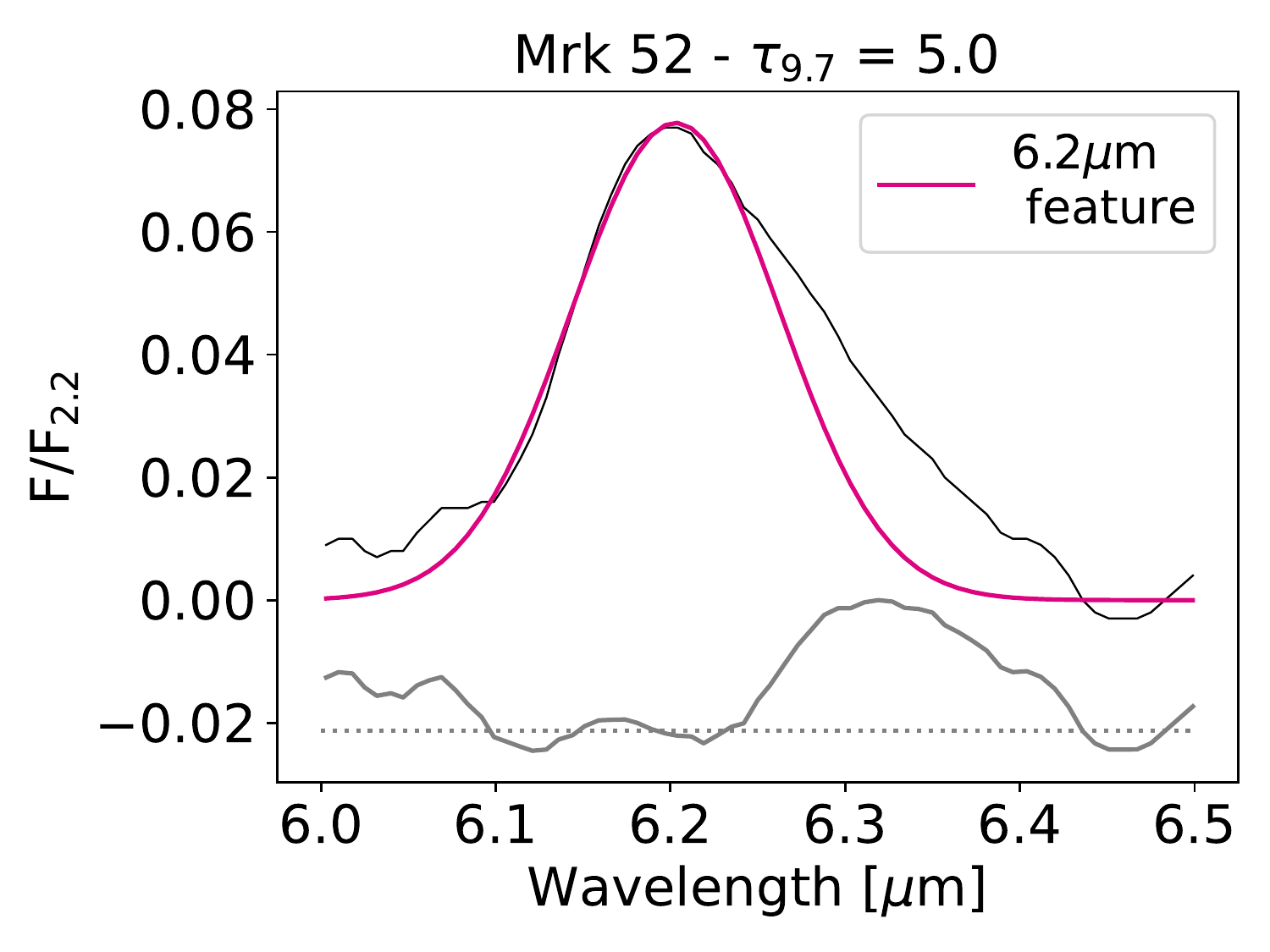}
    \includegraphics[scale=0.38]{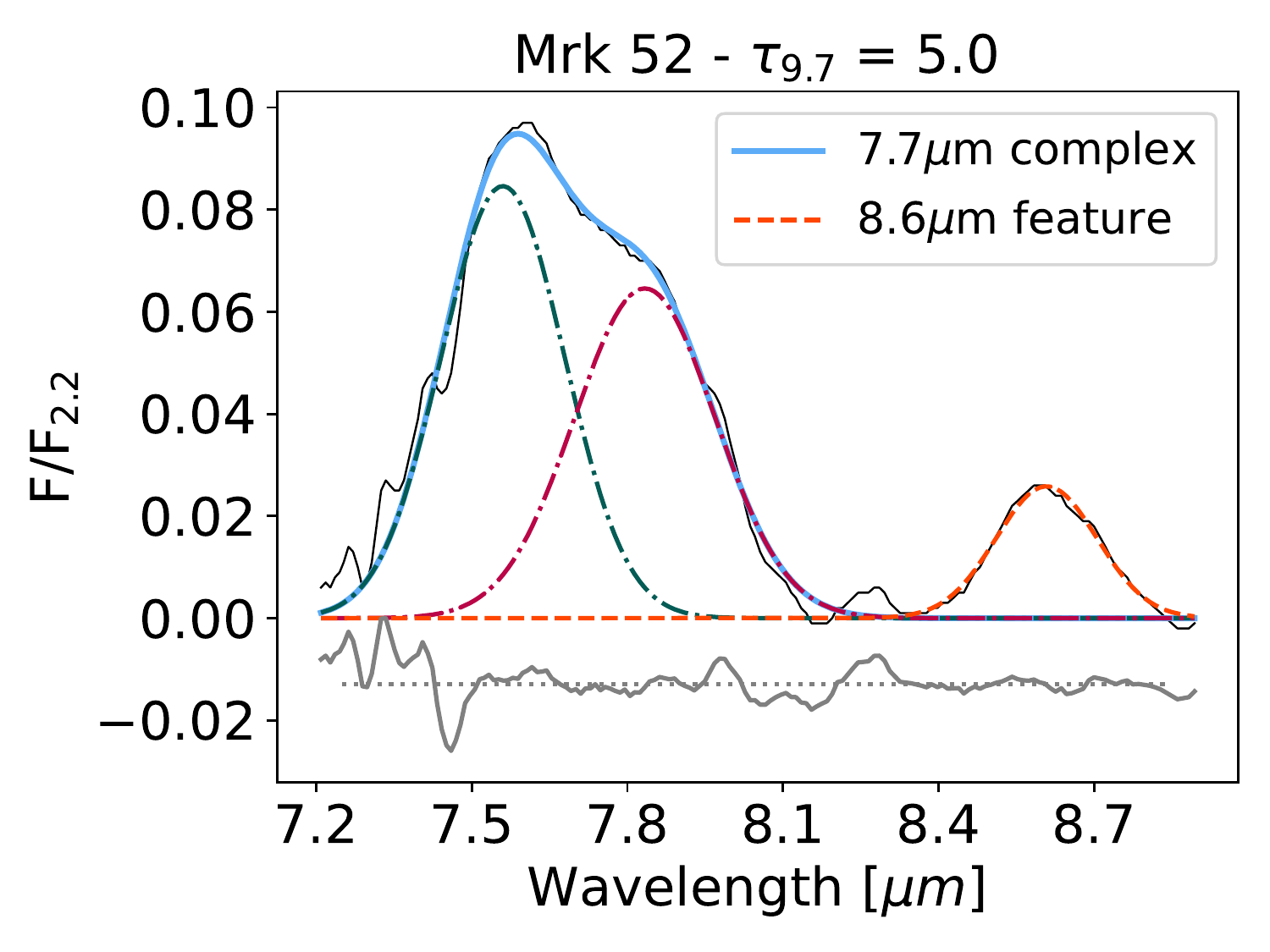}
    
    \includegraphics[scale=0.38]{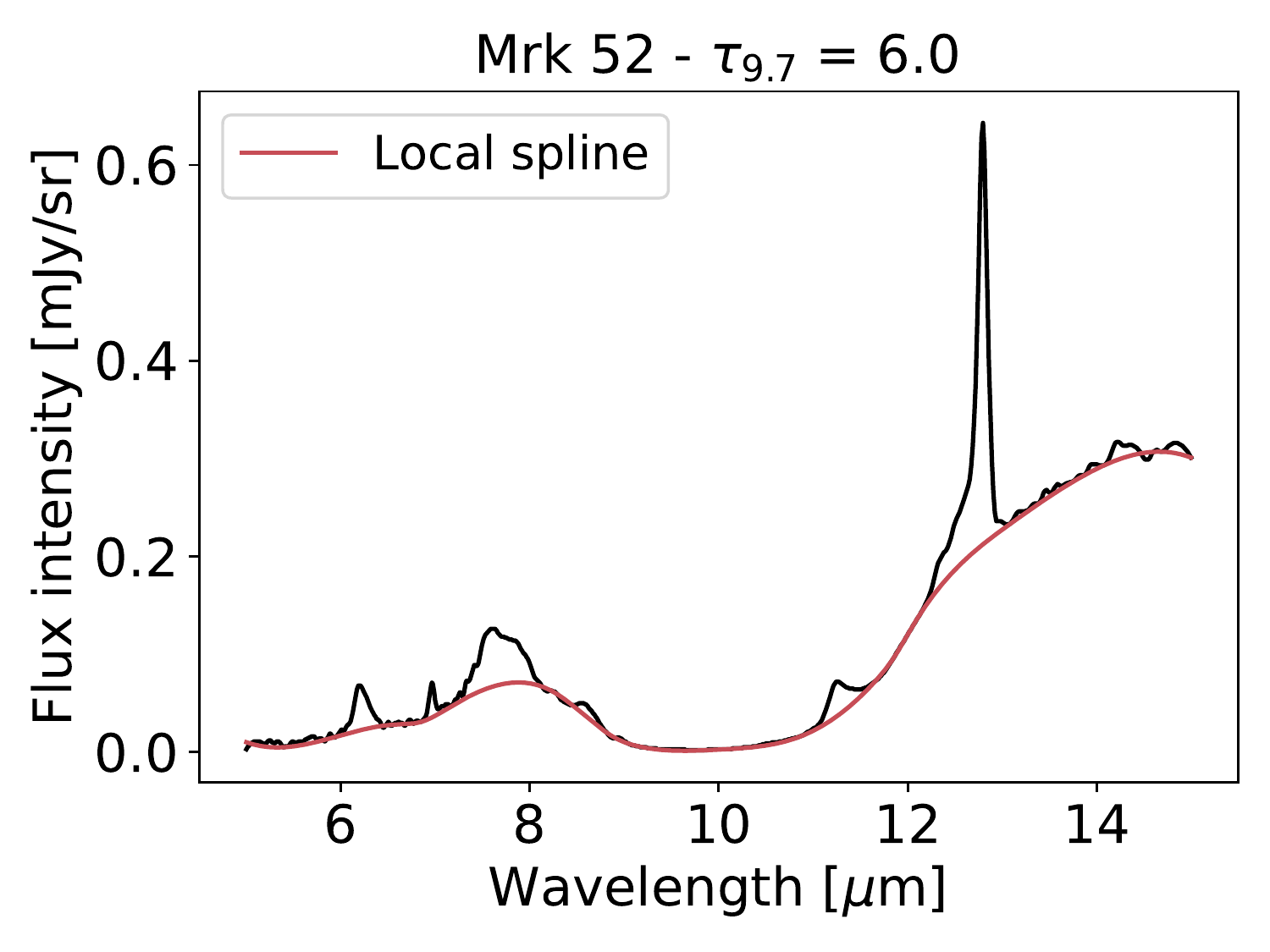}
    \includegraphics[scale=0.38]{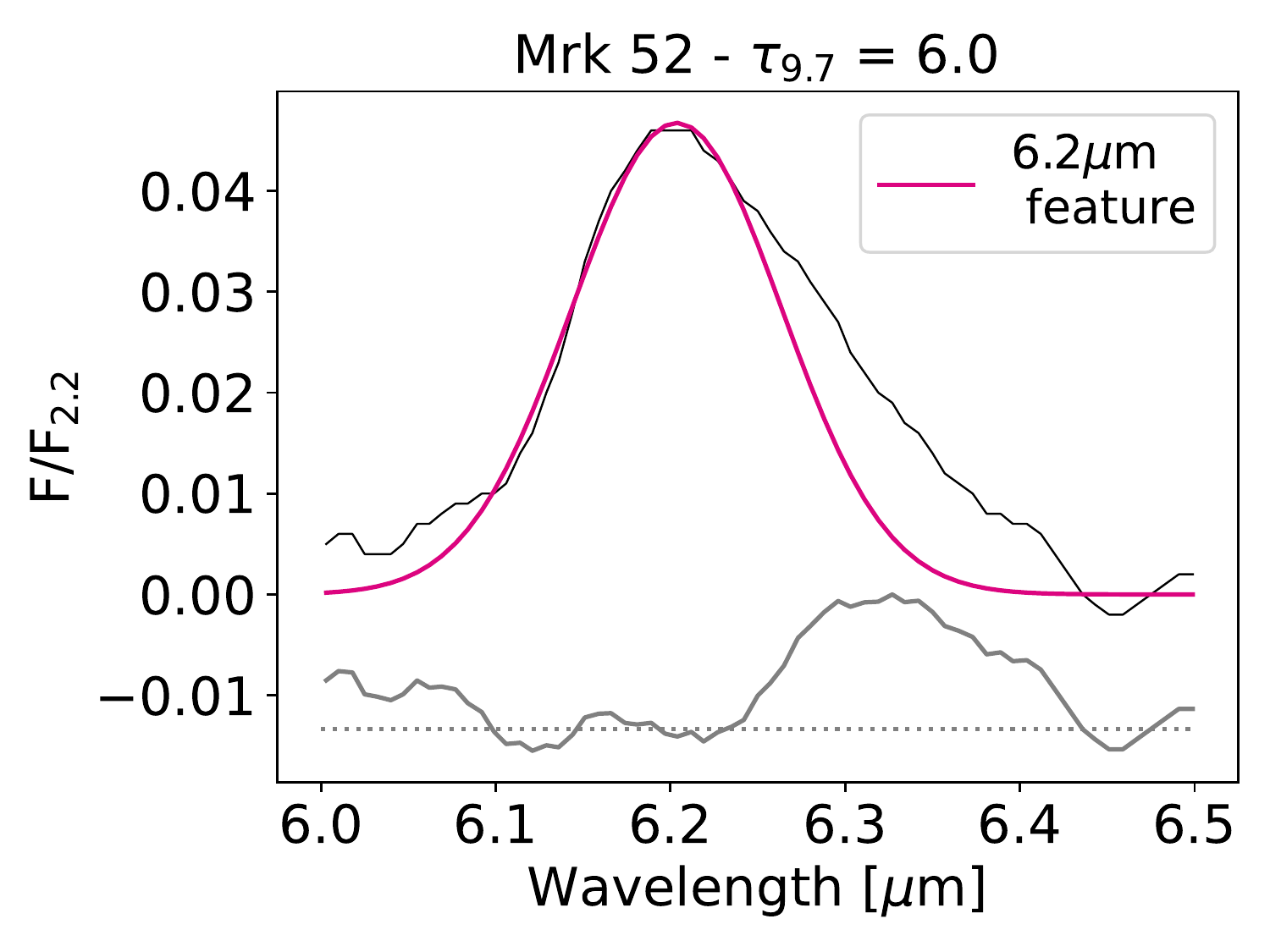}
    \includegraphics[scale=0.38]{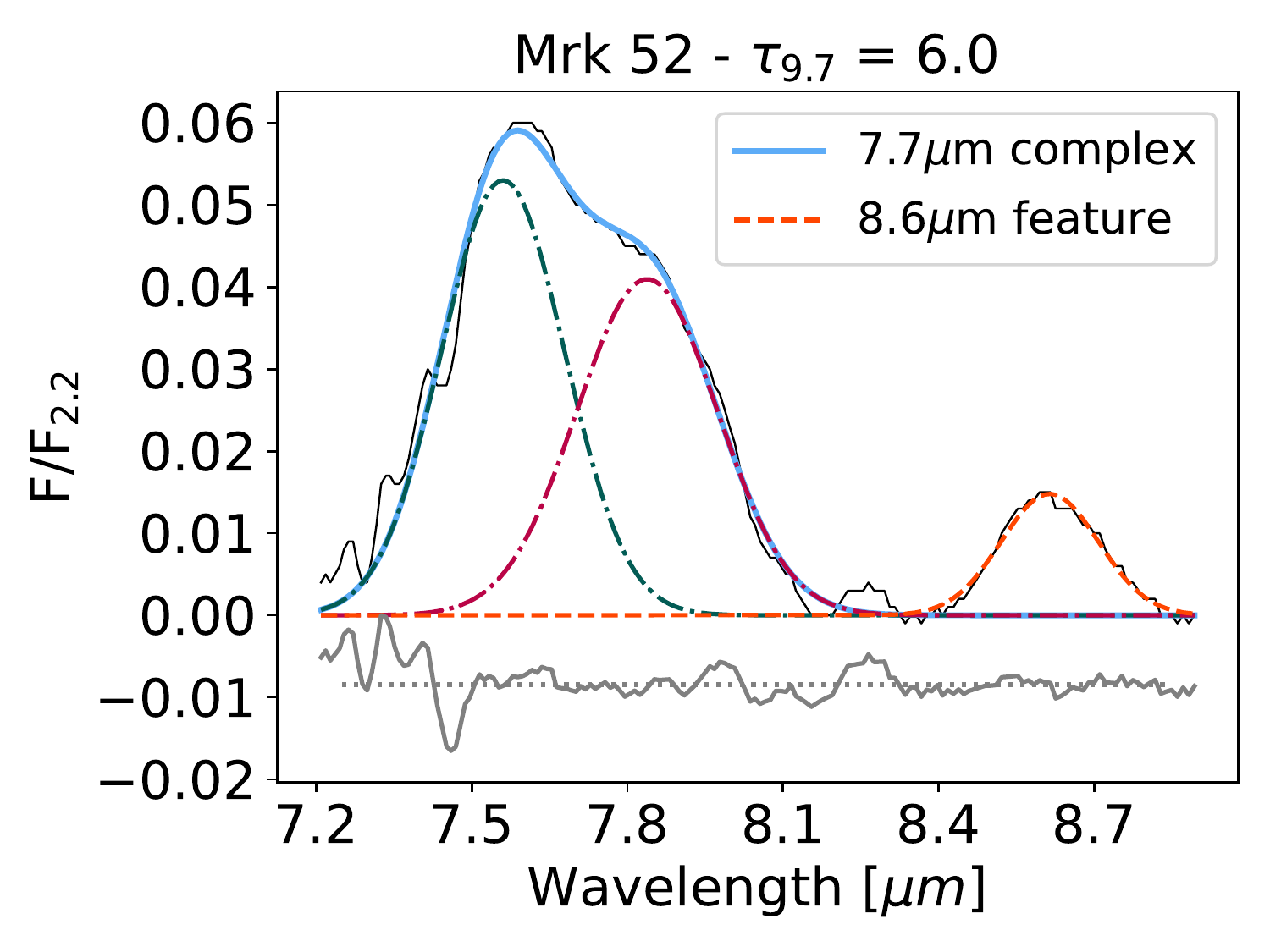}

    \contcaption{ }  
    \end{figure*}

\begin{figure}
    \centering
    \includegraphics[width=\columnwidth,keepaspectratio]{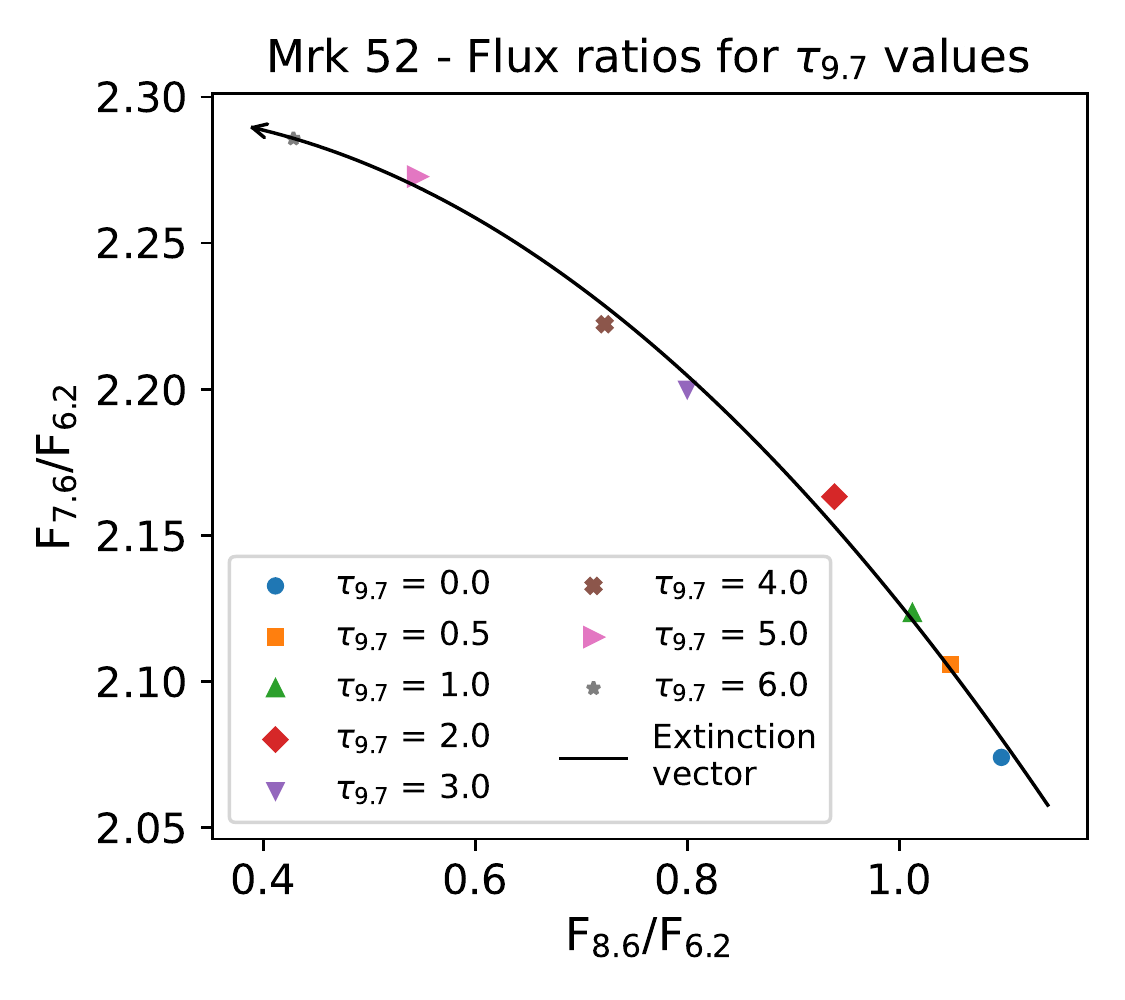}
    \caption{Comparison of the flux intensities normalised to the 6.2$\mu$m band obtained for each $\tau_{9.7}$ values. The extinction vector is represented by the black line and arrow and is  described by Equation~\ref{eq:vector}. } 
    \label{fig:ext-vector}
\end{figure}

The amplitudes and, consequently, the integrated fluxes of the bands seem to be the most sensitive  parameters to the extinction. Therefore, we integrated the fluxes and studied the F$_{7.6}$/F$_{6.2}$ and F$_{8.6}$/F$_{6.2}$ ratios to analyse such influence. Table~\ref{tab:fluxes-ext} and Figure~\ref{fig:ext-vector} show the results. In general, the fluxes of the three bands reduce with higher extinction, as expected. As already mentioned, the 8.6~$\mu$m is the most affected by the extinction, which produces lower F$_{8.6}$/F$_{6.2}$ ratios at higher $\tau_{9.7}$ values. In sequence, the 6.2~$\mu$m band is the second most affected, which leads to higher F$_{7.6}$/F$_{6.2}$ ratios at  higher $\tau_{9.7}$ values. 

To better analyse the effect of the extinction to the F$_{7.6}$/F$_{6.2}$ and F$_{8.6}$/F$_{6.2}$ ratios, we decided to  estimate an extinction vector, which can give an idea of the variability in the flux ratios according to the extinction. We did not take into account the uncertainties in the ratios because this is just a simplified model of the extinction vector and we are not considering the observed flux errors. Our ratios were better fitted with a quadratic function described by Equation~\ref{eq:vector}. This extinction vector is also shown in Figure~\ref{fig:ext-vector}.

\begin{equation}
   \frac{F_{7.6}}{F_{6.2}} \left (\frac{F_{8.6}}{F_{6.2}}  \right ) = - 0.3 \left (\frac{F_{8.6}}{F_{6.2}}  \right )^{2}  + 0.15 \left (\frac{F_{8.6}}{F_{6.2}}  \right )  + 2.28
   \label{eq:vector}
\end{equation}

\begin{table}
\centering
\caption{Integrated flux intensities for the 6.2, 7.7 and 8.6~$\mu$m PAH bands. The values are in mJy/sr. The respective $\tau_{9.7}$ values are also shown.}
\label{tab:fluxes-ext}

\begin{tabular}{ccccc}
\hline
$\tau_{9.7}$ & F$_{6.2}$	&	F$_{7.6}$	&	F$_{7.8}$ &	F$_{8.6}$ \\
\hline
0.0 & 0.135 & 0.280 & 0.237 & 0.148 \\
0.5 & 0.104 & 0.219 & 0.184 & 0.109 \\
1.0 & 0.081 & 0.172 & 0.146 & 0.082 \\
2.0 & 0.049 & 0.106 & 0.090 & 0.046 \\
3.0 & 0.030 & 0.066 & 0.056 & 0.024 \\
4.0 & 0.018 & 0.040 & 0.035 & 0.013 \\
5.0 & 0.011 & 0.025 & 0.022 & 0.006 \\
6.0 & 0.007 & 0.016 & 0.014 & 0.003 \\
\hline
\end{tabular}
\end{table}

\section{Sources -- Fit results}

\begin{table*}
    \centering
    \caption{Best-fit results for the 7.7 and 8.6~$\mu$m bands (Section \ref{sub-analysis}). A is the amplitude, $\lambda\_c$ is the central wavelength and FWHM is the full width at half maximum. The full table is available online.}
    \label{tab:fit7788}
    \begin{tabular}{lcccccc}
    \hline
    Source & $\lambda\_c$ & Err & A & Err & FWHM & Err \\
     & ($\mu$m) & ($\mu$m) & (mJy/sr) & (mJy/sr) & & \\
    \hline
    3C293 & 7.609 & 0.006 & 1.476 & 0.051 & 0.280 & --- \\
     & 7.862 & 0.007 & 1.389 & 0.051 & 0.320 & --- \\
     & 8.613 & 0.005 & 0.529 & 0.021 & 0.261 & 0.013 \\
    3C31 & 7.596 & 0.007 & 2.055 & 0.091 & 0.280 & --- \\
     & 7.829 & 0.008 & 1.887 & 0.094 & 0.320 & --- \\
     & 8.615 & 0.007 & 1.821 & 0.094 & 0.335 & 0.023 \\
    AGN15 & 7.645 & 0.012 & 1.292 & 0.070 & 0.280 & --- \\
     & 7.913 & 0.015 & 1.380 & 0.071 & 0.320 & --- \\
     & 8.660 & 0.006 & 0.229 & 0.015 & 0.200 & 0.017 \\
    Arp220 & 7.586 & 0.008 & 79.727 & 3.712 & 0.280 & --- \\
     & 7.867 & 0.009 & 82.679 & 4.126 & 0.320 & --- \\
     & 8.630 & 0.004 & 8.468 & 0.354 & 0.189 & 0.009 \\
    E12-G21 & 7.562 & 0.010 & 23.021 & 1.191 & 0.280 & --- \\
     & 7.842 & 0.010 & 23.322 & 1.168 & 0.320 & --- \\
     & 8.607 & 0.010 & 23.720 & 1.706 & 0.418 & 0.040 \\
    EIRS-2 & 7.587 & 0.016 & 0.520 & 0.059 & 0.280 & --- \\
     & 7.837 & 0.033 & 0.349 & 0.057 & 0.320 & --- \\
     & 8.684 & 0.028 & 0.456 & 0.103 & 0.348 & 0.095 \\
    GN26 & 7.671 & 0.015 & 0.509 & 0.037 & 0.280 & --- \\
     & 7.950 & 0.028 & 0.352 & 0.040 & 0.320 & --- \\
     & 8.652 & 0.015 & 0.148 & 0.030 & 0.154 & 0.036 \\
    IC342 & 7.583 & 0.009 & 104.704 & 6.068 & 0.280 & --- \\
     & 7.812 & 0.009 & 102.705 & 6.013 & 0.320 & --- \\
     & 8.603 & 0.003 & 60.704 & 1.439 & 0.283 & 0.009 \\
    IRAS02021-2103 & 7.596 & 0.009 & 3.860 & 0.216 & 0.280 & --- \\
     & 7.850 & 0.010 & 4.395 & 0.220 & 0.320 & --- \\
     & 8.620 & 0.007 & 3.024 & 0.158 & 0.353 & 0.024 \\
    \vdots & \vdots & \vdots & \vdots & \vdots & \vdots \\
    UGC7064 & 7.611 & 0.015 & 21.836 & 1.693 & 0.280 & --- \\
     & 7.903 & 0.025 & 16.155 & 1.490 & 0.320 & --- \\
     & 8.572 & 0.004 & 14.587 & 0.466 & 0.331 & 0.014 \\
    \hline 
    \end{tabular}

\end{table*}

\begin{table*}
    \centering
    \caption{Integrated flux intensities for the 6.2, 7.7 and 8.6~$\mu$m PAH bands. The values are in mJy/sr. The full table is available online.}
    \label{tab:fluxes}    
    \begin{tabular}{lcccccccc}
    \hline
    Source & F$_{6.2}$	&	Err	&	F$_{7.6}$	&	Err	&	F$_{7.8}$	&	Err	&	F$_{8.6}$	&	Err	\\
    \hline
    3C293	&	0.629	&	0.020	&	1.476	&	0.051	&	1.380	&	0.049	&	0.529	&	0.021	\\
    3C31	&	1.405	&	0.014	&	2.054	&	0.091	&	1.881	&	0.092	&	1.812	&	0.086	\\
    AGN15	&	0.368	&	0.005	&	1.292	&	0.070	&	1.356	&	0.062	&	0.229	&	0.015	\\
    Arp220	&	32.624	&	0.751	&	79.680	&	3.720	&	82.089	&	3.973	&	8.468	&	0.354	\\
    E12-G21	&	13.416	&	0.430	&	22.994	&	1.197	&	23.224	&	1.138	&	23.143	&	1.254	\\
    EIRS-2	&	0.176	&	0.008	&	0.520	&	0.059	&	0.348	&	0.055	&	0.448	&	0.074	\\
    GN26	&	0.139	&	0.004	&	0.509	&	0.037	&	0.340	&	0.032	&	0.148	&	0.030	\\
    IC342	&	38.532	&	0.978	&	104.637	&	6.081	&	102.484	&	5.947	&	60.651	&	1.413	\\
    IRAS02021-2103	&	2.853	&	0.055	&	3.858	&	0.216	&	4.373	&	0.213	&	2.999	&	0.139	\\
    \vdots & \vdots & \vdots & \vdots & \vdots & \vdots  & \vdots & \vdots & \vdots \\
    UGC7064	&	9.497	&	0.091	&	21.830	&	1.695	&	15.922	&	1.323	&	14.511	&	0.436 \\
    \hline
    \end{tabular}

\end{table*}


\begin{table}
    \centering
    \caption{Distribution of the galaxies into the Peeters' classes for three PAH bands. The classification for the 6.2~$\mu$m band was extracted from  \citet{Canelo18}. The full table is available online.}
    \label{tab:results-classes}
    \begin{tabular}{lccc}
    \hline
    Source & 6.2 $\mu$m & 7.7 $\mu$m & 8.6 $\mu$m \\
      & Class & Class & Class \\
    \hline
    3C293 & B & A & B \\
    3C31 & A & A & B \\
    AGN15 & B & B & B \\
    Arp220 & A & B & B \\
    E12-G21 & A & B & B \\
    EIRS-2 & A & A & B \\
    GN26 & B & A & B \\
    IC342 & A & A & B \\
    IRAS02021-2103 & B & B & B \\
    \vdots & \vdots & \vdots & \vdots \\
    UGC7064 & B & A & A\\
    \hline
    \end{tabular}   

\end{table}



\bsp	
\label{lastpage}
\end{document}